\documentclass[10pt,twocolumn,showpacs,preprintnumbers,amsmath,amssymb,aps,prb,
superscriptaddress,floatfix,footinbib]{revtex4-1}

\usepackage{graphicx}
\usepackage{amsmath}
\usepackage{amssymb}
\usepackage{hyperref}

\usepackage[active]{srcltx}
\usepackage{lipsum}
\usepackage{color}

\newcommand{\matrg}[1]{\underline{\boldsymbol{#1}}}
\newcommand{\SIGTEN}{\mbox{$\matrg \sigma$}}
\newcommand{\TORTEN}{\mbox{$\matrg t$}}
\newcommand{\EETEN}{\mbox{$\matrg p$}}

\begin{document}


\title{Chirality-induced linear response properties in non-coplanar Mn$_3$Ge}
\author{Sebastian Wimmer, Sergiy Mankovsky, Hubert Ebert} 
\affiliation{%
  Department  Chemie,  Physikalische  Chemie,  Universit\"at  M\"unchen,
  Butenandtstr. 5-13, 81377  M\"unchen, Germany\\}


\begin{abstract}
Taking the non-collinear antiferromagnetic hexagonal 
Heusler compound Mn$_3$Ge as a reference system, the contributions to linear response 
phenomena arising solely from the chiral coplanar and non-coplanar spin 
configurations are investigated. Orbital moments, X-ray absorption, anomalous 
and spin Hall effects, as well as corresponding spin-orbit torques and Edelstein 
polarizations are studied depending on a continuous variation of the polar angle 
relative to the Kagome planes of corner-sharing triangles between the 
non-collinear antiferromagnetic and the ferromagnetic limits. By scaling the 
speed of light from the relativistic Dirac case to the non-relativistic limit 
the chirality-induced or topological contributions can be identified by
suppressing the spin-orbit coupling.
\end{abstract}

\maketitle

\section{Introduction \label{sec:Intro}}

Chiral magnetic order, its origins and consequences, continues to be a fascinating area of 
current solid state science.\cite{BHB+07,TGK+08,Bra12,SBH+17,SGP+18} 
Particularly intriguing is the occurrence of mesoscopic quasiparticles formed by 
a continuously varying non-coplanar spin texture with defined topology, 
so-called Skyrmions.\cite{Sky62,MBJ+09,NPB+09} Their properties, creation, as 
well as detection and manipulation is a very active field of 
research,\cite{HBM+11,DHPH14,RKH+15,MMR+16,WLK+16,NKM+17} motivated by potential 
future applicability in magnetic storage.\cite{YKZ+12,SCR+13,FCS13,TMZ+14} 
Electric-field-induced transport plays an important role in this context, as 
corresponding charge and spin currents can be utilized to detect and manipulate 
Skyrmions\cite{YKZ+12,SCR+13,FRC17} and Antiskyrmions.\cite{KN16} Due to the 
non-coplanar spin texture a so-called emergent electromagnetic field arises, 
that leads to chirality-induced or \emph{topological}, in the sense of arising from the 
real-space topology of the spin configuration, contributions to phenomena 
commonly associated with spin-orbit coupling. The most fundamental of them are 
the occurrence of topological orbital 
moments\cite{SN01,NIF03,HWD+15,HFN+16,DBB+16a,HFBM17} and of the related 
topological Hall effect (THE).\cite{TOY+01,TK02,ON02,BDT04,TK10,SRB+12,NT13,FFB+14} A 
corresponding spin Hall effect arising from the real-space topology of the spin 
texture\cite{YLB+15} is of particular interest in antiferromagnetic skyrmions, 
where the THE vanishes.\cite{BFBM17}
%

A second intensively investigated type of chiral magnetic order is that of bulk 
antiferromagnets with non-collinear spin arrangements. The anomalous Hall effect 
(AHE) has been studied extensively in such compounds both 
theoretically\cite{SN01,TK10,CNM14,KF14a,ZSY+17} and 
experimentally.\cite{BW12,SFWL14,NKH15,KTN16,NFS+16,SWA+17,LZL+17} Its relation 
to the magneto-optic Kerr effect of chiral magnets\cite{OM13,FGZ+15} and the 
X-ray circular dichroism\cite{DBB+16a,WMM+18}, both connected to the optical 
conductivity tensor, suggests an alternative, magneto-optical approach to 
non-collinear magnetic order. Of particular relevance to the field are the 
hexagonal Heusler compounds Mn$_3X$ with $X = $Ga, Ge, and Sn, in which the AHE 
has recently been confirmed experimentally.\cite{NKH15,KTN16,NFS+16} Its spin-polarized 
counterpart, the spin Hall effect (SHE), in these and other non-collinear 
antiferromagnets also stimulated theoretical 
efforts\cite{Gom15,ZSY+17,ZZFY17a,ZZS+17} and has been measured in the achiral 
cubic system Mn$_3$Ir.\cite{MCA+14,ZHY+16a} In Ref.~\onlinecite{ZHY+16a} it has 
been furthermore shown, that the SHE contributes to the so-called spin-orbit 
torque (SOT), the current-induced magnetic torque that can be utilized to 
efficiently switch the magnetization. Thermally-induced analogues to the AHE and 
SHE, the anomalous and spin Nernst effects have been studied in Mn$_{3}X$ ($X$ = 
Sn, Ge, Ga) from first principles using the Berry curvature approach and a 
Mott-like formula.\cite{GW17} The anomalous Nernst effect could in fact be measured recently in 
Mn$_3$Sn.\cite{ITK+17}

The merger of these two fields called \emph{topological antiferromagnetic
spintronics}\cite{SMYM18} aims to explore the potential of topologically 
protected quasiparticles with non-trivial real- or momentum-space topology.
This work contributes to this by a first-principles study on the chirality-induced or 
topological contributions to orbital moments, X-ray absorption spectra, anomalous 
and spin Hall effect, as well as to spin-orbit torques and the closely related 
Edelstein effect (EE). Two coplanar non-collinear antiferromagnetic spin 
structures in Mn$_3$Ge, one chiral and the other achiral, will be used as basis 
for investigations on the impact of non-coplanarity by rotating the magnetic 
moments out of the Kagome planes. Scaling the speed of light allows assessing 
the topological contributions to the various effects in absence of spin-orbit 
coupling. This will be accompanied by an analysis of the corresponding 
symmetry-restricted response tensor shapes.
%

The manuscript is organized as follows: In section~\ref{sec:Methods} the 
underlying methods used for obtaining the results in section~\ref{sec:Results} 
will be outlined. The crystallographic and magnetic structures will be discussed 
in section~\ref{ssec:SYM}, including the corresponding symmetry-restricted 
tensor shapes for electrical and spin conductivity as well as spin-orbit 
torkance and Edelstein polarization. Topological orbital moments and their 
signatures in X-ray absorption spectra are the subjects of 
sections~\ref{ssec:OMTs} and \ref{ssec:XAS}, respectively. The 
chirality-induced contributions to the anomalous and spin Hall effect will be 
discussed in section~\ref{ssec:TXHE}, corresponding results for the 
spin-orbit torque and the Edelstein effect will be presented in 
section~\ref{ssec:TSOTnTEE}. Finally, hypothetical non-coplanar antiferromagnets 
will be investigated in section~\ref{ssec:ncpAFM}. A brief summary and outlook 
will be made at the end (\ref{sec:Concl}), additional information can be found 
in Appendix ~\ref{sec:AppA}.

\section{Methods \label{sec:Methods}}

The space-time symmetry analysis of the linear response tensors for 
charge\cite{Kle66,SKWE15a} and spin conductivity\cite{SKWE15a}, spin-orbit 
torque\cite{WCS+16a} and Edelstein polarization\cite{WCE18} performed in this 
work is based on the magnetic space group determined using the software 
FINDSYM\cite{ISOTROPY,SH05}.
Calculations for the corresponding linear response quantities have been
done on the basis of Kubo's linear response formalism
\cite{Kub56,Kub57,But85,CB01a,KCE15}.
These were done in a fully relativistic way using spin-polarized
relativistic Korringa-Kohn-Rostoker (SPR-KKR) \cite{SPRKKR} electronic
structure method within framework of the local spin
density approximation (LSDA).
Explicite expressions used for the calculations of the response
functions can be found in the literature (charge conductivity
\cite{Kub56,Kub57,But85}, spin conductivity
\cite{CB01a,KCE15}, spin-orbit torcance  \cite{WCS+16a}, Edelstein
effect\cite{WCE18}, X-ray absorption formalism\cite{Ebe96})
For the calculation of electric-field induced
response properties the Kubo-St\v{r}eda\cite{Str82} formula has 
been used throughout. To study the impact of the spin texture in absence of 
spin-orbit coupling, the non-relativistic limit of the Dirac formalism has been 
explored by scaling the speed of light. 


\section{Results \label{sec:Results}}

\subsection{Magnetic structure and symmetry \label{ssec:SYM}}

\begin{figure}[hbt]
\begin{center}
  {\includegraphics[angle=0,width=0.9\linewidth,clip]{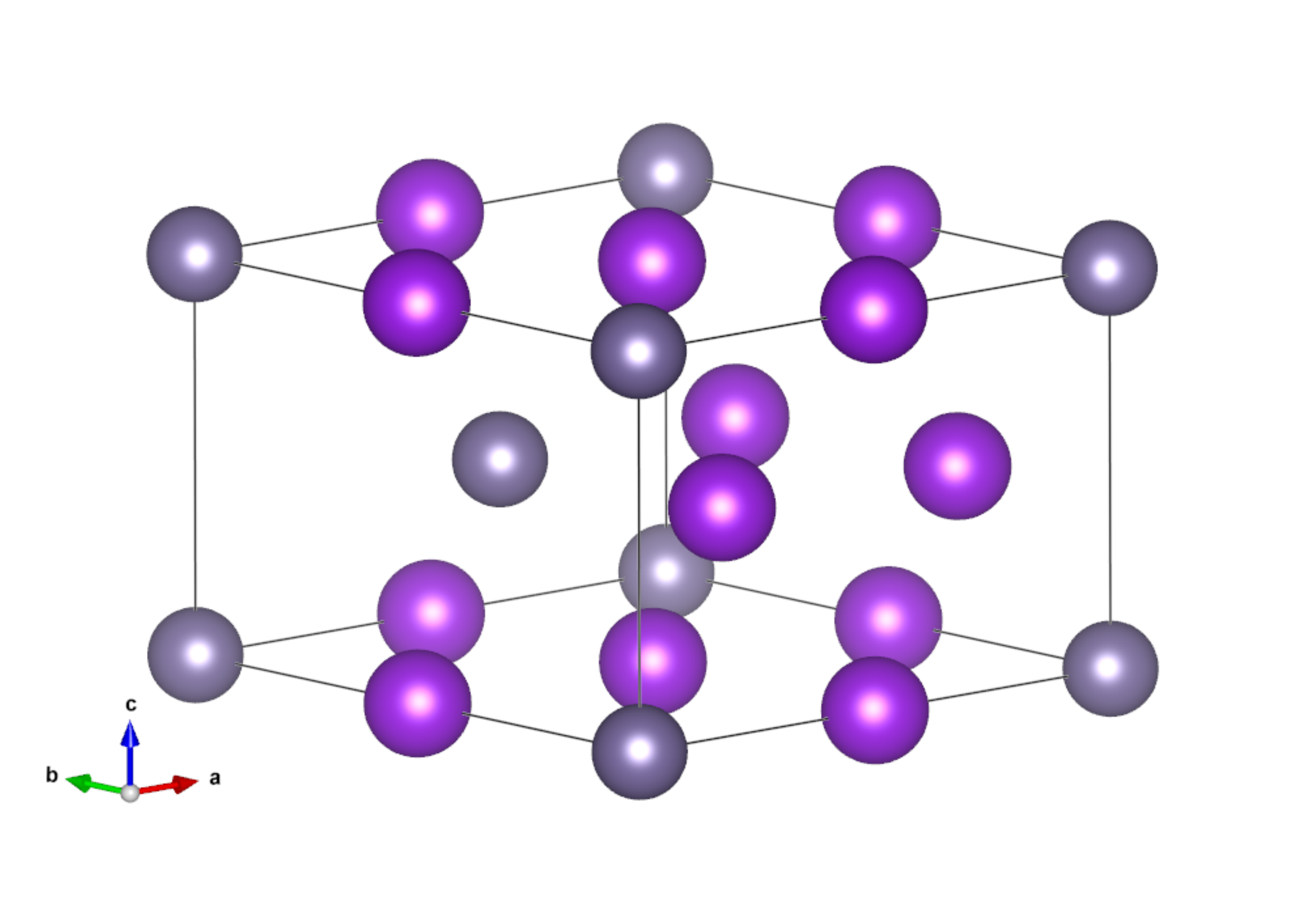}}
 \end{center}
 \caption{\label{fig:Mn3Ge} Hexagonal unit cell of Mn$_3$Ge with 
space group $P6_3/mmc$ (labeled NM in the following). The Mn atoms on the 
Wyckoff positions $6h$ are colored in magenta (dark gray) and Ge atoms (Wyckoff 
positions $2h$) are colored in light gray.\cite{VESTA}}
\end{figure}

\begin{figure}[hbt]
\begin{center}
  {\includegraphics[angle=0,width=0.9\linewidth,clip]{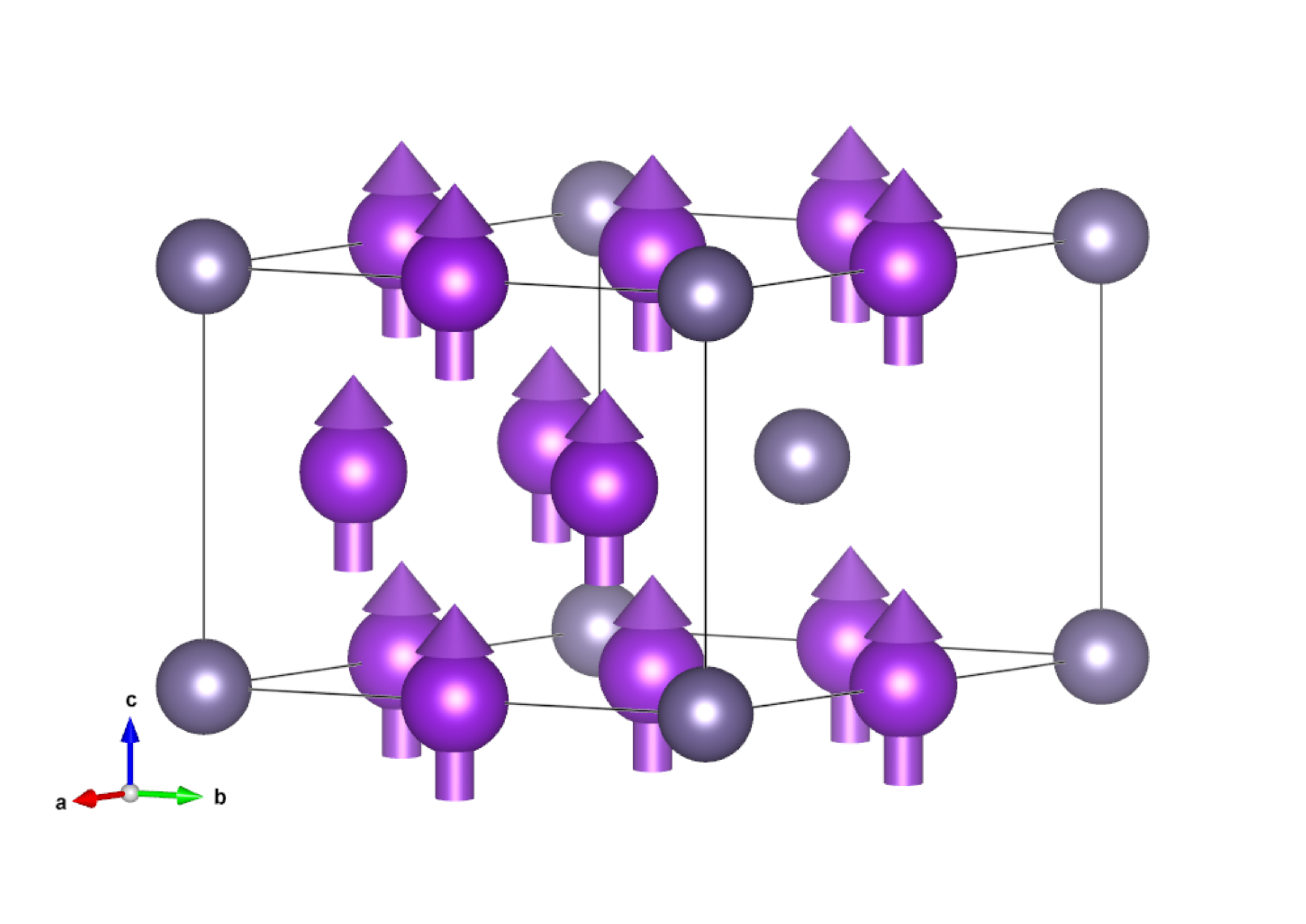}}
  \end{center}
 \caption{\label{fig:FM} Unit cell of hexagonal ferromagnetic 
(FM) Mn$_3$Ge with magnetic space group $P6_3/mm'c'$. Use of colors as in 
Fig.~\ref{fig:Mn3Ge}, magnetic moments on Mn sites are indicated as vectors.\cite{VESTA}}
\end{figure}

The hexagonal Mn$_3$Ge compound crystallizes, as its siblings Mn$_3$Sn and 
Mn$_3$Ga, in the D0$_{19}$ structure with space group $P6_3/mmc$. The non-magnetic 
unit cell is shown in Fig.~\ref{fig:Mn3Ge} and will be labeled NM in the 
following. The Mn atoms on the Wyckoff positions $6h$ in the \{0001\} planes 
colored in magenta (dark gray) form triangular, so-called Kagome lattices, 
stacked alternatingly along the [0001] ($z$) direction. Ge atoms occupying 
the Wyckoff positions $2h$ are colored in light gray.
%
%
%
Figure~\ref{fig:FM} shows the situation of a field-aligned ferromagnetic
structure with all moments (only shown for Mn sites) oriented along the [0001] or $z$
direction ($c$ axis of the unit cell). The corresponding magnetic space group is
$P6_3/mm'c'$. This structure will be labeled FM in the following.
%
%

\begin{figure}[hbt]
  \begin{center}
    {\includegraphics[angle=0,width=0.825\linewidth,clip]{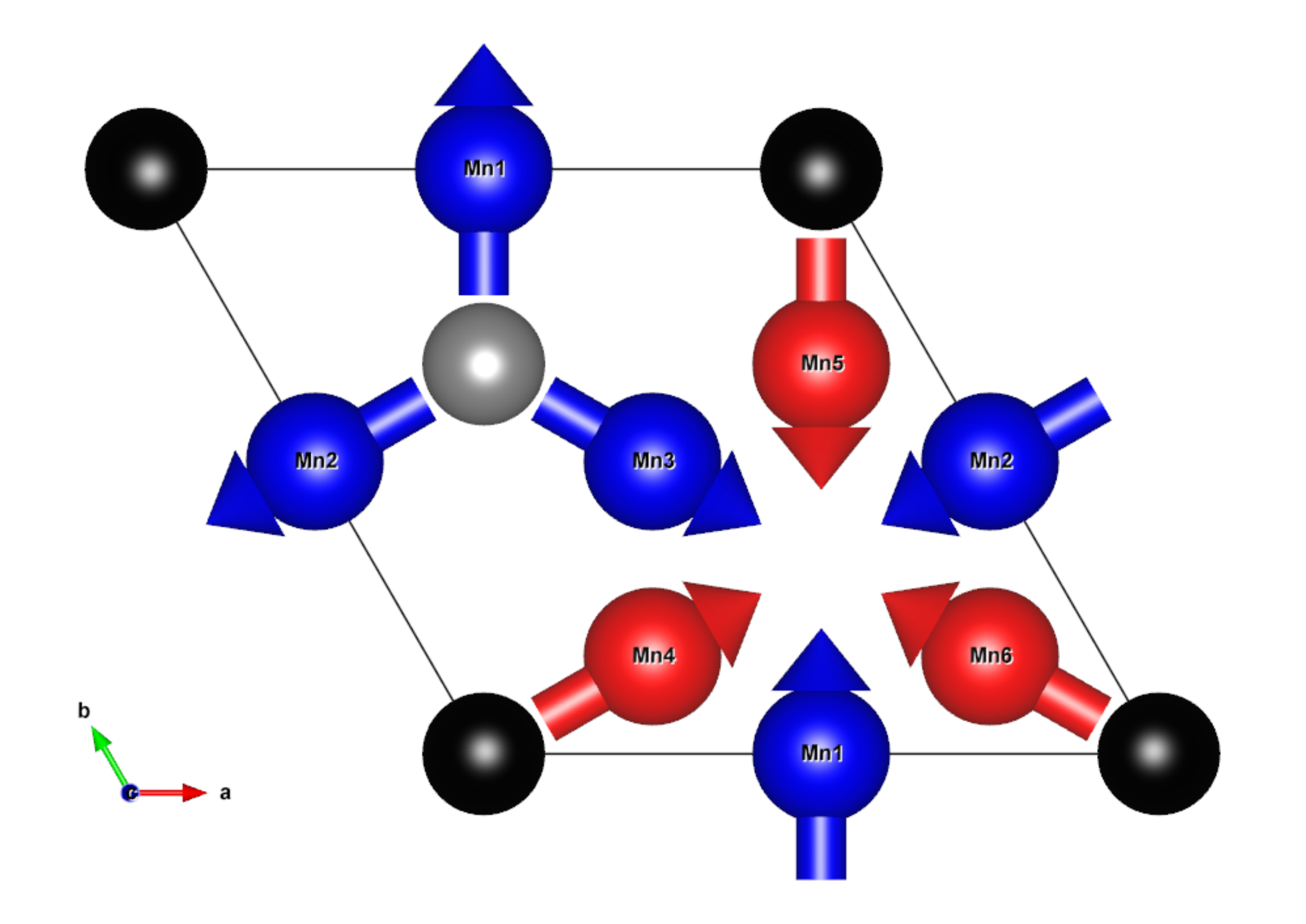}}
     {\includegraphics[angle=0,width=0.825\linewidth,clip]{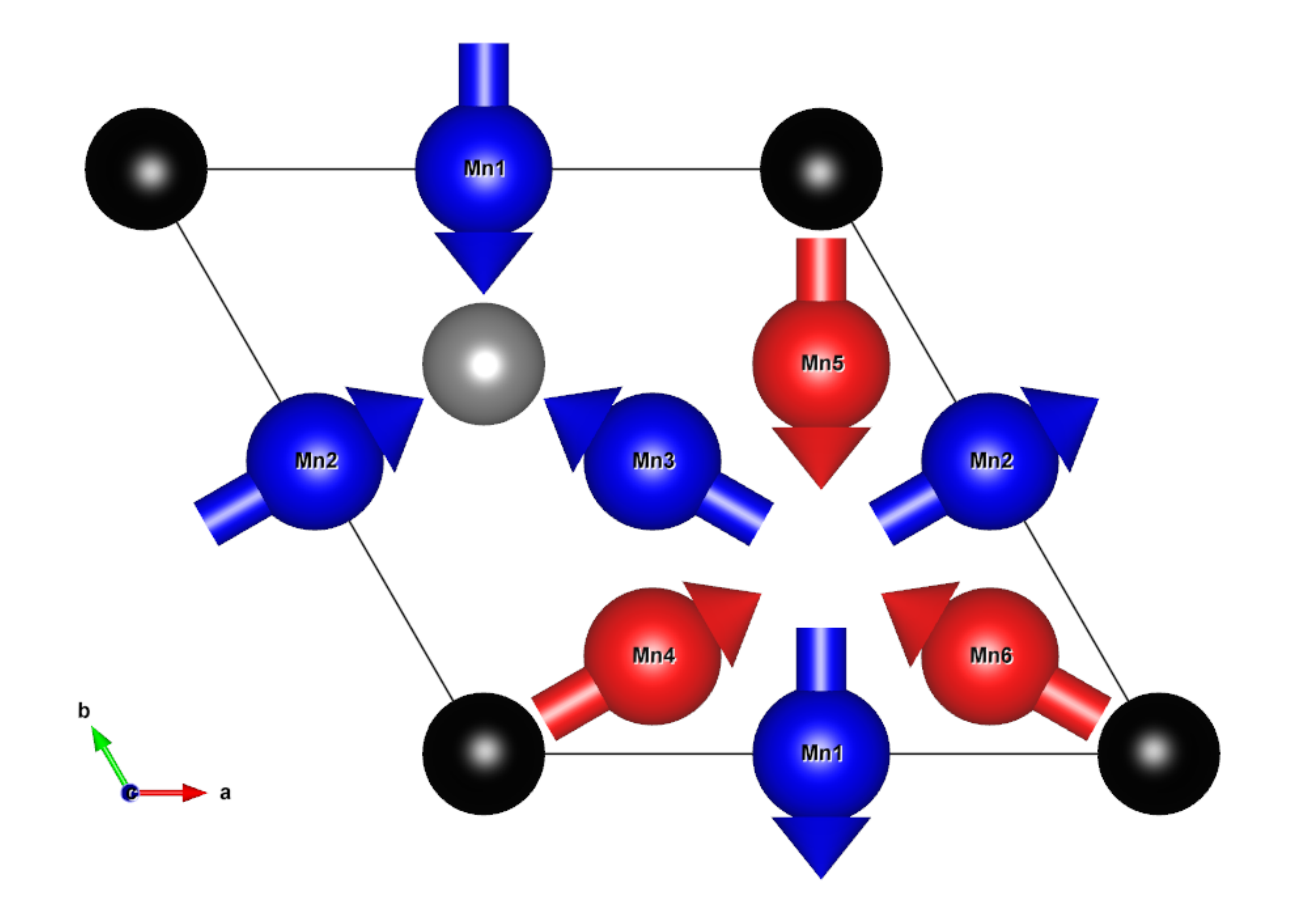}}
\end{center}
   \caption{\label{fig:ncAFMx} Non-collinear coplanar (nc) antiferromagnetic 
reference structures of Mn$_3$Ge, ncAFM0 (top) and ncAFM9 (bottom).  
The coplanar Mn moments in alternating Kagome planes are colored red and blue. The 
achiral structure ncAFM9 in the lower panel is obtained from ncAFM0 by 
reversing all moments in one plane (blue).\cite{VESTA}}
\end{figure}

A number of non-collinear but coplanar antiferromagnetic alignments of the
moments have been discussed for Mn$_3$Ge and related compounds in the
literature (cf. Ref.~\onlinecite{ZYW+13} and references therein). Recently an
overview on the properties of actual and hypothetical spin-compensated configurations has been
given by the present authors.\cite{WMM+18} Two coplanar ones of these structures discussed
therein, both hypothetical, are shown in Fig.~\ref{fig:ncAFMx}. The one in the
upper panel, labeled ncAFM0, has the moments in the two alternating Kagome
planes, indicated by different colors (red and blue), pointing towards the
center of the triangles formed by the Mn atoms.
The two magnetic sub-lattices are connected, e.g., by a $6_3$
screw rotation about an axis going through the center of both triangles, but
also by inversion with respect to the midpoint between these centers followed by time reversal.
Reversing all moments in one layer (here blue), one obtains the structure
ncAFM9 shown in the bottom panel. Here the operation connecting the two
sub-lattices involves an additional time-reversal ($6_3'$), leading to a
centrosymmetric or achiral structure with an inversion center half-way along
$c$. Both structures will serve as references for the investigations on the
consequences of non-coplanarity of the Mn moments in this work.
%
%

\begin{figure}
\begin{center}
{\includegraphics[angle=0,width=0.8\linewidth,clip]{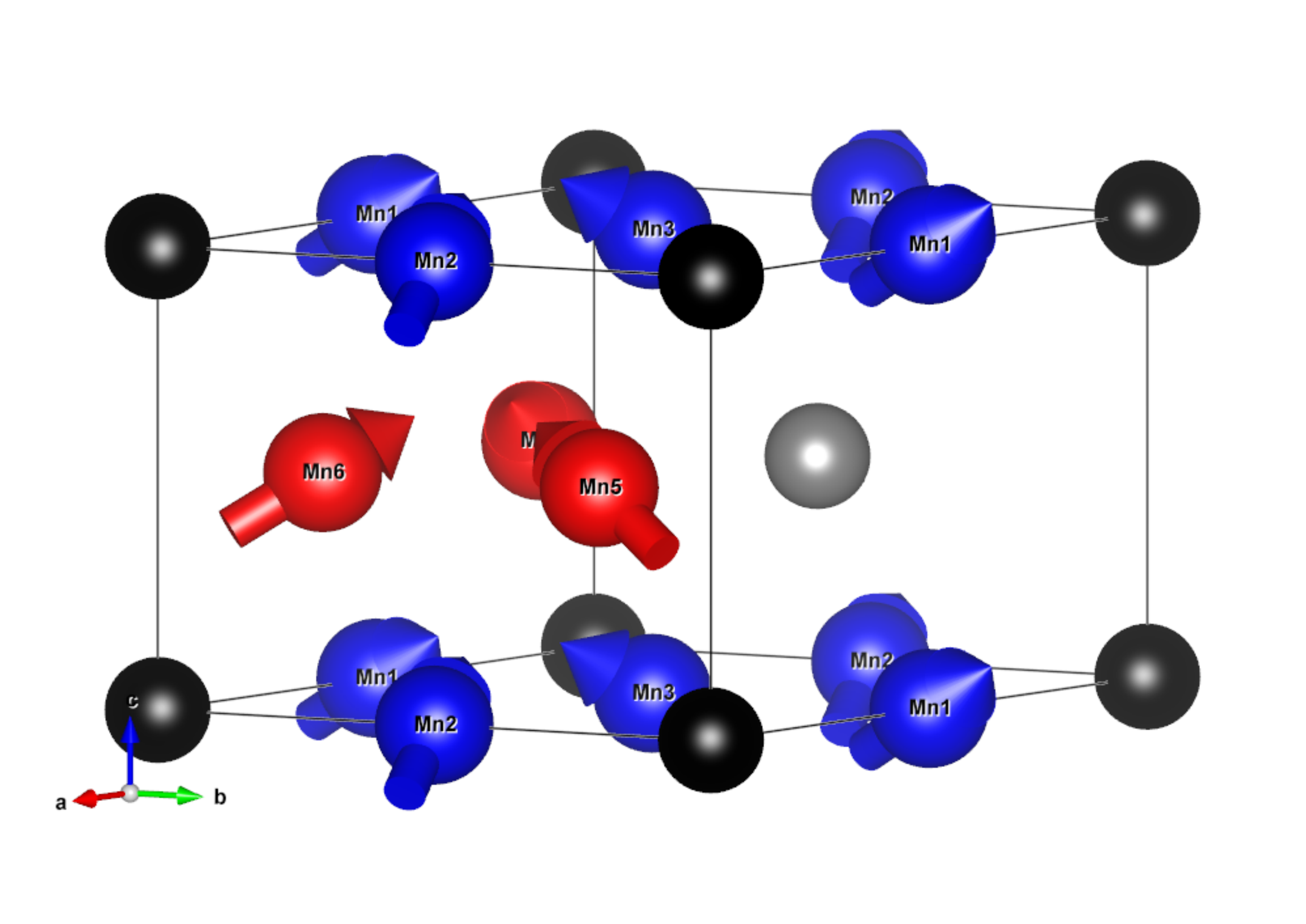}}
{\includegraphics[angle=0,width=0.9\linewidth,clip]{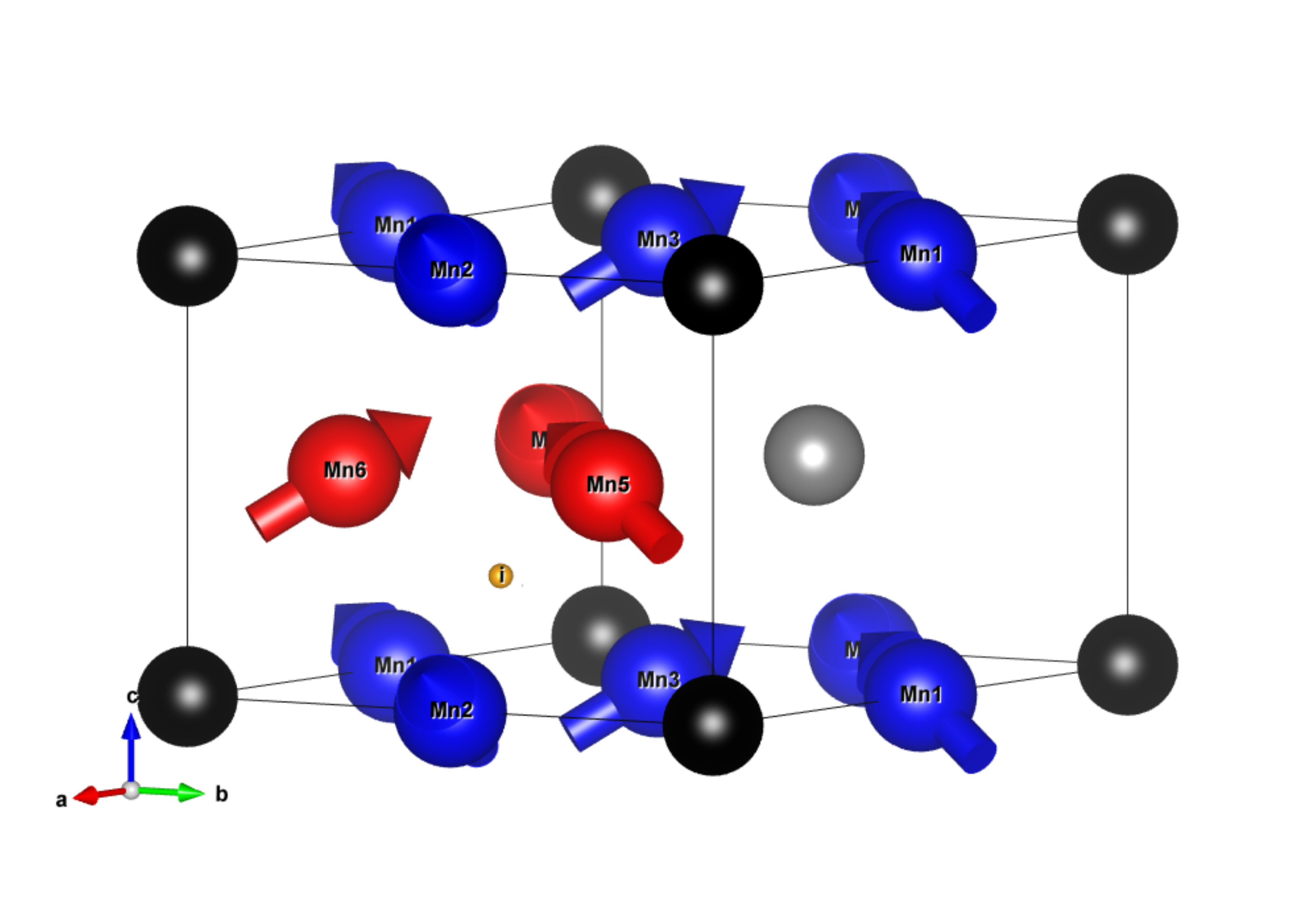}}
\end{center}
   \caption{\label{fig:ncpMx} Non-coplanar (ncp) magnetic
	structures of Mn$_3$Ge derived from ncAFM0 and ncAFM9 (see
	Fig.~\ref{fig:ncAFMx}) by rotating the moments out of the Kagome planes
        by $\theta = 45^\circ$. While the one in the top panel, labeled ncpM0, is 
        chiral, the structure ncpM9 in the bottom panel has an inversion center 
        that is indicated by an orange dot.\cite{VESTA}}
\end{figure}

Rotation of the moments out of both Kagome planes by the same polar angle
$\theta$ between the [0001] direction and the \{0001\} planes leads for $\theta
= 45^\circ$ to the non-coplanar spin arrangements depicted in
Fig.~\ref{fig:ncpMx}. The one in the upper panel, ncpM0 derived from ncAFM0, is
obviously still chiral. The structure derived from ncAFM9, labeled ncpM9 and
shown in the lower panel of Fig.~\ref{fig:ncpMx} accordingly remains achiral, the
inversion center connecting the two sub-lattices is indicated by an orange dot.
%
%

The main aim of the present work is the numerical study of chirality-induced or
\emph{topological} effects in transport and related properties. While the
individual Kagome sub-lattices in the antiferromagnetic structures just
discussed are chiral, as they have a finite vector spin chirality $\vec{S}_i
\times \vec{S}_j + \vec{S}_j \times \vec{S}_k + \vec{S}_k \times \vec{S}_i$,
the anomalous Hall conductivity arising from this is vanishing
globally.
This can be unambiguously derived from the space-time symmetry properties of the
current-current correlation function in terms of the Kubo formula behaviour for the
electrical conductivity.\cite{Kle66} The transformation under all symmetry
operations of the so-called magnetic Laue group (see Ref.~\onlinecite{SKWE15a}
for its definition used here) is sufficient to derive the symmetry restricted
tensor shape. The magnetic space and Laue groups for all spin configurations
discussed in this work are given in Table.~\ref{tb:mlgsMn3Ge}. For convenience
the Laue group is given also according to the older definition used by 
Kleiner\cite{Kle66}.
%
\begin{table}
\begin{tabular}{c|c|c|c}
label & MSG & MPG & MLG\\
\toprule\\[-0.2cm]
NM     & $P6_3/mmc1'$     & $6/mmm1'$     & $6/mmm1'$     $(6221')$ \\
FM     & $P6_3/mm'c'$     & $6/mm'm'$     & $6/mm'm'$     $(62'2')$ \\[0.1cm]
ncAFM0 & $P6_3/m'm'c'$    & $6/m'm'm'$    & $6/mmm1'$     $(6221')$ \\
ncAFM9 & $P6_3'/m'm'c$    & $6'/m'm'm$    & $6'/m'm'm$    $(6'22')$ \\[0.1cm]
ncpM0  & $P6_3m'c'$       & $6m'm'$       & $6/mm'm'$     $(62'2')$ \\
ncpM9  & $P\bar{3}m'1$    & $\bar{3}m'1$  & $\bar{3}m'1$  $(32')$ \\[0.1cm]
ncpAFM0 & $P\bar{3}'1m'$ & $\bar{3}'1m'$ & $\bar{3}1m1'$ $(3'2)$ \\
ncpAFM9 & $P6_3'm'c$     & $6'm'm$       & $6'/m'm'm$    $(6'2'2)$ \\
\end{tabular}
	\caption{\label{tb:mlgsMn3Ge} Magnetic space (MSG), point (MPG) and
	Laue groups (MLG) of the magnetic structures shown in
	Figs.~\ref{fig:ncAFMx} and \ref{fig:ncpMx} with nc and ncp
  standing for non-collinear and non-coplanar, respectively. The Laue groups are given
	following the standard definition used by Seemann \emph{et al.}\cite{SKWE15a} 
        as well as the older one used by Kleiner\cite{Kle66} (in parentheses). 
        The conventional setting concerning the sequence of generators
	is used for the space groups and carried over to the point
  and Laue groups.\cite{ITCA}}
\end{table}
%

The electrical conductivity tensor shapes derived from the corresponding
magnetic Laue group are as follows:\cite{Kle66,SKWE15a}
%
\begin{equation}
  \label{eq:sigma-Mn3Ge_PhNM}
  \SIGTEN^{\rm NM} =
  \left(
    \begin{array}{ccc}
      \sigma_{xx} & 0           & 0 \\
      0           & \sigma_{xx} & 0 \\
      0           & 0           & \sigma_{zz}
    \end{array}
  \right) = \underline{\sigma}^{\rm ncAFM0,9} = \underline{\sigma}^{\rm ncpAFM0,9}
\end{equation}
%
%
\begin{equation}
  \label{eq:sigma-Mn3Ge_PhFM}
  \SIGTEN^{\rm FM} =
  \left(
    \begin{array}{ccc}
      \sigma_{xx} & \sigma_{xy} & 0 \\
     -\sigma_{xy}  & \sigma_{xx} & 0 \\
      0           & 0           & \sigma_{zz}
    \end{array}
	\right) = \underline{\sigma}^{\rm ncpM0,9} \;\mathrm{.}
\end{equation}
%
As stated above, the non-collinear coplanar antiferromagnetic structures in 
Fig.~\ref{fig:ncAFMx} have the same conductivity tensor shape as the 
non-magnetic one. This applies as well to the non-coplanar antiferromagnetic 
structures that will be discussed in Section~\ref{ssec:ncpAFM}. The 
non-coplanar magnetic structures of Fig.~\ref{fig:ncpMx} on the other hand have the same shape of 
\SIGTEN\ as for the ferromagnetic (FM) case. However, as will be shown below, a 
chirality-induced contribution to the anomalous Hall conductivity $\sigma_{xy} = 
-\sigma_{yx}$ can be identified here.

The corresponding spin conductivity tensor shapes for polarization along the $z$ 
or [0001] direction are:\cite{SKWE15a}\\
%
\begin{equation}
  \label{eq:zpolsigma-Mn3Ge_PhNM}
  \SIGTEN^{z,\rm NM} =
  \left(
    \begin{array}{ccc}
      0              & \sigma_{xy}^z & 0 \\
      -\sigma_{xy}^z & 0 \\
      0              & 0           & 0
    \end{array}
  \right) = \underline{\sigma}^{z,\rm ncAFM0,9} = \underline{\sigma}^{z,\rm ncpAFM0,9}
\end{equation}
%
%
\begin{equation}
  \label{eq:zpolsigma-Mn3Ge_PhFM}
  \SIGTEN^{z,\rm FM} =
  \left(
    \begin{array}{ccc}
      \sigma_{xx}^z & \sigma_{xy}^z & 0 \\
     -\sigma_{xy}^z & \sigma_{xx}^z & 0 \\
      0             & 0             & \sigma_{zz}^z
    \end{array}
  \right) = \underline{\sigma}^{z, \rm ncpM0,9} \;\mathrm{.}
\end{equation}
%
The corresponding tensor for polarization along the $x$- and
$y$0direction can be found in Ref.\ \onlinecite{SKWE15a}.
The antiferromagnetic structures, regardless whether coplanar or non-coplanar, 
chiral or achiral, show only one independent non-zero element, namely the spin Hall 
conductivity $\sigma_{xy}^z = -\sigma_{yx}^z$ as in the non-magnetic case. Note 
however, that the tensors for the other two polarization directions differ for 
the structures ncAFM9, ncpAFM0, and ncpAFM9 (see 
Ref.~\onlinecite{SKWE15a}). The non-coplanar magnetic structures ncpM0
and ncpM9  have the same tensor 
shape for \SIGTEN$^z$ as the ferromagnetic one. While ncpM0 has the same 
magnetic Laue group ($6/mm'm'$) and accordingly the same tensor shapes for all
\SIGTEN$^k$ as the FM structure, the other two polarization directions, $x$ and $y$,
behave again differently for ncpM9. Also here a sizable chirality-induced
contribution will be shown to exist.

The Edelstein polarization tensor shapes for the non-centrosymmetric spin 
configurations ncAFM0, ncpM0, ncpAFM0, and ncpAFM9 are found as 
follows:\cite{WCE18}\\
%
\begin{equation}
  \label{eq:pEE-ncAFM0}
  \EETEN^{\rm ncAFM0} =
  \left(
    \begin{array}{ccc}
      p_{xx} & 0 & 0 \\
      0 & p_{xx} & 0 \\
      0 & 0 & p_{zz}
    \end{array}
  \right)
\end{equation}
%
%
\begin{equation}
  \label{eq:pEE-ncpM0}
  \EETEN^{\rm ncpM0} =
  \left(
    \begin{array}{ccc}
      p_{xx} & p_{xy} & 0 \\
     -p_{xy} & p_{xx} & 0 \\
      0      & 0      & p_{zz}
    \end{array}
  \right)
\end{equation}
%
%
\begin{equation}
  \label{eq:pEE-ncpAFM0}
  \EETEN^{\rm ncpAFM0} =
  \left(
    \begin{array}{ccc}
      p_{xx} & 0      & 0 \\
      0      & p_{xx} & 0 \\
      0      & 0      & p_{zz}
    \end{array}
  \right)
\end{equation}
%
%
\begin{equation}
  \label{eq:pEE-ncpAFM9}
  \EETEN^{\rm ncpAFM9} =
  \left(
    \begin{array}{ccc}
      0      & p_{xy} & 0 \\
     -p_{xy} & 0      & 0 \\
      0      & 0      & 0
    \end{array}
  \right) \;\mathrm{.}
\end{equation}
%

Finally, the shapes of the spin-orbit torkance tensors \TORTEN\ are identical to 
the ones given for \EETEN\ in 
Eqs.~(\ref{eq:pEE-ncAFM0})-(\ref{eq:pEE-ncpAFM9})\cite{WCS+16a}.
 It should be stressed that for these two response properties the 
tensor shape is determined by the magnetic point group and not by the magnetic 
Laue group.


\subsection{Orbital moments \label{ssec:OMTs}}

\begin{figure}
 \begin{center}
 \includegraphics[angle=0,width=0.9\linewidth,clip]{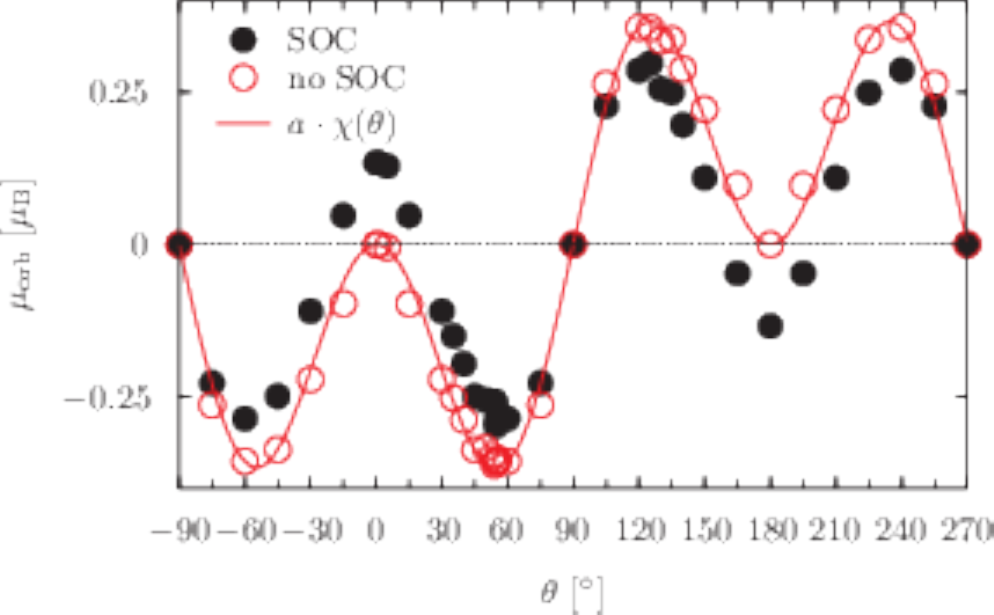}
\vspace*{-0.25cm}
\end{center}
 \caption{\label{fig:OMT-AFM0} Orbital moment as a function of 
polar angle $\theta$ in the non-coplanar chiral magnet ncpM0. Results 
including spin-orbit coupling (SOC) are shown as full (black) circles, those for 
vanishing SOC are given as open (red) circles. A fit of the latter to the scalar 
spin chirality $\chi(\theta)$ (see text) is shown as solid (red) line.}
\end{figure}

The occurrence of chirality-induced orbital moments in non-coplanar spin
arrangements has been predicted already quite some time ago.\cite{SN01,NIF03}
First-principles calculations in, e.g., atomic-scale spin
lattices\cite{HWD+15,HFN+16}, tri-atomic clusters of ferromagnetic
$3d$-elements on a surface\cite{DBB+16a}, and bulk $\gamma$-FeMn\cite{HFBM17}
could verify these in the limit of vanishing spin-orbit coupling. The
(non-)coplanarity between three spins can be expressed compactly by the
so-called scalar spin chirality $\chi_{ijk} = \vec{S}_i \cdot (\vec{S}_j \times
\vec{S}_k)$. If the volume of the parallelepiped spanned by the three spin vectors
is non-zero, they obviously are non-coplanar. In Figure~\ref{fig:OMT-AFM0} the
orbital moment $\mu_{orb}$\footnote{The sum over all sites is shown, which leaves only the
z-component non-zero.} is shown for ncpM0 as a function of the polar angle
$\theta$ between the [0001] direction and the \{0001\} planes.
Here $\mu_{orb}$ is defined as the modulus of the vector sum over all
sites in the unit cell that leads to the effective orbital moment along
the $z$ direction.
In case of
vanishing spin-orbit coupling (no SOC, red open circles) the remaining
chirality induced contribution to the orbital moment can
indeed be fairly well fitted with a function $\propto \cos(\theta)
\sin^2(\theta)$ reflecting the scalar spin chirality
$\chi_{ijk}(\theta)$. The zeros of this function correspond to the
ferromagnetic 
state ($\theta = 0^\circ,180^\circ$) and the non-collinear antiferromagnetic
state ($\theta = -90^\circ,90^\circ,270^\circ$). The extremal values are found
for integer multiples of $\theta = \arccos(1/\sqrt{3}) \approx 54.7356^\circ$,
i.e., for the \emph{magic angle}.

%
%
%
\begin{figure}
 \begin{center}
 \includegraphics[angle=0,width=0.95\linewidth,clip]{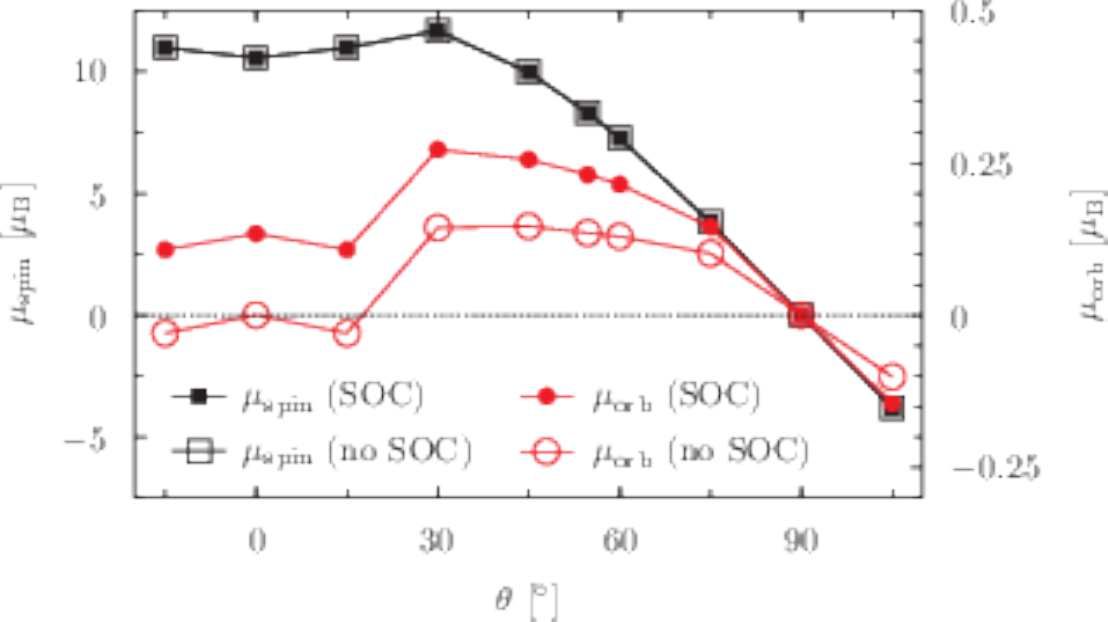}
\end{center}
 \caption{\label{fig:OMT-AFM9} Spin and orbital moment as a 
function of polar angle $\theta$ in the non-coplanar achiral magnet ncpM9. 
Results including spin-orbit coupling (SOC) are shown as full symbols, those for 
vanishing SOC are given as open symbols.}
\end{figure}
%

Figure~\ref{fig:OMT-AFM9} shows spin and orbital moments for the structure
ncpM9 as a function of polar angle $\theta$, again with and without spin-orbit
coupling. For the spin moment (black squares) spin-orbit coupling is, as to be
expected, of negligible relevance. The orbital moment (red circles) again has a
large chirality-induced component, whose angular dependence however does not
appear to be simply proportional to the scalar spin chirality. While the even
symmetry about $\theta = 0^\circ$ and the odd symmetry around $\theta =
90^\circ$ is obeyed, the behavior in-between seems to be more complicated.
Note that, despite the global inversion symmetry connecting the two Kagome
sub-lattices, these have, as in the case of ncpM0, the same finite scalar spin
chirality.

\subsection{X-ray absorption spectra \label{ssec:XAS}}

X-ray absorption spectroscopy has a long and successful history concerning its 
application as a local probe to magnetic systems. In particular the so-called 
XMCD sum rules\cite{ES75,TCSL92,CTAW93} allow for example to deduce from the 
integrated L$_{2,3}$-spectra of 3d-transition metals their spin and orbital 
magnetic moments. In line with the sum rules an angular dependence according to 
$\cos(\hat m \cdot \hat q)$ is normally assumed, where $\hat m$ and $\hat q$ are 
the orientation of the local moment probed by XMCD and of the X-ray
beam, respectively. This 
simple relation implies that in spin-compensated antiferromagnetic systems the 
XMCD should vanish. 
However, both XMCD\cite{WMM+18}  as well as the magneto-optic Kerr effect 
(MOKE)\cite{OM13,CNM14} are in fact, due to their relation to the 
frequency-dependent conductivity tensor,\cite{Ebe96} expected to be observable 
in any magnetic structure that allows for a finite anomalous Hall conductivity. 

\begin{figure}
 \begin{center}
 \includegraphics[angle=0,width=0.9\linewidth,clip]{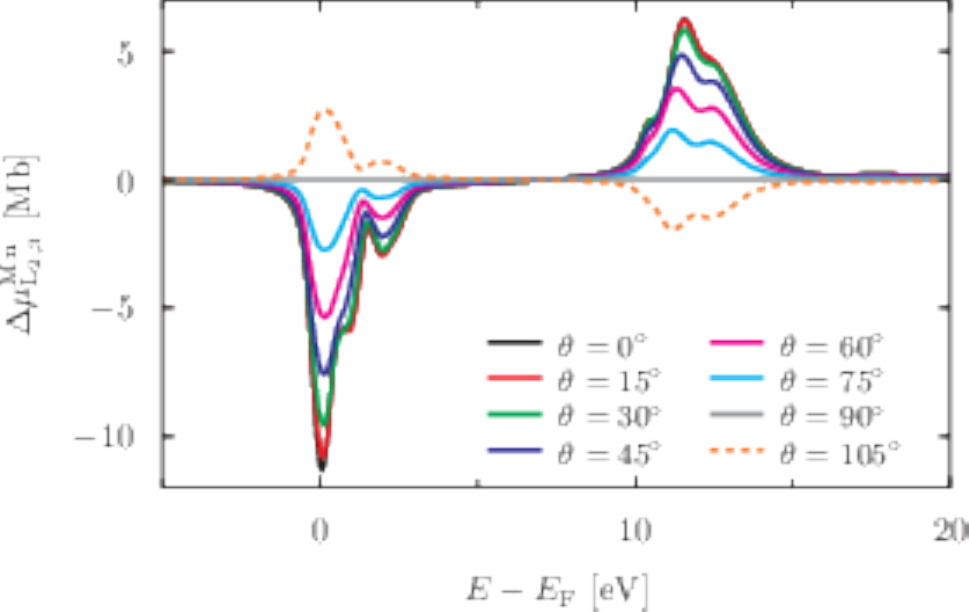}\\
 \includegraphics[angle=0,width=0.9\linewidth,clip]{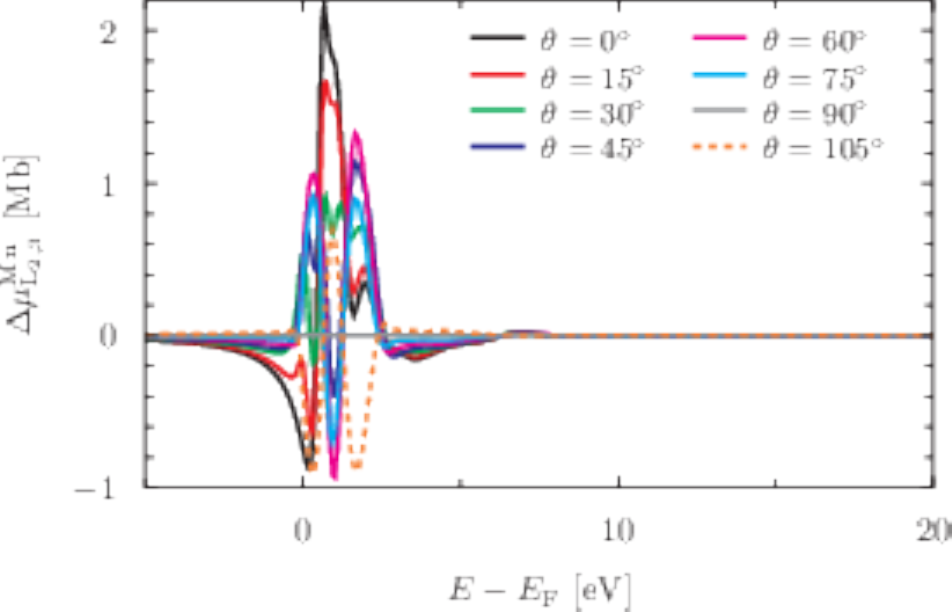}
\end{center}
 \caption{\label{fig:XAS_ncpM0} X-ray magnetic circular dichroism 
(XMCD) spectra $\Delta\mu^{\rm Mn}_{\rm L_{2,3}}$ at the Mn L$_{2,3}$-edge in the non-coplanar chiral magnetic 
structure ncpM0 with (top) and without (bottom) inclusion of spin-orbit 
coupling (SOC). The polar angle $\theta$ gives the tilt of the moments w.r.t.\ 
the [0001] direction, that coincides with the direction of the X-ray
beam.}
\end{figure}

In order to elucidate whether also the chirality-induced orbital moment discussed 
in Section~\ref{ssec:OMTs} can be deduced from X-ray absorption as suggested by 
Dos Santos Dias \emph{et al.}\cite{DBB+16a}, we perform first-principles 
calculations of XAS spectra as a function of polar angle $\theta$. The XMCD 
signals in the non-coplanar magnetic structure ncpM0 in- and excluding 
spin-orbit coupling is shown in Figure~\ref{fig:XAS_ncpM0} in the top and bottom 
panels, respectively.
%
%
%
\begin{figure}
 \begin{center}
 \includegraphics[angle=0,width=0.9\linewidth,clip]{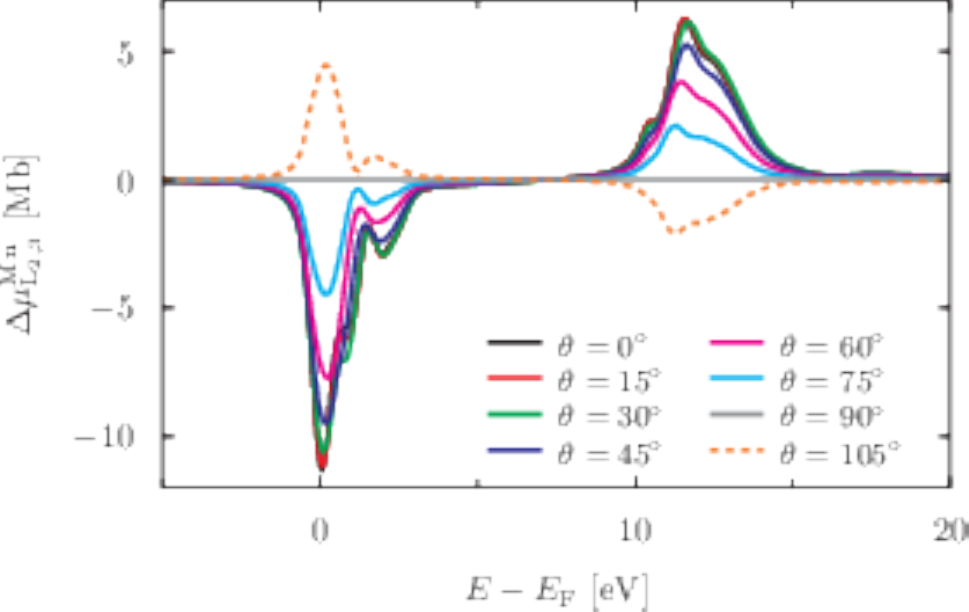}
 \includegraphics[angle=0,width=0.9\linewidth,clip]{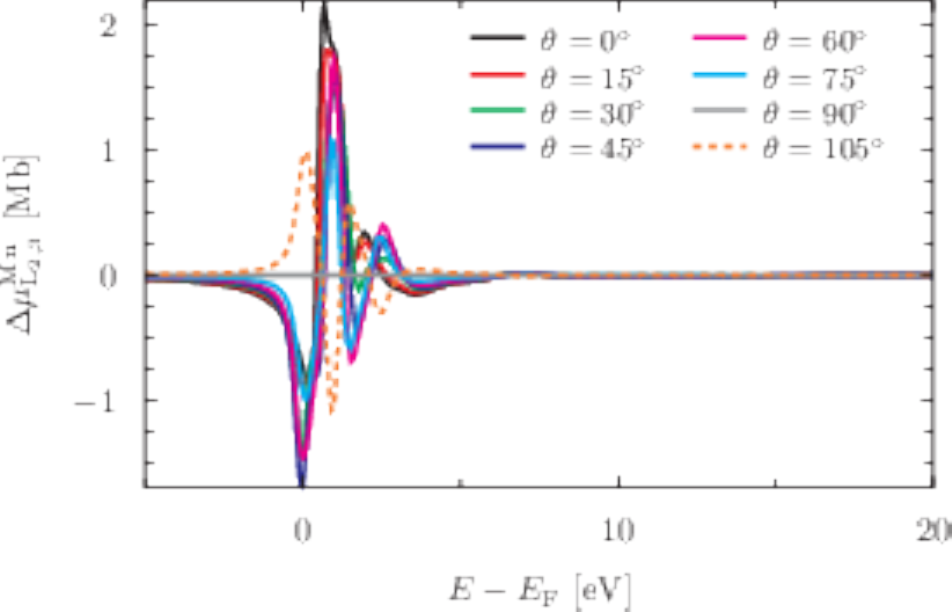}
 \end{center}
 \caption{\label{fig:XAS_ncpM9} X-ray magnetic circular dichroism 
(XMCD) spectra $\Delta\mu^{\rm Mn}_{\rm L_{2,3}}$ at the Mn L$_{2,3}$-edge in the non-coplanar achiral magnetic 
structure ncpM9 with (top) and without (bottom) inclusion of spin-orbit 
coupling (SOC). The polar angle $\theta$ gives the tilt of the moments w.r.t.\ 
the [0001] direction, that coincides with the direction of the X-ray
beam.}
\end{figure}
%
The absorption for incidence along the [0001] direction is calculated for the 
L$_{2,3}$-edge of Mn and summed over all sites of the unit cell. Suppressing 
spin-orbit coupling obviously leads to a degeneracy of the 2$p$ initial
states, i.e. it removes in particular the spin-orbit splitting into
$2p_{1/2}$- and  $2p_{3/2}$-shells. Accordingly only one edge is visible
in the lower panel, that nevertheless
shows an XMCD signal.
In both cases the strength of 
the signal is decreasing with increasing $\theta$ and anti-symmetric w.r.t.\ 
reversal of the global $z$ component of the magnetization
The same applies to the 
XMCD spectra for the achiral structure ncpM9 in Fig.~\ref{fig:XAS_ncpM9}. 
Here the fine structure at the L$_{2}$-edge is slightly different from that in
ncpM0 for the fully relativistic spectra in the top panel and quite so for the 
non-relativistic ones in the bottom panel.

\begin{figure}
 \begin{center}
 \includegraphics[angle=0,width=0.9\linewidth,clip]{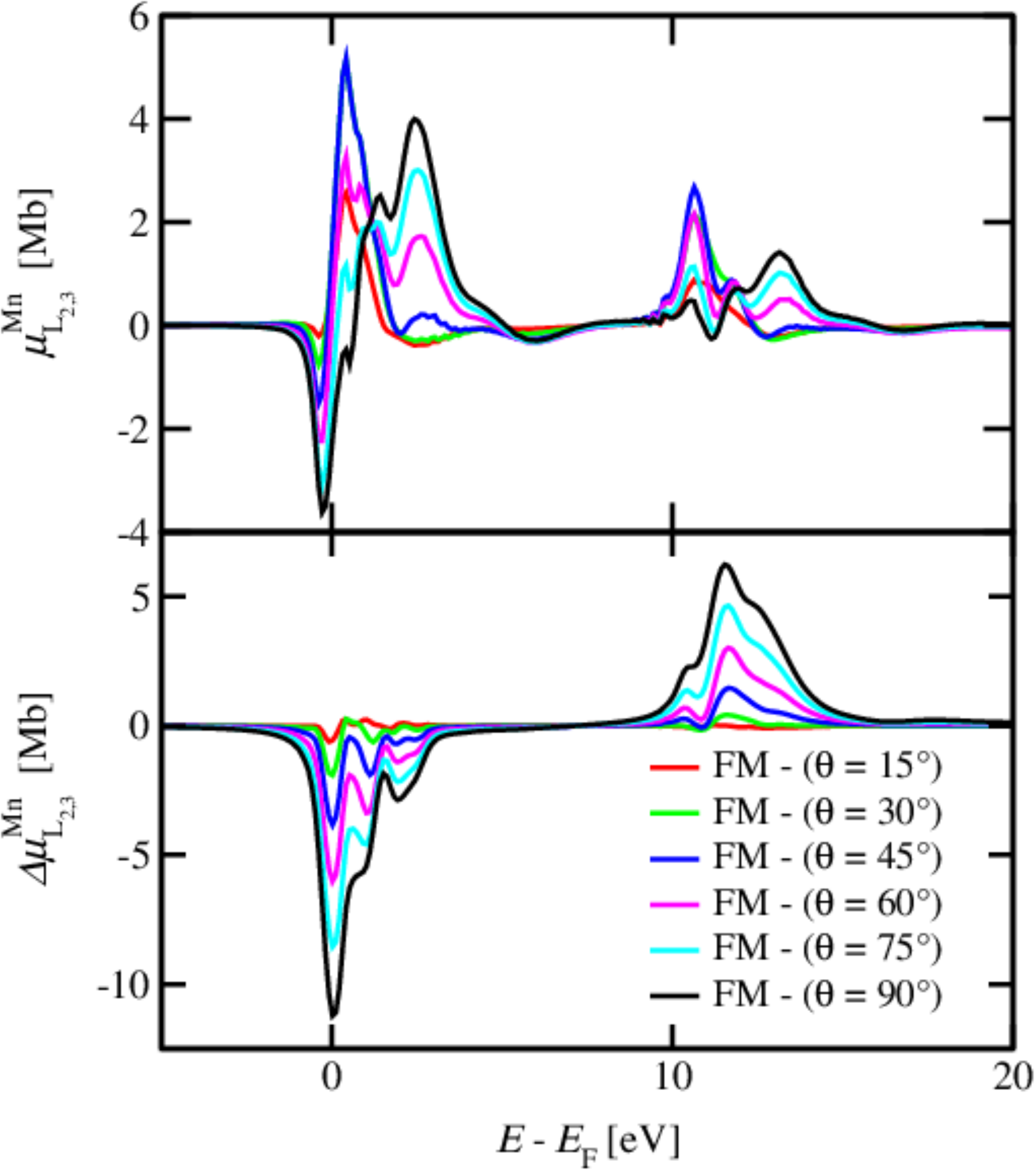}
 \end{center}
 \caption{\label{fig:XAS_FM-ncpM0} Difference between 
ferromagnetic and non-collinear polarization-averaged XAS spectra (top) and 
X-ray magnetic circular dichroism (XMCD) spectra (bottom) at the Mn 
L$_{2,3}$-edge in the chiral magnetic structure ncpM0. The polar angle 
$\theta$ gives the tilt of the moments w.r.t.\ the [0001] direction.} 
\end{figure}

\begin{figure}
 \begin{center}
 \includegraphics[angle=0,width=0.9\linewidth,clip]{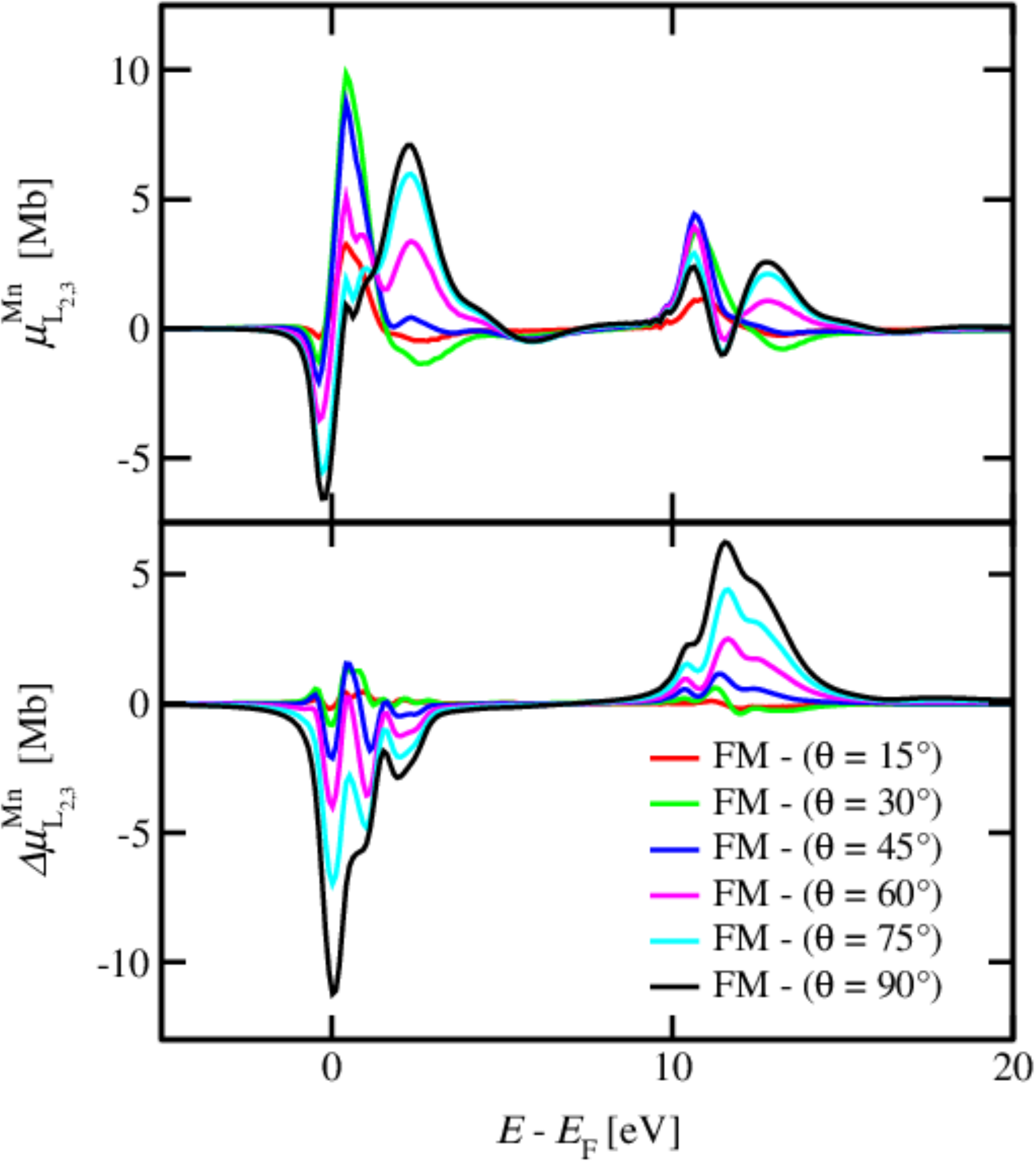}
 \end{center}
 \caption{\label{fig:XAS_FM-ncpM9} Difference between 
ferromagnetic and non-collinear polarization-averaged XAS spectra (top) and 
X-ray magnetic circular dichroism (XMCD) spectra (bottom) at the Mn 
L$_{2,3}$-edge in the achiral magnetic structure ncpM9. The polar angle 
$\theta$ gives the tilt of the moments w.r.t.\ the [0001] direction.} 
\end{figure}

As the XMCD signal is determined by both spin and orbital magnetic
moment, a clear-cut decomposition is desirable in 
order to assess the chirality-induced contribution to the latter by X-ray 
absorption spectroscopy . Since the 
standard XMCD sum rules cannot be applied here and their generalization to 
non-collinear magnetic order is still on open issue, an approximate scheme 
following the proposal in Ref.~\citenum{DBB+16a} has been employed. 
Figures~\ref{fig:XAS_FM-ncpM0} and \ref{fig:XAS_FM-ncpM9} show the difference 
between the average XAS and the XMCD signals for the field-aligned 
(ferromagnetic, FM, $\theta = 0^\circ$) limit and the non-collinear structures with 
$\theta \neq 0^\circ$ in ncpM0 and ncpM9, respectively.
%
%
%
%
As stated above, this follows the proposed protocol of Dos Santos Dias \emph{et 
al.},\cite{DBB+16a} devised for magneto-optical experiments on skyrmionic 
systems. A remaining obstacle is however the assessment of the 
spin-moment-induced contribution for which a linear and spin texture-independent 
relation to the polar angle has been assumed by these authors. Obviously an 
unambiguous separation into spin- and orbital as well as spin-orbit- and 
chirality-induced contributions is duly needed. Note that due to the simplified 
assumption of a collinear arrangement of moments the standard XMCD rules 
certainly have to be revised in order to make full use of the proposed 
procedure.

\subsection{Transport results: (T)AHE and (T)SHE \label{ssec:TXHE}}

\begin{figure}
 \begin{center}
 \includegraphics[angle=0,width=0.9\linewidth,clip]{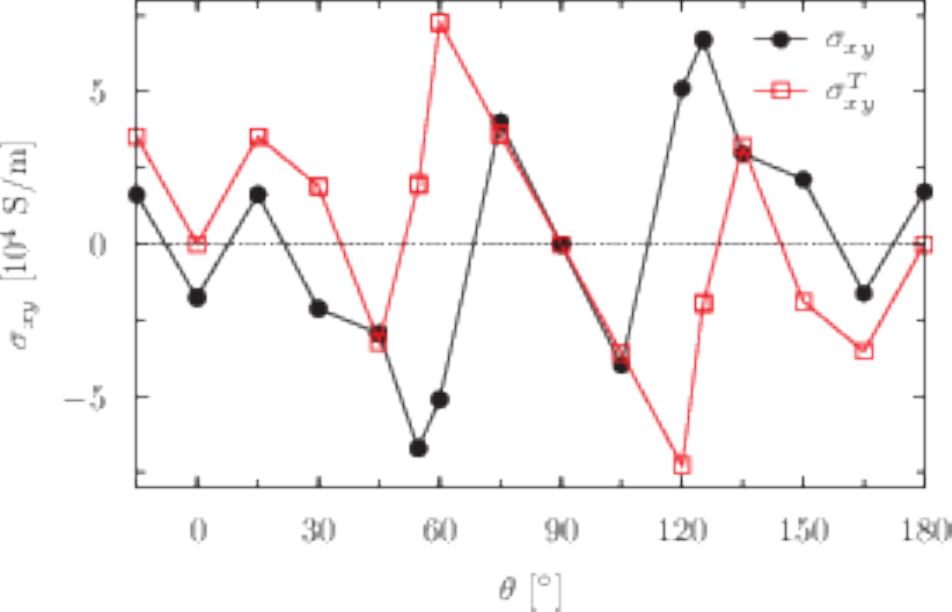}
 \includegraphics[angle=0,width=0.9\linewidth,clip]{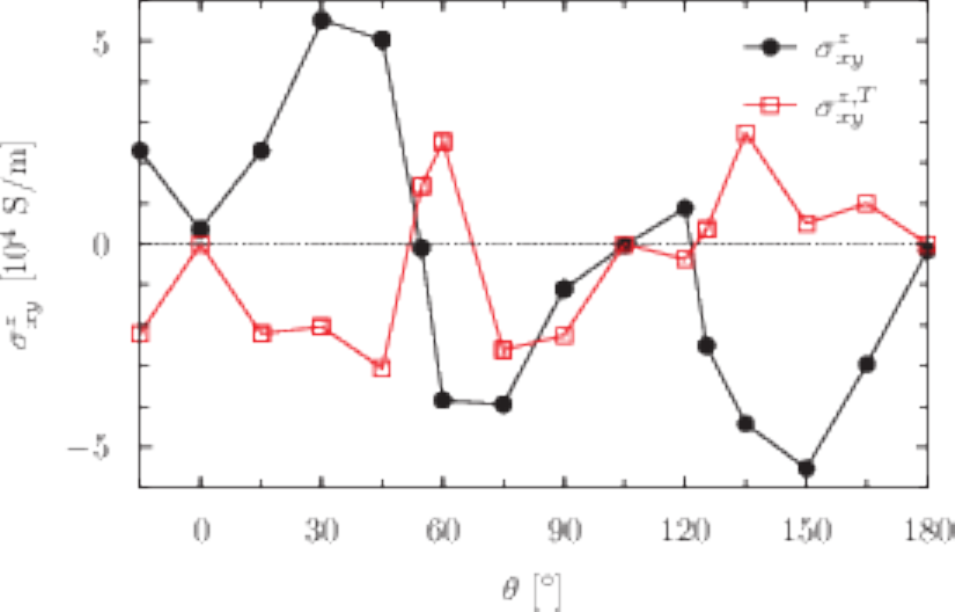}
\end{center}
\caption{\label{fig:TAnSHE-AFM0} Top: Anomalous Hall conductivity 
$\sigma_{xy}$ and its chirality-induced or topological contribution 
$\sigma_{xy}^T$ as functions of polar angle $\theta$ in ncpM0. Bottom: 
Corresponding results for the (topological) spin Hall conductivity $\sigma_{xy}$ 
($\sigma_{xy}^{z,T}$).}
\end{figure}

Results for the anomalous Hall conductivity $\sigma_{xy}$  as a function of $\theta$
are shown
in the top panel of Fig.\ \ref{fig:TAnSHE-AFM0} for ncpM0. As can be seen 
both AHC and the chirality-induced or topological contribution $\sigma_{xy}^T$ 
obtained in the non-relativistic limit $c_0/c \rightarrow 0$ (see 
Appendix~\ref{sec:AppA}) are anti-symmetric or odd w.r.t.\ magnetization 
reversal around $\theta = 90^\circ$. The chirality-induced component is clearly 
not simply proportional to the scalar spin chirality (see 
Fig.~\ref{fig:OMT-AFM0}), similar to the observation made by Hanke \emph{et 
al.}\cite{HFBM17} for $\gamma$-FeMn. The largest values for $\sigma_{xy}^T$ are 
for example found for $\theta = 60^\circ$ and 120$^\circ$ and not for the magic angle. 
In addition we observe two sign changes between the coplanar antiferromagnetic 
structure ($\theta = 90^\circ$) and the ferromagnetic states at $\theta = 0^\circ$ and 
180$^\circ$. In these limits $\sigma_{xy}^T$ vanishes and the AHC is purely 
spin-orbit-induced. Note, that the calculations were performed for the finite 
temperature $T = 300$\,K that was accounted for by uncorrelated lattice displacements via 
the so-called alloy analogy model (AAM).\cite{EMC+15} This was done in order to 
circumvent the numerical difficulties arising for the $\vec{k}$-space 
integration for perfectly ordered systems. However, as it turns out, the conductivities are almost 
entirely intrinsic in nature in the sense of negligible impact of the so-called vertex 
corrections associated with the thermally-induced disorder. Accordingly it is 
the magnetic band structure that determines the angular dependence of the AHC, 
or expressed alternatively its Berry curvature \cite{Ber84,XCN10} as skew scattering 
contributions arising from the presence of impurities\cite{NK14} or locally 
correlated fluctuating spins\cite{IN18} are not considered here.
%
%
The spin Hall conductivity in the relativistic and non-relativistic limits,
$\sigma_{xy}^z$ and $\sigma_{xy}^{z,T}$, given in the lower panel of
Fig.~\ref{fig:TAnSHE-AFM0} is even w.r.t.\ magnetization reversal around
antiferromagnetic configuration at $\theta = 90^\circ$.  It is strongly
dependent on the non-coplanar spin texture and quite differently so for its
spin-orbit- and chirality-induced contributions.  These can be of the same or
of different sign, leading to partial or even nearly complete cancellation as
for $\theta \approx 54.7356^\circ$. Quite interestingly, at $\theta =
105^\circ$ both appear to vanish. Note that in the ferromagnetic limit at
$\theta = 0^\circ$ and 180$^\circ$ the total value is small but non-zero, while
the topological contribution $\sigma_{xy}^{z,T}$ vanishes.

\begin{figure}
 \begin{center}
 \includegraphics[angle=0,width=0.9\linewidth,clip]{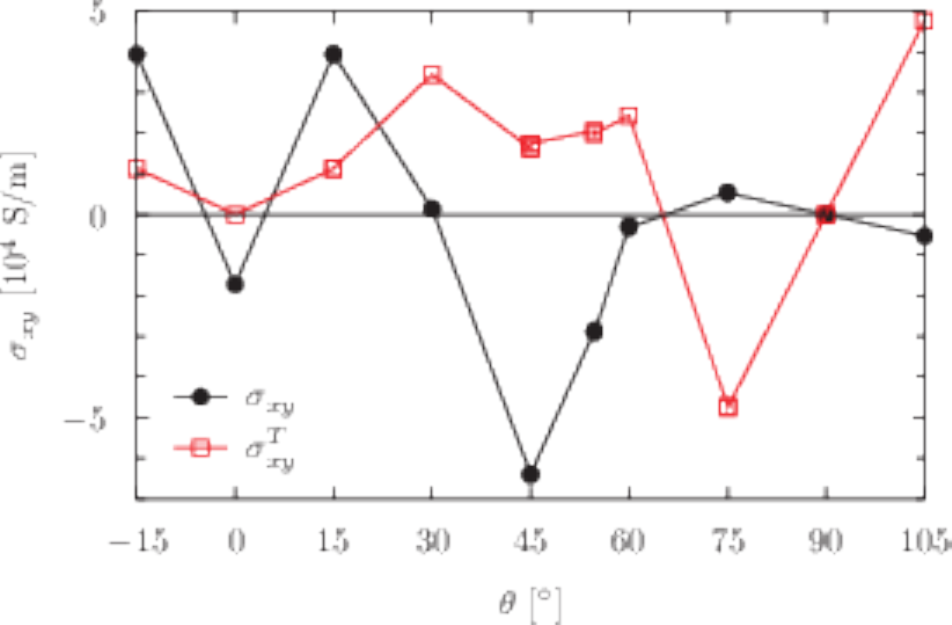}
 \includegraphics[angle=0,width=0.9\linewidth,clip]{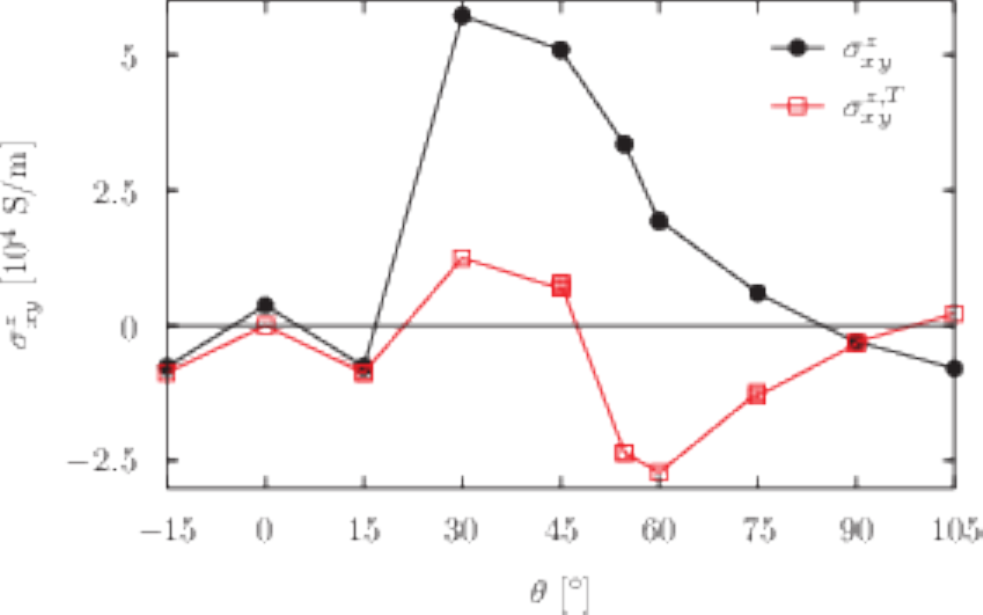}
\end{center}
	\caption{\label{fig:TAnSHE-AFM9} Top: Anomalous Hall conductivity $\sigma_{xy}$
	and its chirality-induced or topological contribution $\sigma_{xy}^T$ as functions of polar angle
	$\theta$ in ncpM9. Bottom: Corresponding results for the (topological)
        spin Hall conductivity $\sigma_{xy}$ ($\sigma_{xy}^{z,T}$).}
\end{figure}

Quite similar observations can be made in Fig.\ \ref{fig:TAnSHE-AFM9} for the 
achiral spin structure ncpM9 shown for the range $\theta = -15-105^\circ$. 
As can be seen, the AHC in the top panel is found to be anti-symmetric (odd) 
w.r.t.\ magnetization reversal whereas the SHC in the bottom panel is 
symmetric. The detailed angular dependence is distinct from that in ncpM0 
for both quantities, i.e., the two hypothetical structures could be 
distinguished experimentally. We propose that the abundance of assumed spin 
configurations in hexagonal Mn$_3X$ compounds could be confirmed or discarded 
via corresponding transport measurements rotating an applied magnetic field 
supported together with accompanying first-principles calculations.
%
%

The longitudinal charge transport is even w.r.t.\ magnetization reversal and 
anisotropic, i.e., $\sigma_{xx} = \sigma_{yy} \neq \sigma_{zz}$. In the absence 
of spin-orbit coupling and the associated anisotropic magneto-resistance, the 
anisotropy of the spin texture as well as the bare crystal-induced anisotropy 
already present in the non-magnetic case remain. Similar observations can be 
reported for the longitudinal spin conductivities $\sigma_{xx}^z = \sigma_{yy}^z 
\neq \sigma_{zz}^z$, that however are not fully even w.r.t.\ magnetization 
reversal. The spin conductivity tensor elements for the other spin
polarizations either show large chirality-induced  
contributions as well ($\sigma_{xy}^k = -\sigma_{yx}^k$ with $k = \{x,y\}$) or 
are exclusively SOC-induced ($\sigma_{iz}^k \neq -\sigma_{zi}^k$ with $i \neq k 
= \{x,y\}$). Note that all of them are even w.r.t.\ magnetization reversal.


\subsection{Spinorbitronic effects: (T)SOT and (T)EE \label{ssec:TSOTnTEE}}

\begin{figure}
 \begin{center}
 \includegraphics[angle=0,width=0.9\linewidth,clip]{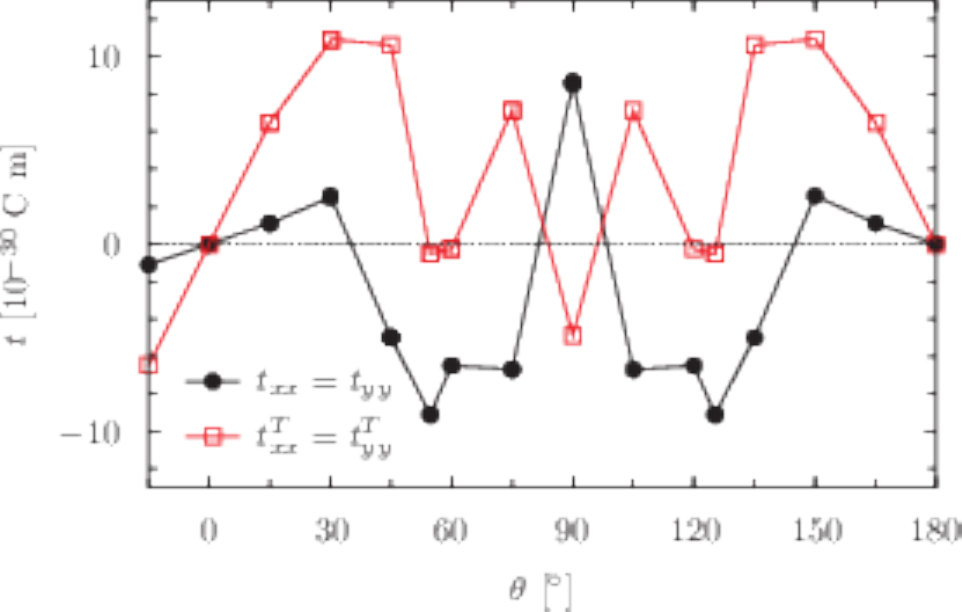}\\[0.1cm]
 \includegraphics[angle=0,width=0.9\linewidth,clip]{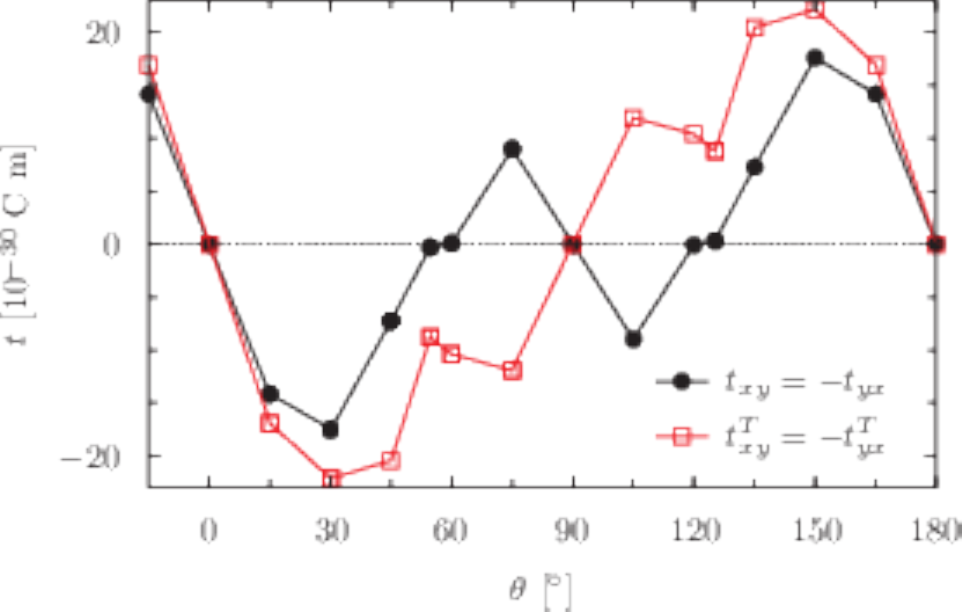}\\[0.1cm]
 \includegraphics[angle=0,width=0.9\linewidth,clip]{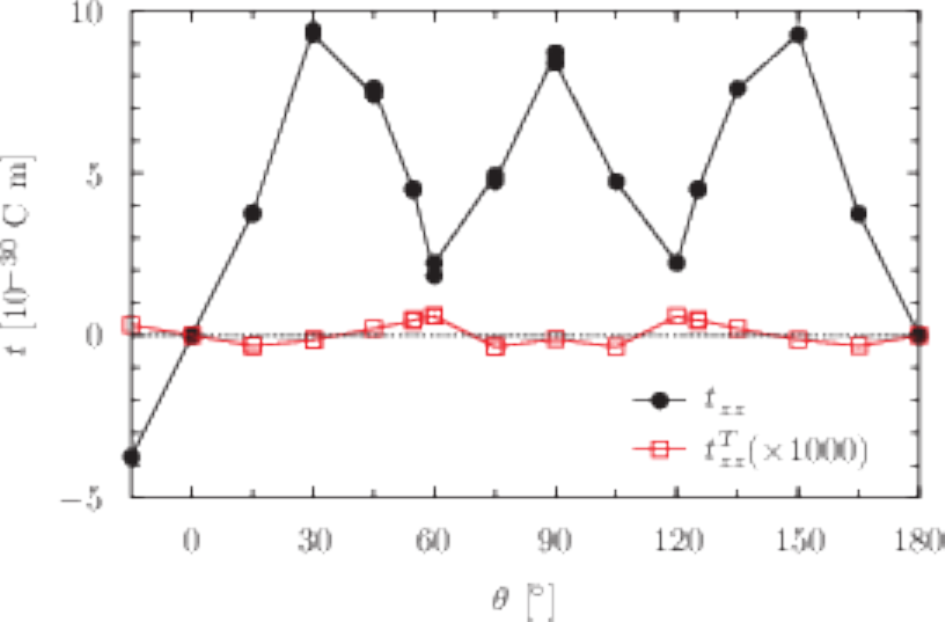}
 \end{center}
	\caption{\label{fig:TSOT-AFM0} Spin-orbit torkances $t_{xx} = t_{yy}$ (top),
	$t_{xy} = -t_{yx}$ (middle), and $t_{zz}$ (bottom) as functions of polar angle $\theta$ in ncpM0.
	The chirality-induced contributions $t_{ij}^T$ are given as red open squares.}
\end{figure}

Naturally the question arises whether the so-called spinorbitronic phenomena 
spin-orbit torque (SOT) and Edelstein effect (EE) also exhibit chirality-induced 
contributions leading to finite values in the absence of spin-orbit coupling. 
Employing the same Kubo linear response framework used for the charge and spin 
transport calculations in the previous section, but exchanging the operator for 
the response, the (spin) current density operators, by either the magnetic 
torque operator\cite{WCS+16a} or the spin magnetization operator\cite{WCE18}, 
the torkances $t_{ij}$ and Edelstein polarizations $p_{ij}$ can be computed from 
first principles. Figure~\ref{fig:TSOT-AFM0} shows the polar-angle dependence of 
the torkance tensor elements $t_{xx} = t_{yy}$ (top), $t_{xy} = -t_{yx}$ 
(middle), and $t_{zz}$ (bottom) in the chiral compound ncpM0.
%
%
The diagonal torkances in the top and bottom panels are obviously even w.r.t.\ 
magnetization reversal, i.e., anti-symmetric w.r.t.\ $\theta = 90^\circ$, while 
the off-diagonal anti-symmetric element $t_{xy} = -t_{yx}$ in the middle panel 
is odd. For this as well as for the $t_{xx} = t_{yy}$ indeed a sizable 
chirality-induced contribution is found that appears to be largest at $\theta = 
30^\circ$ and 150$^\circ$. The diagonal torkance $t_{zz}$ in the bottom panel, 
corresponding to a rotation of the moments about the [0001] or $z$ axis 
coinciding with the direction of the applied electric field, is almost 
exclusively spin-orbit-driven. In all three cases the full torkances vanish in 
the ferromagnetic limit ($\theta = 0^\circ$ and 180$^\circ$) due to inversion 
symmetry.

\begin{figure}
 \begin{center}
 \includegraphics[angle=0,width=0.9\linewidth,clip]{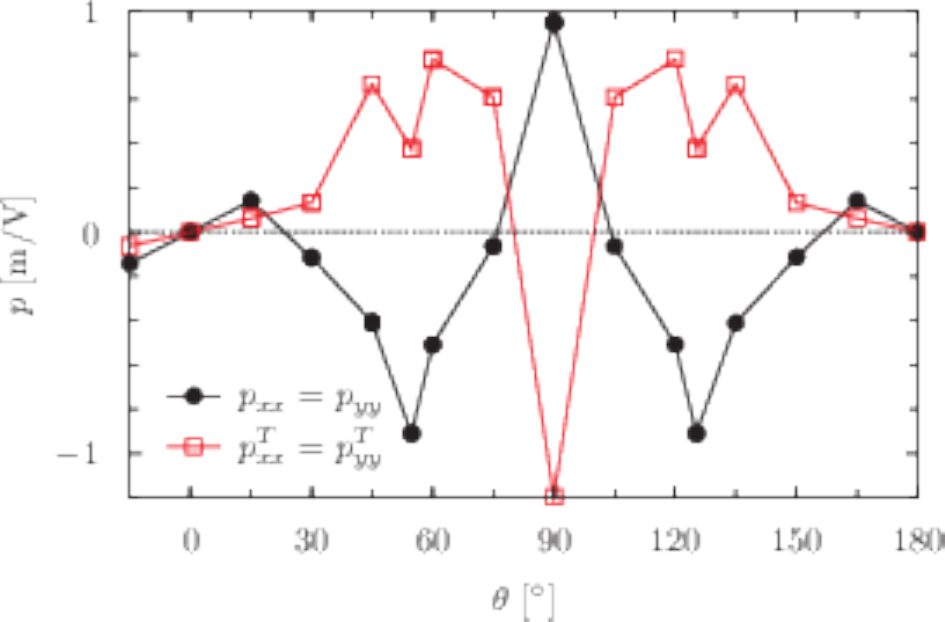}\\[0.1cm]
 \includegraphics[angle=0,width=0.9\linewidth,clip]{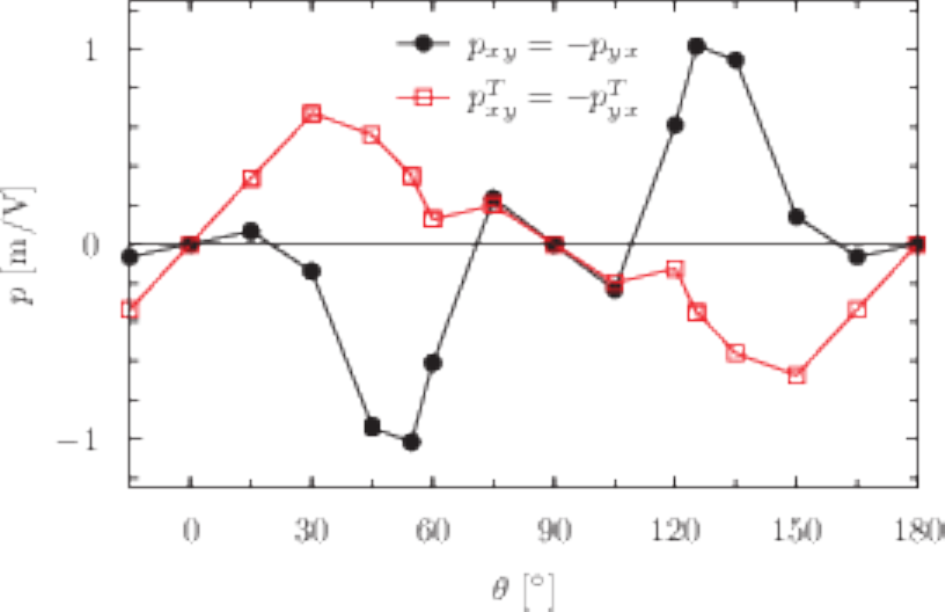}\\[0.1cm]
 \includegraphics[angle=0,width=0.9\linewidth,clip]{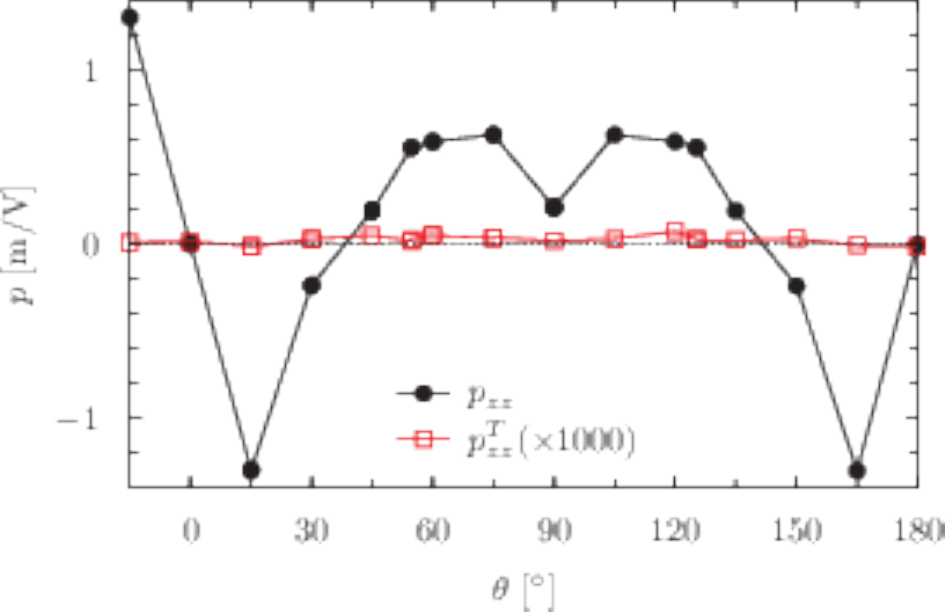} 
\end{center}
	\caption{\label{fig:TEE-AFM0} Edelstein polarization tensor elements  $p_{xx} = p_{yy}$ (top),
	$p_{xy} = -p_{yx}$ (middle), and $p_{zz}$ (bottom) as functions
        of the polar angle $\theta$ in ncpM0.
	The small chirality-induced contributions $p_{ij}^T$ scaled by
        the factor 1000 are given as red open squares.}
\end{figure}

The Edelstein polarization is one of the two microscopic mechanisms 
usually discussed as a source for the SOT, namely the (Rashba-)Edelstein torque, 
while the spin-Hall torque is attributed to the spin-transfer-torque-like action 
of a spin-polarized current on the local magnetization. The elements of
the corresponding Edelstein polarization 
tensor \EETEN\ are shown in Fig.~\ref{fig:TEE-AFM0} as a function of the polar angle 
$\theta$.
%
%
The elements $p_{ij}$ are found to behave very similar to the corresponding 
elements of \TORTEN\,, i.e., the diagonal elements are even, the off-diagonal 
ones are odd, and $p_{xx} = p_{yy}$ (top) as well as $p_{xy} = -p_{yx}$ (middle) 
are overall chirality-dominated while $p_{zz}$ in the bottom panel is again 
essentially SOC-induced. Note however, that the correspondence between $t_{ij}$ 
and $p_{ij}$ is not trivial, as an additional crossproduct with the local 
magnetization is involved for the operator representing the response in
the case of the torkance.
This leads for 
example for the odd torkance $t_{xy} = -t_{yx}$ in the middle panel of 
Fig.~\ref{fig:TSOT-AFM0} to a different angular dependence as compared to 
$p_{xy} = -p_{yx}$ in particular close to the ferromagnetic limits at the left 
and right ends. For the topological contributions this is even more pronounced. 
While the odd Edelstein polarization is chirality-dominated close to the 
antiferromagnetic configuration at $\theta = 90^\circ$, the SOT- and 
chirality-induced torkances are even of different sign here. The achiral 
coplanar and non-coplanar structures ncAFM9 and ncpM9 are found to be 
numerically zero as demanded by the inversion symmetry (see 
Section~\ref{ssec:SYM}).

\subsection{Non-coplanar antiferromagnets \label{ssec:ncpAFM}}

\begin{figure}[hbt]
 \begin{center}
{\includegraphics[angle=0,width=\linewidth,clip]{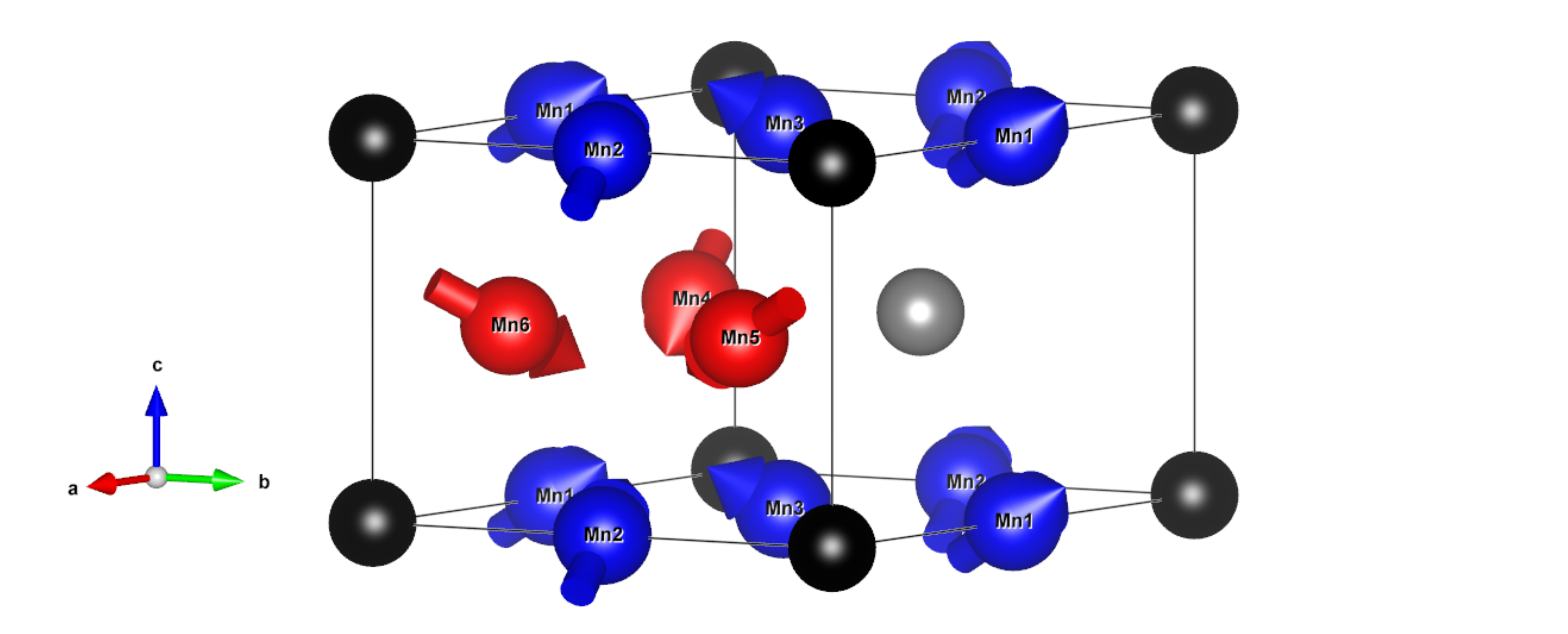}}
{\includegraphics[angle=0,width=\linewidth,clip]{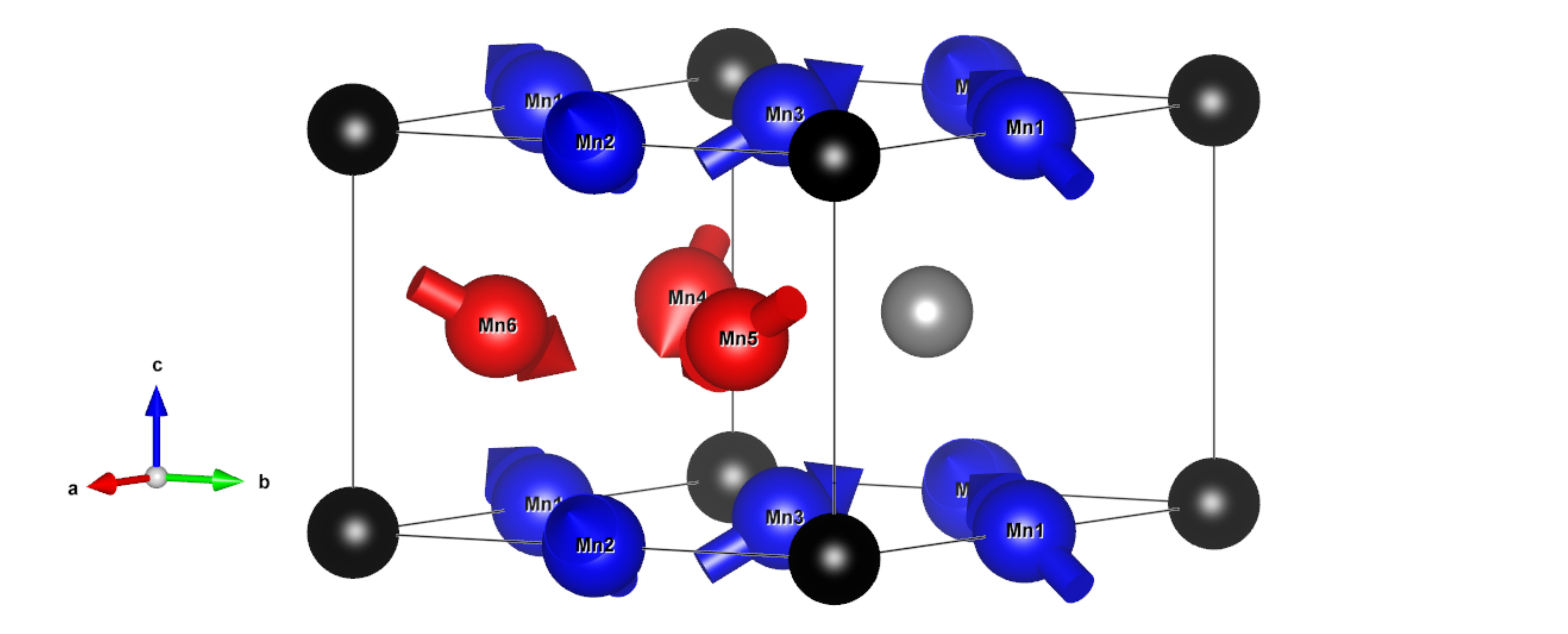}}
\end{center}
 \caption{\label{fig:ncpAFM} Non-coplanar antiferromagnetic
	structures obtained from Fig.~\ref{fig:ncAFMx} by rotating the moments
	in the two Kagome planes into opposite directions by the same angle.
	The structure in the top panel has the same chirality in both
	sub-lattices (ncpAFM0), while in the lower panel they are of opposite
	sign (ncpAFM9).\cite{VESTA}}
\end{figure}

By rotating the moments in the two Kagome planes in opposite directions by the 
same angle $\theta$, non-coplanar antiferromagnetic structures as shown in 
Fig.~\ref{fig:ncpAFM} for $\theta = \pm 45^\circ$ are obtained. The upper panel 
is derived from the co-planar AFM structure ncAFM0, while the ncpAFM9 
structure in the lower panel is obtained from the achiral ncAFM9 structure. 
As $\theta$ differs for both magnetic sub-lattices, the inversion symmetry is 
obviously broken, i.e., a chiral structure results. Note, that for ncpAFM0 
inversion combined with time-reversal ($\bar{1}'$) is still a symmetry 
operation.
%
%

The tensor shapes for charge and $z$-polarized spin conductivity correspond for 
both structures to the non-magnetic case, i.e., they are diagonal with $\sigma_{xx} = 
\sigma_{yy} \neq \sigma_{zz}$ and only one independent element of $\SIGTEN^z$, 
the spin Hall conductivity $\sigma_{xy}^z = -\sigma_{yx}^z$. The shapes of 
$\SIGTEN^x$ and $\SIGTEN^y$ for ncpAFM0 and ncpAFM9, however, differ 
from each other as well as from the ones for the NM structure.\cite{SKWE15a} For 
the torkance and the Edelstein polarization the tensor shapes are given in 
Eqs.~(\ref{eq:pEE-ncpAFM0}) and (\ref{eq:pEE-ncpAFM9}). Obviously there is a 
finite Edelstein polarization as well as spin-orbit torkance present for both 
structures. However, the corresponding tensor shapes differ from each other, as 
the magnetic point group has to be considered here.\cite{WCS+16a,WCE18}

\begin{figure}[hbt]
 \begin{center}
 \includegraphics[angle=0,width=0.9\linewidth,clip]{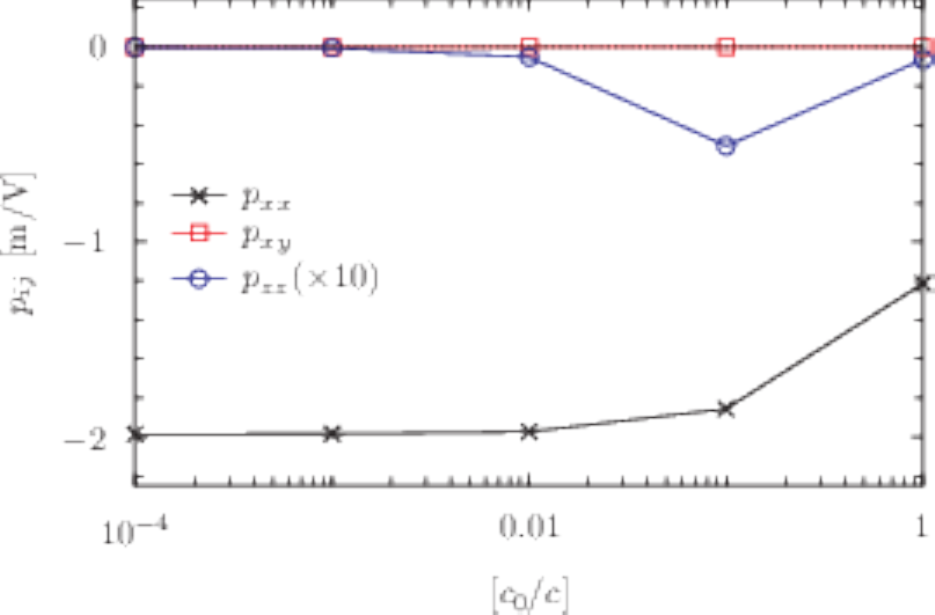}\\[0.1cm]
 \includegraphics[angle=0,width=0.9\linewidth,clip]{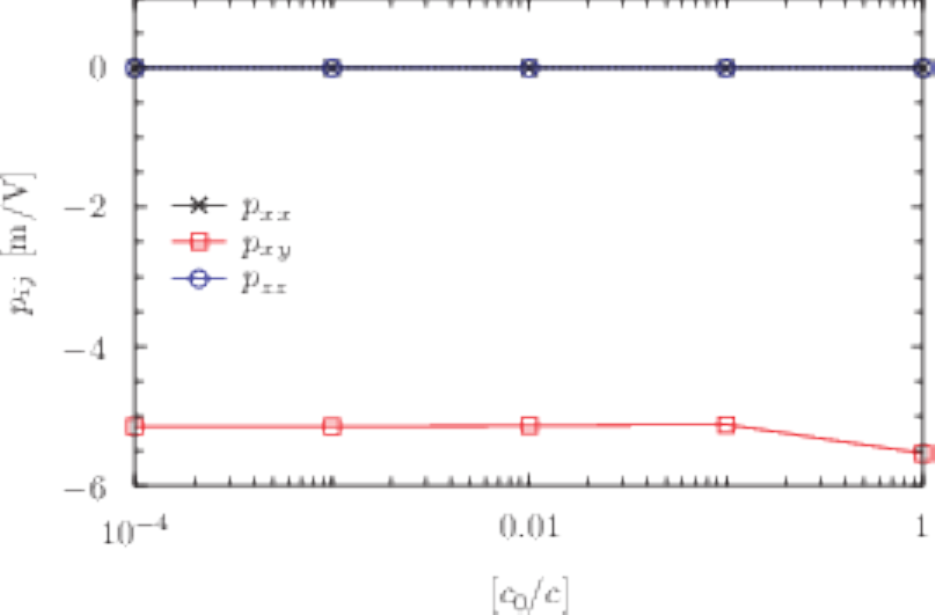}
\end{center}
 \caption{\label{fig:Cscaling_EE_ncpAFMx} Edelstein polarizations
$p_{xx}$, $p_{xy}$, and $p_{zz}$ in ncpAFM0 (top) and ncpAFM9 (bottom) 
as function of $c_0/c$. For both structures the polar angle $\theta$ 
is $45^\circ$. The relativistic limit is on the right ($c_0/c = 1$), 
the non-relativistic one on the left ($c_0/c \rightarrow 0$).}
\end{figure}

Figure~\ref{fig:Cscaling_EE_ncpAFMx} shows the Edelstein polarization tensor 
elements $p_{xx}$, $p_{xy}$, and $p_{zz}$ for both structures at $\theta = \pm 
45^\circ$ as a function of a scaled speed of light $c$ (see 
Appendix~\ref{sec:AppA} for details).
%
%
Confirming the tensor shapes in Eqs.~(\ref{eq:pEE-ncpAFM0}) and (\ref{eq:pEE-ncpAFM9}),
in the upper panel only the diagonal elements are non-zero, while in the lower
panel only $p_{xy}$ is non-vanishing. Furthermore it can be stated that for ncpAFM0 again 
$p_{zz}$ is smaller than $p_{xx}$ and vanishes in the non-relativistic limit 
($c_0/c \rightarrow 0$), while $p_{xx}$ has a large chirality-induced
contribution. The anti-symmetric Edelstein polarization $p_{xy}$ for ncpAFM9 
shown in the bottom panel of Fig.~\ref{fig:Cscaling_EE_ncpAFMx} is even 
almost exclusively arising from the spin texture.

\begin{figure}[hbt]
 \begin{center}
 \includegraphics[angle=0,width=0.9\linewidth,clip]{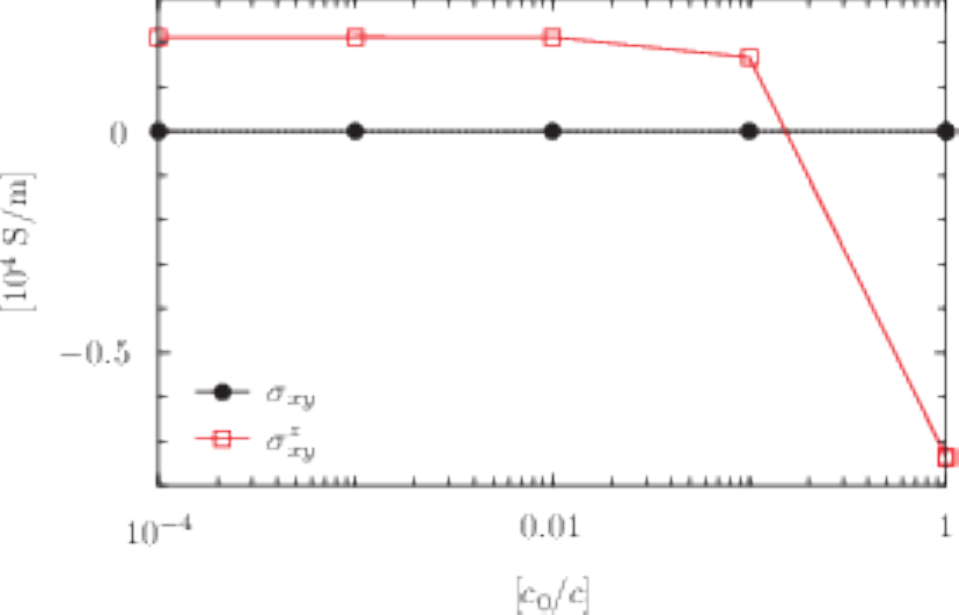}\\[0.1cm]
 \includegraphics[angle=0,width=0.875\linewidth,clip]{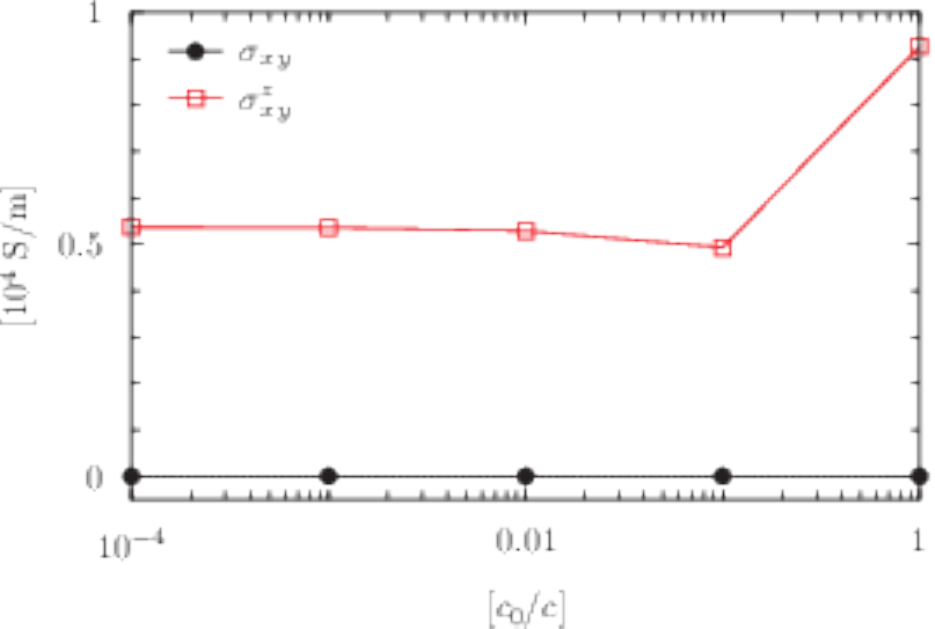}
\end{center}
 \caption{\label{fig:TAHEnSHE_ncpAFM} Anomalous and spin Hall
conductivities, $\sigma_{xy}$ and $\sigma_{xy}^z$, respectively, in ncpAFM0 
(top) and ncpAFM9 (bottom) as functions of $c_0/c$. 
For both structures the polar angle $\theta$ is $45^\circ$. The relativistic limit 
is on the right ($c_0/c = 1$), the non-relativistic one on the left 
($c_0/c \rightarrow 0$).}
\end{figure}

In agreement with the absence of off-diagonal anti-symmetric conductivity tensor
elements the XMCD signals of the two chiral magnetic sub-lattices cancel each 
other numerically exactly (not shown here). The same applies to the anomalous Hall 
conductivity shown in Fig.~\ref{fig:TAHEnSHE_ncpAFM}, while the spin Hall 
conductivities are found to be finite with sizable topological contributions.
%
%


\section{Conclusions \label{sec:Concl}}


To summarize, the effect of a non-collinear and non-coplanar spin texture on 
orbital moments, X-ray absorption, charge and spin transport as well as 
spin-orbit torque and Edelstein polarization has been investigated by 
first-principles calculations for hexagonal Mn$_3$Ge. By smoothly varying the 
polar angle w.r.t.\ to the Kagome planes of corner-sharing triangles in two 
hypothetical reference structures, one globally chiral one achiral, the 
chirality-induced or topological contributions are compared to the 
spin-orbit-induced parts. To obtain the former in absence of the latter, the 
non-relativistic limit has been taken by scaling the speed of light $c$. The key 
findings are first of all the occurrence of topological orbital moments in 
presence and absence of global inversion symmetry, in the latter case following 
the angular dependence expected from the scalar spin chirality. A proposal of 
its experimental verification by XMCD measurements in a rotating external 
magnetic field is supported by a comparison of spectra for the field-aligned 
ferromagnetic case with those of non-coplanar spin configurations. Also here 
the limit of vanishing spin-orbit coupling has been investigated, conclusive 
statements could however not yet been made due to limitations of the standard 
XMCD sum rules. Furthermore the presence, angular dependence, as well as 
magnitude of the chirality-induced contributions to the anomalous and spin Hall 
conductivities has been demonstrated. Similar calculations of the spin-orbit 
torkance and the Edelstein polarization reveal sizable topological contributions 
also here, that can, depending on which quantity and which tensor element is 
considered as well as on the polar angle, enhance or suppress the 
spin-orbit-induced effects and be either dominating or vanishing.

Future studies on realistic non-collinear antiferromagnets of the hexagonal Mn$_3X$ 
type with $X =$ Ga, Ge, or Sn based on the experimentally assumed or 
theoretically proposed spin structures could help determining the actual 
configuration and the relevance of chirality-induced contributions in measured 
response properties. A proposed extension of the XMCD sum rules to non-collinear 
magnetic order and the absence of spin-orbit coupling should be able to support 
experimental efforts on the quantification of topological orbital moments.





\begin{acknowledgments}
	Financial support by the DFG via SFB~1277 (\emph{Emer-
	gente relativistische Effekte in der Kondensierten Ma-
	terie}) is gratefully acknowledged.
\end{acknowledgments}


\appendix
\section{Manipulating the spin-orbit coupling \label{sec:AppA}}

The topological contributions to the various response quantities discussed were 
determined by either setting the spin-orbit coupling explicitly to zero in the 
self-consistent calculations (when dealing with the orbital magnetic
moment) or by scaling the speed of light  
$c$ in the X-ray absorption and Kubo linear response calculations. The limit 
$c_0/c \rightarrow 0$ with the speed of light in vacuum $c_0$,  or equivalently 
$c/c_0 \rightarrow \infty$, corresponds to the non-relativistic case. Obviously not 
only the spin-orbit coupling is affected this way, but also the so-called 
scalar-relativistic effects. However, for the properties relevant to this work 
the spin-orbit coupling is the relevant relativistic correction.\\

\begin{figure}[htb]
 \begin{center}
 \includegraphics[angle=0,width=0.9\linewidth,clip]{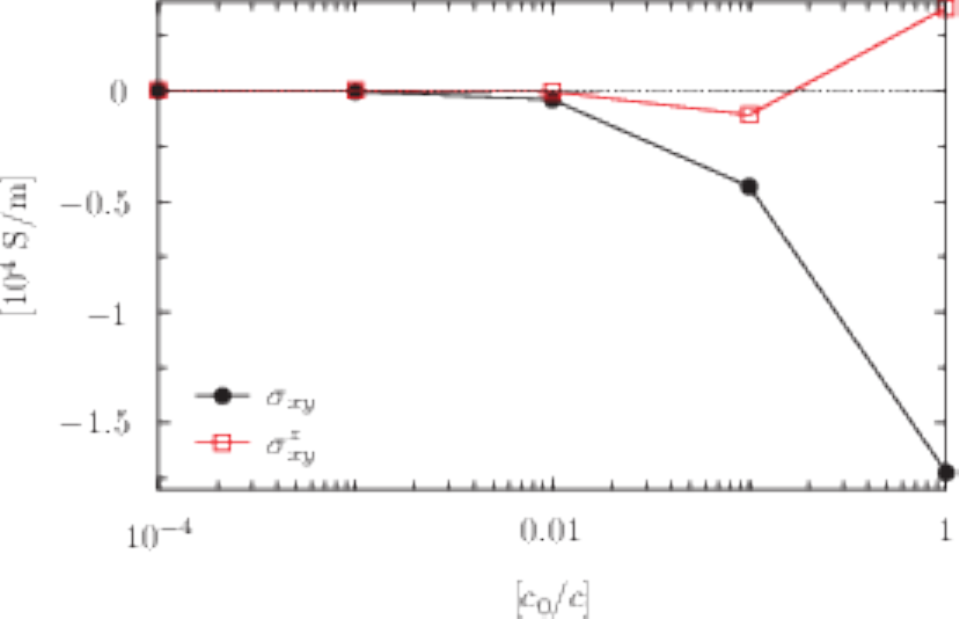}\\
 \includegraphics[angle=0,width=0.9\linewidth,clip]{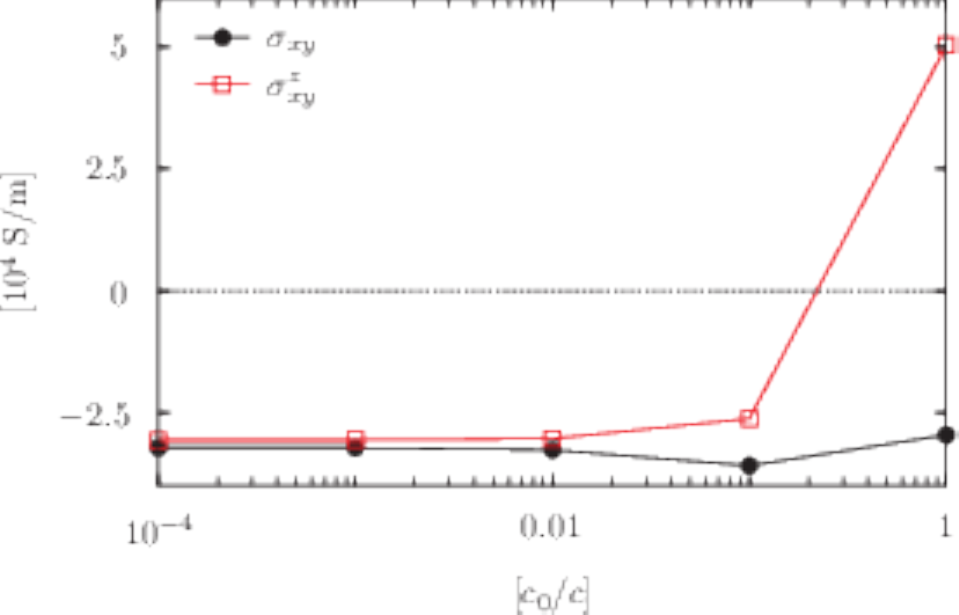}\\
 \includegraphics[angle=0,width=0.875\linewidth,clip]{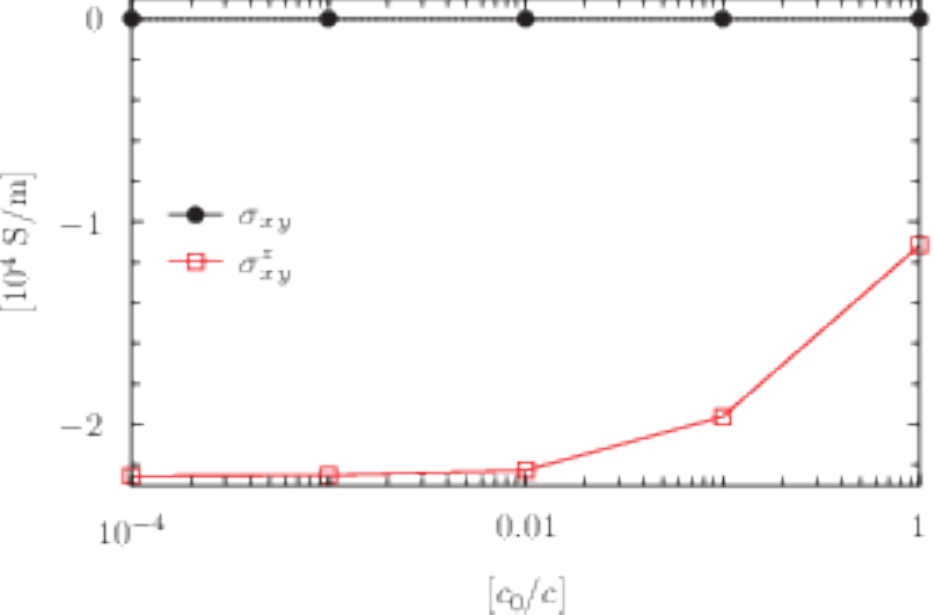}
\end{center}
 \caption{\label{fig:Cscaling_AHCnSHC_AFM0} Anomalous and spin 
Hall conductivity as a function of $c_0/c$ in the ferromagnetic state 
(top), the non-coplanar chiral magnetic state ncpM0 (middle) for $\theta = 
45^\circ$, and the co-planar non-collinear chiral state ncAFM0 (bottom). 
The relativistic limit is on the right ($c_0/c = 1$), the 
non-relativistic one on the left ($c_0/c \rightarrow 0$).}
\end{figure}

Figure~\ref{fig:Cscaling_AHCnSHC_AFM0} shows the anomalous (black symbols) and 
spin Hall (red symbols) conductivities as a function of the scaled speed of 
light $c_0/c$ for the ferromagnetic state (top), the non-coplanar magnetic state 
ncpM0 depicted in the top panel of Fig.~\ref{fig:ncpMx} (middle), and the 
coplanar non-collinear antiferromagnetic state ncAFM0 (bottom).
%
%
Going from the relativistic limit on the right side to the non-relativistic on 
the left, both $\sigma_{xy}$ and $\sigma_{xy}^z$ vanish for the ferromagnet, 
revealing their purely SOC-driven nature. In the non-coplanar magnetic state 
($\theta = 45^\circ$) in the middle panel both are finite for $c_0/c 
\rightarrow 0$, i.e., they exhibit a chirality-induced or topological 
contribution arising from the spin texture. For both quantities this is sizable, 
the so-called topological Hall effect (THE) is even dominating in the 
relativistic limit, while the topological spin Hall effect (TSHE) is of opposite 
sign when compared to the SOC-induced contribution and about half as large. For 
the non-collinear antiferromagnetic structure ncAFM0 in the bottom panel the 
anomalous Hall effect vanishes due to symmetry (see Table~\ref{tb:mlgsMn3Ge} and 
Eq.~\ref{eq:sigma-Mn3Ge_PhNM}), while the TSHE is even larger than the total 
SHE. This means that here the contribution due to spin-orbit coupling is again of 
opposite sign and but now the chirality-induced part is about twice as large.

\begin{figure}[htb]
 \begin{center}
   \includegraphics[angle=0,width=0.9\linewidth,clip]{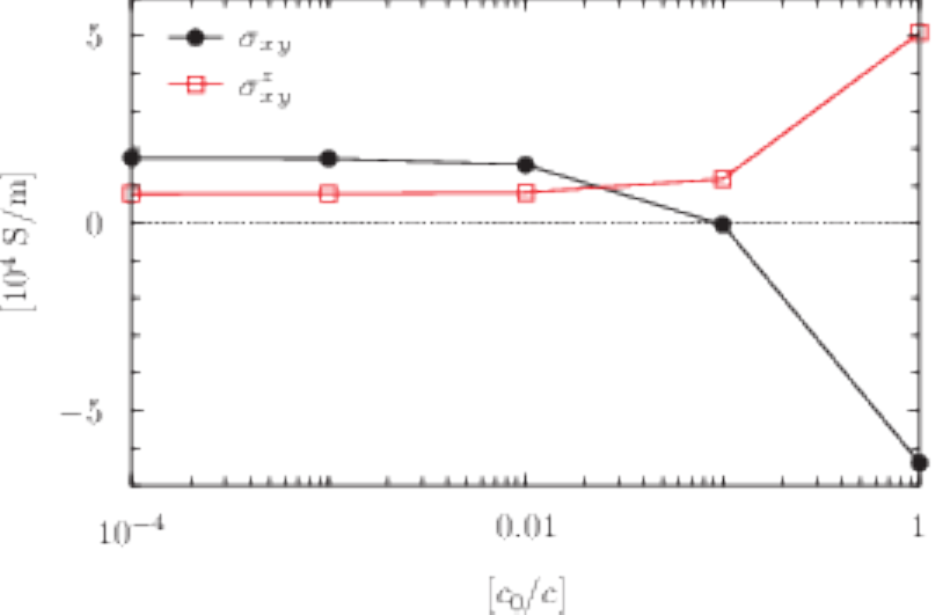}\\
   \includegraphics[angle=0,width=0.925\linewidth,clip]{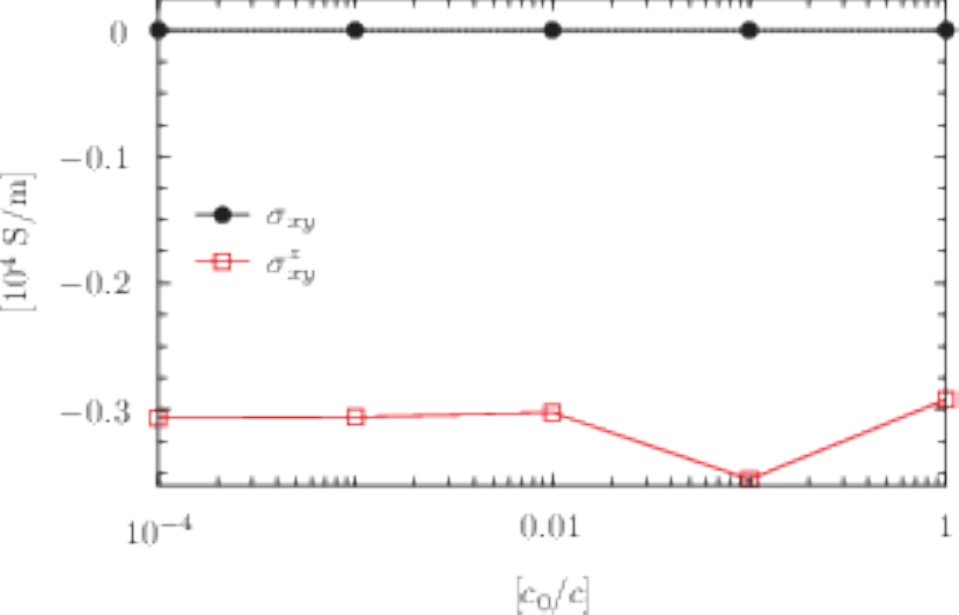}
\end{center}
 \caption{\label{fig:Cscaling_AHCnSHC_AFM9} Anomalous and spin
	Hall conductivity as a function of $c_0/c$ in the
	non-coplanar achiral magnetic state ncpM9 (top) for $\theta = 
        45^\circ$, and the co-planar non-collinear achiral state ncAFM9 (bottom). 
        The relativistic limit is on the right ($c_0/c = 1$), the 
        non-relativistic one on the left ($c_0/c \rightarrow 0$).}
\end{figure}

Corresponding results for the structures ncpM9 (again for $\theta = 
45^\circ$) and ncAFM9 are presented in 
Fig.~\ref{fig:Cscaling_AHCnSHC_AFM9}. Here the relation of chirality- and 
SOC-induced contributions is somewhat different. The former is smaller and of 
opposite sign for the AHE, while of the same sign for the SHE in the 
non-coplanar magnetic structure (top). In the coplanar antiferromagnetic state (bottom)
the AHE is again vanishing, whereas the SHE is predominantly chirality-induced.
%


\begin{thebibliography}{84}%
\makeatletter
\providecommand \@ifxundefined [1]{%
 \@ifx{#1\undefined}
}%
\providecommand \@ifnum [1]{%
 \ifnum #1\expandafter \@firstoftwo
 \else \expandafter \@secondoftwo
 \fi
}%
\providecommand \@ifx [1]{%
 \ifx #1\expandafter \@firstoftwo
 \else \expandafter \@secondoftwo
 \fi
}%
\providecommand \natexlab [1]{#1}%
\providecommand \enquote  [1]{``#1''}%
\providecommand \bibnamefont  [1]{#1}%
\providecommand \bibfnamefont [1]{#1}%
\providecommand \citenamefont [1]{#1}%
\providecommand \href@noop [0]{\@secondoftwo}%
\providecommand \href [0]{\begingroup \@sanitize@url \@href}%
\providecommand \@href[1]{\@@startlink{#1}\@@href}%
\providecommand \@@href[1]{\endgroup#1\@@endlink}%
\providecommand \@sanitize@url [0]{\catcode `\\12\catcode `\$12\catcode
  `\&12\catcode `\#12\catcode `\^12\catcode `\_12\catcode `\%12\relax}%
\providecommand \@@startlink[1]{}%
\providecommand \@@endlink[0]{}%
\providecommand \url  [0]{\begingroup\@sanitize@url \@url }%
\providecommand \@url [1]{\endgroup\@href {#1}{\urlprefix }}%
\providecommand \urlprefix  [0]{URL }%
\providecommand \Eprint [0]{\href }%
\providecommand \doibase [0]{http://dx.doi.org/}%
\providecommand \selectlanguage [0]{\@gobble}%
\providecommand \bibinfo  [0]{\@secondoftwo}%
\providecommand \bibfield  [0]{\@secondoftwo}%
\providecommand \translation [1]{[#1]}%
\providecommand \BibitemOpen [0]{}%
\providecommand \bibitemStop [0]{}%
\providecommand \bibitemNoStop [0]{.\EOS\space}%
\providecommand \EOS [0]{\spacefactor3000\relax}%
\providecommand \BibitemShut  [1]{\csname bibitem#1\endcsname}%
\let\auto@bib@innerbib\@empty
\bibitem [{\citenamefont {Bode}\ \emph {et~al.}(2007)\citenamefont {Bode},
  \citenamefont {Heide}, \citenamefont {von Bergmann}, \citenamefont
  {Ferriani}, \citenamefont {Heinze}, \citenamefont {Bihlmayer}, \citenamefont
  {Kubetzka}, \citenamefont {Pietzsch}, \citenamefont {Bl{\"u}gel},\ and\
  \citenamefont {Wiesendanger}}]{BHB+07}%
  \BibitemOpen
  \bibfield  {author} {\bibinfo {author} {\bibfnamefont {M.}~\bibnamefont
  {Bode}}, \bibinfo {author} {\bibfnamefont {M.}~\bibnamefont {Heide}},
  \bibinfo {author} {\bibfnamefont {K.}~\bibnamefont {von Bergmann}}, \bibinfo
  {author} {\bibfnamefont {P.}~\bibnamefont {Ferriani}}, \bibinfo {author}
  {\bibfnamefont {S.}~\bibnamefont {Heinze}}, \bibinfo {author} {\bibfnamefont
  {G.}~\bibnamefont {Bihlmayer}}, \bibinfo {author} {\bibfnamefont
  {A.}~\bibnamefont {Kubetzka}}, \bibinfo {author} {\bibfnamefont
  {O.}~\bibnamefont {Pietzsch}}, \bibinfo {author} {\bibfnamefont
  {S.}~\bibnamefont {Bl{\"u}gel}}, \ and\ \bibinfo {author} {\bibfnamefont
  {R.}~\bibnamefont {Wiesendanger}},\ }\href
  {http://dx.doi.org/10.1038/nature05802} {\bibfield  {journal} {\bibinfo
  {journal} {Nature}\ }\textbf {\bibinfo {volume} {447}},\ \bibinfo {pages}
  {190} (\bibinfo {year} {2007})}\BibitemShut {NoStop}%
\bibitem [{\citenamefont {Train}\ \emph {et~al.}(2008)\citenamefont {Train},
  \citenamefont {Gheorghe}, \citenamefont {Krstic}, \citenamefont {Chamoreau},
  \citenamefont {Ovanesyan}, \citenamefont {Rikken}, \citenamefont {Gruselle},\
  and\ \citenamefont {Verdaguer}}]{TGK+08}%
  \BibitemOpen
  \bibfield  {author} {\bibinfo {author} {\bibfnamefont {C.}~\bibnamefont
  {Train}}, \bibinfo {author} {\bibfnamefont {R.}~\bibnamefont {Gheorghe}},
  \bibinfo {author} {\bibfnamefont {V.}~\bibnamefont {Krstic}}, \bibinfo
  {author} {\bibfnamefont {L.-M.}\ \bibnamefont {Chamoreau}}, \bibinfo {author}
  {\bibfnamefont {N.~S.}\ \bibnamefont {Ovanesyan}}, \bibinfo {author}
  {\bibfnamefont {G.~L. J.~A.}\ \bibnamefont {Rikken}}, \bibinfo {author}
  {\bibfnamefont {M.}~\bibnamefont {Gruselle}}, \ and\ \bibinfo {author}
  {\bibfnamefont {M.}~\bibnamefont {Verdaguer}},\ }\href
  {http://dx.doi.org/10.1038/nmat2256} {\bibfield  {journal} {\bibinfo
  {journal} {Nature Materials}\ }\textbf {\bibinfo {volume} {7}},\ \bibinfo
  {pages} {729} (\bibinfo {year} {2008})}\BibitemShut {NoStop}%
\bibitem [{\citenamefont {Braun}(2012)}]{Bra12}%
  \BibitemOpen
  \bibfield  {author} {\bibinfo {author} {\bibfnamefont {H.-B.}\ \bibnamefont
  {Braun}},\ }\href {\doibase 10.1080/00018732.2012.663070} {\bibfield
  {journal} {\bibinfo  {journal} {Advances in Physics}\ }\textbf {\bibinfo
  {volume} {61}},\ \bibinfo {pages} {1} (\bibinfo {year} {2012})}\BibitemShut
  {NoStop}%
\bibitem [{\citenamefont {Stein}\ \emph {et~al.}(2017)\citenamefont {Stein},
  \citenamefont {Baum}, \citenamefont {Holbein}, \citenamefont {Finger},
  \citenamefont {Cronert}, \citenamefont {T\"olzer}, \citenamefont
  {Fr\"ohlich}, \citenamefont {Biesenkamp}, \citenamefont {Schmalzl},
  \citenamefont {Steffens}, \citenamefont {Lee},\ and\ \citenamefont
  {Braden}}]{SBH+17}%
  \BibitemOpen
  \bibfield  {author} {\bibinfo {author} {\bibfnamefont {J.}~\bibnamefont
  {Stein}}, \bibinfo {author} {\bibfnamefont {M.}~\bibnamefont {Baum}},
  \bibinfo {author} {\bibfnamefont {S.}~\bibnamefont {Holbein}}, \bibinfo
  {author} {\bibfnamefont {T.}~\bibnamefont {Finger}}, \bibinfo {author}
  {\bibfnamefont {T.}~\bibnamefont {Cronert}}, \bibinfo {author} {\bibfnamefont
  {C.}~\bibnamefont {T\"olzer}}, \bibinfo {author} {\bibfnamefont
  {T.}~\bibnamefont {Fr\"ohlich}}, \bibinfo {author} {\bibfnamefont
  {S.}~\bibnamefont {Biesenkamp}}, \bibinfo {author} {\bibfnamefont
  {K.}~\bibnamefont {Schmalzl}}, \bibinfo {author} {\bibfnamefont
  {P.}~\bibnamefont {Steffens}}, \bibinfo {author} {\bibfnamefont {C.~H.}\
  \bibnamefont {Lee}}, \ and\ \bibinfo {author} {\bibfnamefont
  {M.}~\bibnamefont {Braden}},\ }\href {\doibase
  10.1103/PhysRevLett.119.177201} {\bibfield  {journal} {\bibinfo  {journal}
  {Phys. Rev. Lett.}\ }\textbf {\bibinfo {volume} {119}},\ \bibinfo {pages}
  {177201} (\bibinfo {year} {2017})}\BibitemShut {NoStop}%
\bibitem [{\citenamefont {Spencer}\ \emph {et~al.}(2018)\citenamefont
  {Spencer}, \citenamefont {Gayles}, \citenamefont {Porter}, \citenamefont
  {Sugimoto}, \citenamefont {Aslam}, \citenamefont {Kinane}, \citenamefont
  {Charlton}, \citenamefont {Freimuth}, \citenamefont {Chadov}, \citenamefont
  {Langridge}, \citenamefont {Sinova}, \citenamefont {Felser}, \citenamefont
  {Bl\"ugel}, \citenamefont {Mokrousov},\ and\ \citenamefont
  {Marrows}}]{SGP+18}%
  \BibitemOpen
  \bibfield  {author} {\bibinfo {author} {\bibfnamefont {C.~S.}\ \bibnamefont
  {Spencer}}, \bibinfo {author} {\bibfnamefont {J.}~\bibnamefont {Gayles}},
  \bibinfo {author} {\bibfnamefont {N.~A.}\ \bibnamefont {Porter}}, \bibinfo
  {author} {\bibfnamefont {S.}~\bibnamefont {Sugimoto}}, \bibinfo {author}
  {\bibfnamefont {Z.}~\bibnamefont {Aslam}}, \bibinfo {author} {\bibfnamefont
  {C.~J.}\ \bibnamefont {Kinane}}, \bibinfo {author} {\bibfnamefont {T.~R.}\
  \bibnamefont {Charlton}}, \bibinfo {author} {\bibfnamefont {F.}~\bibnamefont
  {Freimuth}}, \bibinfo {author} {\bibfnamefont {S.}~\bibnamefont {Chadov}},
  \bibinfo {author} {\bibfnamefont {S.}~\bibnamefont {Langridge}}, \bibinfo
  {author} {\bibfnamefont {J.}~\bibnamefont {Sinova}}, \bibinfo {author}
  {\bibfnamefont {C.}~\bibnamefont {Felser}}, \bibinfo {author} {\bibfnamefont
  {S.}~\bibnamefont {Bl\"ugel}}, \bibinfo {author} {\bibfnamefont
  {Y.}~\bibnamefont {Mokrousov}}, \ and\ \bibinfo {author} {\bibfnamefont
  {C.~H.}\ \bibnamefont {Marrows}},\ }\href {\doibase
  10.1103/PhysRevB.97.214406} {\bibfield  {journal} {\bibinfo  {journal} {Phys.
  Rev. B}\ }\textbf {\bibinfo {volume} {97}},\ \bibinfo {pages} {214406}
  (\bibinfo {year} {2018})}\BibitemShut {NoStop}%
\bibitem [{\citenamefont {Skyrme}(1962)}]{Sky62}%
  \BibitemOpen
  \bibfield  {author} {\bibinfo {author} {\bibfnamefont {T.}~\bibnamefont
  {Skyrme}},\ }\href {\doibase http://dx.doi.org/10.1016/0029-5582(62)90775-7}
  {\bibfield  {journal} {\bibinfo  {journal} {Nuclear Physics}\ }\textbf
  {\bibinfo {volume} {31}},\ \bibinfo {pages} {556 } (\bibinfo {year}
  {1962})}\BibitemShut {NoStop}%
\bibitem [{\citenamefont {M{\"u}hlbauer}\ \emph {et~al.}(2009)\citenamefont
  {M{\"u}hlbauer}, \citenamefont {Binz}, \citenamefont {Jonietz}, \citenamefont
  {Pfleiderer}, \citenamefont {Rosch}, \citenamefont {Neubauer}, \citenamefont
  {Georgii},\ and\ \citenamefont {B{\"o}ni}}]{MBJ+09}%
  \BibitemOpen
  \bibfield  {author} {\bibinfo {author} {\bibfnamefont {S.}~\bibnamefont
  {M{\"u}hlbauer}}, \bibinfo {author} {\bibfnamefont {B.}~\bibnamefont {Binz}},
  \bibinfo {author} {\bibfnamefont {F.}~\bibnamefont {Jonietz}}, \bibinfo
  {author} {\bibfnamefont {C.}~\bibnamefont {Pfleiderer}}, \bibinfo {author}
  {\bibfnamefont {A.}~\bibnamefont {Rosch}}, \bibinfo {author} {\bibfnamefont
  {A.}~\bibnamefont {Neubauer}}, \bibinfo {author} {\bibfnamefont
  {R.}~\bibnamefont {Georgii}}, \ and\ \bibinfo {author} {\bibfnamefont
  {P.}~\bibnamefont {B{\"o}ni}},\ }\href {\doibase 10.1126/science.1166767}
  {\bibfield  {journal} {\bibinfo  {journal} {Science}\ }\textbf {\bibinfo
  {volume} {323}},\ \bibinfo {pages} {915} (\bibinfo {year}
  {2009})}\BibitemShut {NoStop}%
\bibitem [{\citenamefont {Neubauer}\ \emph {et~al.}(2009)\citenamefont
  {Neubauer}, \citenamefont {Pfleiderer}, \citenamefont {Binz}, \citenamefont
  {Rosch}, \citenamefont {Ritz}, \citenamefont {Niklowitz},\ and\ \citenamefont
  {B\"oni}}]{NPB+09}%
  \BibitemOpen
  \bibfield  {author} {\bibinfo {author} {\bibfnamefont {A.}~\bibnamefont
  {Neubauer}}, \bibinfo {author} {\bibfnamefont {C.}~\bibnamefont
  {Pfleiderer}}, \bibinfo {author} {\bibfnamefont {B.}~\bibnamefont {Binz}},
  \bibinfo {author} {\bibfnamefont {A.}~\bibnamefont {Rosch}}, \bibinfo
  {author} {\bibfnamefont {R.}~\bibnamefont {Ritz}}, \bibinfo {author}
  {\bibfnamefont {P.~G.}\ \bibnamefont {Niklowitz}}, \ and\ \bibinfo {author}
  {\bibfnamefont {P.}~\bibnamefont {B\"oni}},\ }\href {\doibase
  10.1103/PhysRevLett.102.186602} {\bibfield  {journal} {\bibinfo  {journal}
  {Phys. Rev. Lett.}\ }\textbf {\bibinfo {volume} {102}},\ \bibinfo {pages}
  {186602} (\bibinfo {year} {2009})},\ \bibinfo {note} {erratum: Phys. Rev.
  Lett. \textbf{110}, 209902 (2013)}\BibitemShut {NoStop}%
\bibitem [{\citenamefont {Heinze}\ \emph {et~al.}(2011)\citenamefont {Heinze},
  \citenamefont {von Bergmann}, \citenamefont {Menzel}, \citenamefont {Brede},
  \citenamefont {Kubetzka}, \citenamefont {Wiesendanger}, \citenamefont
  {Bihlmayer},\ and\ \citenamefont {Blugel}}]{HBM+11}%
  \BibitemOpen
  \bibfield  {author} {\bibinfo {author} {\bibfnamefont {S.}~\bibnamefont
  {Heinze}}, \bibinfo {author} {\bibfnamefont {K.}~\bibnamefont {von
  Bergmann}}, \bibinfo {author} {\bibfnamefont {M.}~\bibnamefont {Menzel}},
  \bibinfo {author} {\bibfnamefont {J.}~\bibnamefont {Brede}}, \bibinfo
  {author} {\bibfnamefont {A.}~\bibnamefont {Kubetzka}}, \bibinfo {author}
  {\bibfnamefont {R.}~\bibnamefont {Wiesendanger}}, \bibinfo {author}
  {\bibfnamefont {G.}~\bibnamefont {Bihlmayer}}, \ and\ \bibinfo {author}
  {\bibfnamefont {S.}~\bibnamefont {Blugel}},\ }\href {\doibase
  10.1038/nphys2045} {\bibfield  {journal} {\bibinfo  {journal} {Nat. Phys.}\
  }\textbf {\bibinfo {volume} {7}},\ \bibinfo {pages} {713} (\bibinfo {year}
  {2011})}\BibitemShut {NoStop}%
\bibitem [{\citenamefont {Dupe}\ \emph {et~al.}(2014)\citenamefont {Dupe},
  \citenamefont {Hoffmann}, \citenamefont {Paillard},\ and\ \citenamefont
  {Heinze}}]{DHPH14}%
  \BibitemOpen
  \bibfield  {author} {\bibinfo {author} {\bibfnamefont {B.}~\bibnamefont
  {Dupe}}, \bibinfo {author} {\bibfnamefont {M.}~\bibnamefont {Hoffmann}},
  \bibinfo {author} {\bibfnamefont {C.}~\bibnamefont {Paillard}}, \ and\
  \bibinfo {author} {\bibfnamefont {S.}~\bibnamefont {Heinze}},\ }\href
  {http://dx.doi.org/10.1038/ncomms5030} {\bibfield  {journal} {\bibinfo
  {journal} {Nature Communications}\ }\textbf {\bibinfo {volume} {5}},\
  \bibinfo {pages} {4030} (\bibinfo {year} {2014})}\BibitemShut {NoStop}%
\bibitem [{\citenamefont {Reichlov\'a}\ \emph {et~al.}(2015)\citenamefont
  {Reichlov\'a}, \citenamefont {Kriegner}, \citenamefont {Hol\'y},
  \citenamefont {Olejn\'{\i}k}, \citenamefont {Nov\'ak}, \citenamefont
  {Yamada}, \citenamefont {Miura}, \citenamefont {Ogawa}, \citenamefont
  {Takahashi}, \citenamefont {Jungwirth},\ and\ \citenamefont
  {Wunderlich}}]{RKH+15}%
  \BibitemOpen
  \bibfield  {author} {\bibinfo {author} {\bibfnamefont {H.}~\bibnamefont
  {Reichlov\'a}}, \bibinfo {author} {\bibfnamefont {D.}~\bibnamefont
  {Kriegner}}, \bibinfo {author} {\bibfnamefont {V.}~\bibnamefont {Hol\'y}},
  \bibinfo {author} {\bibfnamefont {K.}~\bibnamefont {Olejn\'{\i}k}}, \bibinfo
  {author} {\bibfnamefont {V.}~\bibnamefont {Nov\'ak}}, \bibinfo {author}
  {\bibfnamefont {M.}~\bibnamefont {Yamada}}, \bibinfo {author} {\bibfnamefont
  {K.}~\bibnamefont {Miura}}, \bibinfo {author} {\bibfnamefont
  {S.}~\bibnamefont {Ogawa}}, \bibinfo {author} {\bibfnamefont
  {H.}~\bibnamefont {Takahashi}}, \bibinfo {author} {\bibfnamefont
  {T.}~\bibnamefont {Jungwirth}}, \ and\ \bibinfo {author} {\bibfnamefont
  {J.}~\bibnamefont {Wunderlich}},\ }\href {\doibase
  10.1103/PhysRevB.92.165424} {\bibfield  {journal} {\bibinfo  {journal} {Phys.
  Rev. B}\ }\textbf {\bibinfo {volume} {92}},\ \bibinfo {pages} {165424}
  (\bibinfo {year} {2015})}\BibitemShut {NoStop}%
\bibitem [{\citenamefont {Moreau-Luchaire}\ \emph {et~al.}(2016)\citenamefont
  {Moreau-Luchaire}, \citenamefont {Moutaﬁs}, \citenamefont {Reyren},
  \citenamefont {Sampaio}, \citenamefont {Vaz}, \citenamefont {Van~Horne},
  \citenamefont {Bouzehouane}, \citenamefont {Garcia}, \citenamefont
  {Deranlot}, \citenamefont {Warnicke}, \citenamefont {Wohlh{\"u}ter},
  \citenamefont {George}, \citenamefont {Weigand}, \citenamefont {Raabe},
  \citenamefont {Cros},\ and\ \citenamefont {Fert}}]{MMR+16}%
  \BibitemOpen
  \bibfield  {author} {\bibinfo {author} {\bibfnamefont {C.}~\bibnamefont
  {Moreau-Luchaire}}, \bibinfo {author} {\bibfnamefont {C.}~\bibnamefont
  {Moutaﬁs}}, \bibinfo {author} {\bibfnamefont {N.}~\bibnamefont {Reyren}},
  \bibinfo {author} {\bibfnamefont {J.}~\bibnamefont {Sampaio}}, \bibinfo
  {author} {\bibfnamefont {C.~A.~F.}\ \bibnamefont {Vaz}}, \bibinfo {author}
  {\bibfnamefont {N.}~\bibnamefont {Van~Horne}}, \bibinfo {author}
  {\bibfnamefont {K.}~\bibnamefont {Bouzehouane}}, \bibinfo {author}
  {\bibfnamefont {K.}~\bibnamefont {Garcia}}, \bibinfo {author} {\bibfnamefont
  {C.}~\bibnamefont {Deranlot}}, \bibinfo {author} {\bibfnamefont
  {P.}~\bibnamefont {Warnicke}}, \bibinfo {author} {\bibfnamefont
  {P.}~\bibnamefont {Wohlh{\"u}ter}}, \bibinfo {author} {\bibfnamefont {J.-M.}\
  \bibnamefont {George}}, \bibinfo {author} {\bibfnamefont {M.}~\bibnamefont
  {Weigand}}, \bibinfo {author} {\bibfnamefont {J.}~\bibnamefont {Raabe}},
  \bibinfo {author} {\bibfnamefont {V.}~\bibnamefont {Cros}}, \ and\ \bibinfo
  {author} {\bibfnamefont {A.}~\bibnamefont {Fert}},\ }\href
  {http://dx.doi.org/10.1038/nnano.2015.313} {\bibfield  {journal} {\bibinfo
  {journal} {Nature Nanotechnology}\ }\textbf {\bibinfo {volume} {11}},\
  \bibinfo {pages} {444} (\bibinfo {year} {2016})}\BibitemShut {NoStop}%
\bibitem [{\citenamefont {Woo}\ \emph {et~al.}(2016)\citenamefont {Woo},
  \citenamefont {Litzius}, \citenamefont {Kr{\"u}ger}, \citenamefont {Im},
  \citenamefont {Caretta}, \citenamefont {Richter}, \citenamefont {Mann},
  \citenamefont {Krone}, \citenamefont {Reeve}, \citenamefont {Weigand},
  \citenamefont {Agrawal}, \citenamefont {Lemesh}, \citenamefont {Mawass},
  \citenamefont {Fischer}, \citenamefont {Kl{\"a}ui},\ and\ \citenamefont
  {Beach}}]{WLK+16}%
  \BibitemOpen
  \bibfield  {author} {\bibinfo {author} {\bibfnamefont {S.}~\bibnamefont
  {Woo}}, \bibinfo {author} {\bibfnamefont {K.}~\bibnamefont {Litzius}},
  \bibinfo {author} {\bibfnamefont {B.}~\bibnamefont {Kr{\"u}ger}}, \bibinfo
  {author} {\bibfnamefont {M.-Y.}\ \bibnamefont {Im}}, \bibinfo {author}
  {\bibfnamefont {L.}~\bibnamefont {Caretta}}, \bibinfo {author} {\bibfnamefont
  {K.}~\bibnamefont {Richter}}, \bibinfo {author} {\bibfnamefont
  {M.}~\bibnamefont {Mann}}, \bibinfo {author} {\bibfnamefont {A.}~\bibnamefont
  {Krone}}, \bibinfo {author} {\bibfnamefont {R.~M.}\ \bibnamefont {Reeve}},
  \bibinfo {author} {\bibfnamefont {M.}~\bibnamefont {Weigand}}, \bibinfo
  {author} {\bibfnamefont {P.}~\bibnamefont {Agrawal}}, \bibinfo {author}
  {\bibfnamefont {I.}~\bibnamefont {Lemesh}}, \bibinfo {author} {\bibfnamefont
  {M.-A.}\ \bibnamefont {Mawass}}, \bibinfo {author} {\bibfnamefont
  {P.}~\bibnamefont {Fischer}}, \bibinfo {author} {\bibfnamefont
  {M.}~\bibnamefont {Kl{\"a}ui}}, \ and\ \bibinfo {author} {\bibfnamefont
  {G.~S.~D.}\ \bibnamefont {Beach}},\ }\href
  {http://dx.doi.org/10.1038/nmat4593} {\bibfield  {journal} {\bibinfo
  {journal} {Nature Materials}\ }\textbf {\bibinfo {volume} {15}},\ \bibinfo
  {pages} {501} (\bibinfo {year} {2016})}\BibitemShut {NoStop}%
\bibitem [{\citenamefont {Nayak}\ \emph {et~al.}(2017)\citenamefont {Nayak},
  \citenamefont {Kumar}, \citenamefont {Ma}, \citenamefont {Werner},
  \citenamefont {Pippel}, \citenamefont {Sahoo}, \citenamefont {Damay},
  \citenamefont {R{\"o}{\ss}ler}, \citenamefont {Felser},\ and\ \citenamefont
  {Parkin}}]{NKM+17}%
  \BibitemOpen
  \bibfield  {author} {\bibinfo {author} {\bibfnamefont {A.~K.}\ \bibnamefont
  {Nayak}}, \bibinfo {author} {\bibfnamefont {V.}~\bibnamefont {Kumar}},
  \bibinfo {author} {\bibfnamefont {T.}~\bibnamefont {Ma}}, \bibinfo {author}
  {\bibfnamefont {P.}~\bibnamefont {Werner}}, \bibinfo {author} {\bibfnamefont
  {E.}~\bibnamefont {Pippel}}, \bibinfo {author} {\bibfnamefont
  {R.}~\bibnamefont {Sahoo}}, \bibinfo {author} {\bibfnamefont
  {F.}~\bibnamefont {Damay}}, \bibinfo {author} {\bibfnamefont {U.~K.}\
  \bibnamefont {R{\"o}{\ss}ler}}, \bibinfo {author} {\bibfnamefont
  {C.}~\bibnamefont {Felser}}, \ and\ \bibinfo {author} {\bibfnamefont
  {S.~S.~P.}\ \bibnamefont {Parkin}},\ }\href
  {http://dx.doi.org/10.1038/nature23466} {\bibfield  {journal} {\bibinfo
  {journal} {Nature}\ }\textbf {\bibinfo {volume} {548}},\ \bibinfo {pages}
  {561} (\bibinfo {year} {2017})}\BibitemShut {NoStop}%
\bibitem [{\citenamefont {Yu}\ \emph {et~al.}(2012)\citenamefont {Yu},
  \citenamefont {Kanazawa}, \citenamefont {Zhang}, \citenamefont {Nagai},
  \citenamefont {Hara}, \citenamefont {Kimoto}, \citenamefont {Matsui},
  \citenamefont {Onose},\ and\ \citenamefont {Tokura}}]{YKZ+12}%
  \BibitemOpen
  \bibfield  {author} {\bibinfo {author} {\bibfnamefont {X.~Z.}\ \bibnamefont
  {Yu}}, \bibinfo {author} {\bibfnamefont {N.}~\bibnamefont {Kanazawa}},
  \bibinfo {author} {\bibfnamefont {W.~Z.}\ \bibnamefont {Zhang}}, \bibinfo
  {author} {\bibfnamefont {T.}~\bibnamefont {Nagai}}, \bibinfo {author}
  {\bibfnamefont {T.}~\bibnamefont {Hara}}, \bibinfo {author} {\bibfnamefont
  {K.}~\bibnamefont {Kimoto}}, \bibinfo {author} {\bibfnamefont
  {Y.}~\bibnamefont {Matsui}}, \bibinfo {author} {\bibfnamefont
  {Y.}~\bibnamefont {Onose}}, \ and\ \bibinfo {author} {\bibfnamefont
  {Y.}~\bibnamefont {Tokura}},\ }\href {http://dx.doi.org/10.1038/ncomms1990}
  {\bibfield  {journal} {\bibinfo  {journal} {Nature Communications}\ }\textbf
  {\bibinfo {volume} {3}},\ \bibinfo {pages} {988} (\bibinfo {year}
  {2012})}\BibitemShut {NoStop}%
\bibitem [{\citenamefont {Sampaio}\ \emph {et~al.}(2013)\citenamefont
  {Sampaio}, \citenamefont {Cros}, \citenamefont {Rohart}, \citenamefont
  {Thiaville},\ and\ \citenamefont {Fert}}]{SCR+13}%
  \BibitemOpen
  \bibfield  {author} {\bibinfo {author} {\bibfnamefont {J.}~\bibnamefont
  {Sampaio}}, \bibinfo {author} {\bibfnamefont {V.}~\bibnamefont {Cros}},
  \bibinfo {author} {\bibfnamefont {S.}~\bibnamefont {Rohart}}, \bibinfo
  {author} {\bibfnamefont {A.}~\bibnamefont {Thiaville}}, \ and\ \bibinfo
  {author} {\bibfnamefont {A.}~\bibnamefont {Fert}},\ }\href
  {http://dx.doi.org/10.1038/nnano.2013.210} {\bibfield  {journal} {\bibinfo
  {journal} {Nature Nanotechnology}\ }\textbf {\bibinfo {volume} {8}},\
  \bibinfo {pages} {839} (\bibinfo {year} {2013})}\BibitemShut {NoStop}%
\bibitem [{\citenamefont {Fert}\ \emph {et~al.}(2013)\citenamefont {Fert},
  \citenamefont {Cros},\ and\ \citenamefont {Sampaio}}]{FCS13}%
  \BibitemOpen
  \bibfield  {author} {\bibinfo {author} {\bibfnamefont {A.}~\bibnamefont
  {Fert}}, \bibinfo {author} {\bibfnamefont {V.}~\bibnamefont {Cros}}, \ and\
  \bibinfo {author} {\bibfnamefont {J.}~\bibnamefont {Sampaio}},\ }\href
  {http://dx.doi.org/10.1038/nnano.2013.29} {\bibfield  {journal} {\bibinfo
  {journal} {Nature Nanotechnology}\ }\textbf {\bibinfo {volume} {8}},\
  \bibinfo {pages} {152} (\bibinfo {year} {2013})}\BibitemShut {NoStop}%
\bibitem [{\citenamefont {Tomasello}\ \emph {et~al.}(2014)\citenamefont
  {Tomasello}, \citenamefont {Martinez}, \citenamefont {Zivieri}, \citenamefont
  {Torres}, \citenamefont {Carpentieri},\ and\ \citenamefont
  {Finocchio}}]{TMZ+14}%
  \BibitemOpen
  \bibfield  {author} {\bibinfo {author} {\bibfnamefont {R.}~\bibnamefont
  {Tomasello}}, \bibinfo {author} {\bibfnamefont {E.}~\bibnamefont {Martinez}},
  \bibinfo {author} {\bibfnamefont {R.}~\bibnamefont {Zivieri}}, \bibinfo
  {author} {\bibfnamefont {L.}~\bibnamefont {Torres}}, \bibinfo {author}
  {\bibfnamefont {M.}~\bibnamefont {Carpentieri}}, \ and\ \bibinfo {author}
  {\bibfnamefont {G.}~\bibnamefont {Finocchio}},\ }\href
  {http://dx.doi.org/10.1038/srep06784} {\bibfield  {journal} {\bibinfo
  {journal} {Scientific Reports}\ }\textbf {\bibinfo {volume} {4}},\ \bibinfo
  {pages} {6784} (\bibinfo {year} {2014})}\BibitemShut {NoStop}%
\bibitem [{\citenamefont {Fert}\ \emph {et~al.}(2017)\citenamefont {Fert},
  \citenamefont {Reyren},\ and\ \citenamefont {Cros}}]{FRC17}%
  \BibitemOpen
  \bibfield  {author} {\bibinfo {author} {\bibfnamefont {A.}~\bibnamefont
  {Fert}}, \bibinfo {author} {\bibfnamefont {N.}~\bibnamefont {Reyren}}, \ and\
  \bibinfo {author} {\bibfnamefont {V.}~\bibnamefont {Cros}},\ }\href
  {http://dx.doi.org/10.1038/natrevmats.2017.31} {\bibfield  {journal}
  {\bibinfo  {journal} {Nature Reviews Materials}\ }\textbf {\bibinfo {volume}
  {2}},\ \bibinfo {pages} {17031} (\bibinfo {year} {2017})}\BibitemShut
  {NoStop}%
\bibitem [{\citenamefont {Koshibae}\ and\ \citenamefont
  {Nagaosa}(2016)}]{KN16}%
  \BibitemOpen
  \bibfield  {author} {\bibinfo {author} {\bibfnamefont {W.}~\bibnamefont
  {Koshibae}}\ and\ \bibinfo {author} {\bibfnamefont {N.}~\bibnamefont
  {Nagaosa}},\ }\href {http://dx.doi.org/10.1038/ncomms10542} {\bibfield
  {journal} {\bibinfo  {journal} {Nature Communications}\ }\textbf {\bibinfo
  {volume} {7}},\ \bibinfo {pages} {10542} (\bibinfo {year}
  {2016})}\BibitemShut {NoStop}%
\bibitem [{\citenamefont {Shindou}\ and\ \citenamefont {Nagaosa}(2001)}]{SN01}%
  \BibitemOpen
  \bibfield  {author} {\bibinfo {author} {\bibfnamefont {R.}~\bibnamefont
  {Shindou}}\ and\ \bibinfo {author} {\bibfnamefont {N.}~\bibnamefont
  {Nagaosa}},\ }\href {\doibase 10.1103/PhysRevLett.87.116801} {\bibfield
  {journal} {\bibinfo  {journal} {Phys. Rev. Lett.}\ }\textbf {\bibinfo
  {volume} {87}},\ \bibinfo {pages} {116801} (\bibinfo {year}
  {2001})}\BibitemShut {NoStop}%
\bibitem [{\citenamefont {Nakamura}\ \emph {et~al.}(2003)\citenamefont
  {Nakamura}, \citenamefont {Ito},\ and\ \citenamefont {Freeman}}]{NIF03}%
  \BibitemOpen
  \bibfield  {author} {\bibinfo {author} {\bibfnamefont {K.}~\bibnamefont
  {Nakamura}}, \bibinfo {author} {\bibfnamefont {T.}~\bibnamefont {Ito}}, \
  and\ \bibinfo {author} {\bibfnamefont {A.~J.}\ \bibnamefont {Freeman}},\
  }\href {\doibase 10.1103/PhysRevB.68.180404} {\bibfield  {journal} {\bibinfo
  {journal} {Phys. Rev. B}\ }\textbf {\bibinfo {volume} {68}},\ \bibinfo
  {pages} {180404} (\bibinfo {year} {2003})}\BibitemShut {NoStop}%
\bibitem [{\citenamefont {Hoffmann}\ \emph {et~al.}(2015)\citenamefont
  {Hoffmann}, \citenamefont {Weischenberg}, \citenamefont {Dup\'e},
  \citenamefont {Freimuth}, \citenamefont {Ferriani}, \citenamefont
  {Mokrousov},\ and\ \citenamefont {Heinze}}]{HWD+15}%
  \BibitemOpen
  \bibfield  {author} {\bibinfo {author} {\bibfnamefont {M.}~\bibnamefont
  {Hoffmann}}, \bibinfo {author} {\bibfnamefont {J.}~\bibnamefont
  {Weischenberg}}, \bibinfo {author} {\bibfnamefont {B.}~\bibnamefont
  {Dup\'e}}, \bibinfo {author} {\bibfnamefont {F.}~\bibnamefont {Freimuth}},
  \bibinfo {author} {\bibfnamefont {P.}~\bibnamefont {Ferriani}}, \bibinfo
  {author} {\bibfnamefont {Y.}~\bibnamefont {Mokrousov}}, \ and\ \bibinfo
  {author} {\bibfnamefont {S.}~\bibnamefont {Heinze}},\ }\href {\doibase
  10.1103/PhysRevB.92.020401} {\bibfield  {journal} {\bibinfo  {journal} {Phys.
  Rev. B}\ }\textbf {\bibinfo {volume} {92}},\ \bibinfo {pages} {020401}
  (\bibinfo {year} {2015})}\BibitemShut {NoStop}%
\bibitem [{\citenamefont {Hanke}\ \emph {et~al.}(2016)\citenamefont {Hanke},
  \citenamefont {Freimuth}, \citenamefont {Nandy}, \citenamefont {Zhang},
  \citenamefont {Bl\"ugel},\ and\ \citenamefont {Mokrousov}}]{HFN+16}%
  \BibitemOpen
  \bibfield  {author} {\bibinfo {author} {\bibfnamefont {J.-P.}\ \bibnamefont
  {Hanke}}, \bibinfo {author} {\bibfnamefont {F.}~\bibnamefont {Freimuth}},
  \bibinfo {author} {\bibfnamefont {A.~K.}\ \bibnamefont {Nandy}}, \bibinfo
  {author} {\bibfnamefont {H.}~\bibnamefont {Zhang}}, \bibinfo {author}
  {\bibfnamefont {S.}~\bibnamefont {Bl\"ugel}}, \ and\ \bibinfo {author}
  {\bibfnamefont {Y.}~\bibnamefont {Mokrousov}},\ }\href {\doibase
  10.1103/PhysRevB.94.121114} {\bibfield  {journal} {\bibinfo  {journal} {Phys.
  Rev. B}\ }\textbf {\bibinfo {volume} {94}},\ \bibinfo {pages} {121114}
  (\bibinfo {year} {2016})}\BibitemShut {NoStop}%
\bibitem [{\citenamefont {dos Santos~Dias}\ \emph {et~al.}(2016)\citenamefont
  {dos Santos~Dias}, \citenamefont {Bouaziz}, \citenamefont {Bouhassoune},
  \citenamefont {Bl{\"u}gel},\ and\ \citenamefont {Lounis}}]{DBB+16a}%
  \BibitemOpen
  \bibfield  {author} {\bibinfo {author} {\bibfnamefont {M.}~\bibnamefont {dos
  Santos~Dias}}, \bibinfo {author} {\bibfnamefont {J.}~\bibnamefont {Bouaziz}},
  \bibinfo {author} {\bibfnamefont {M.}~\bibnamefont {Bouhassoune}}, \bibinfo
  {author} {\bibfnamefont {S.}~\bibnamefont {Bl{\"u}gel}}, \ and\ \bibinfo
  {author} {\bibfnamefont {S.}~\bibnamefont {Lounis}},\ }\href
  {http://dx.doi.org/10.1038/ncomms13613} {\bibfield  {journal} {\bibinfo
  {journal} {Nature Communications}\ }\textbf {\bibinfo {volume} {7}},\
  \bibinfo {pages} {13613} (\bibinfo {year} {2016})}\BibitemShut {NoStop}%
\bibitem [{\citenamefont {Hanke}\ \emph {et~al.}(2017)\citenamefont {Hanke},
  \citenamefont {Freimuth}, \citenamefont {Bl{\"u}gel},\ and\ \citenamefont
  {Mokrousov}}]{HFBM17}%
  \BibitemOpen
  \bibfield  {author} {\bibinfo {author} {\bibfnamefont {J.-P.}\ \bibnamefont
  {Hanke}}, \bibinfo {author} {\bibfnamefont {F.}~\bibnamefont {Freimuth}},
  \bibinfo {author} {\bibfnamefont {S.}~\bibnamefont {Bl{\"u}gel}}, \ and\
  \bibinfo {author} {\bibfnamefont {Y.}~\bibnamefont {Mokrousov}},\ }\href
  {http://dx.doi.org/10.1038/srep41078} {\bibfield  {journal} {\bibinfo
  {journal} {Scientific Reports}\ }\textbf {\bibinfo {volume} {7}},\ \bibinfo
  {pages} {41078} (\bibinfo {year} {2017})}\BibitemShut {NoStop}%
\bibitem [{\citenamefont {Taguchi}\ \emph {et~al.}(2001)\citenamefont
  {Taguchi}, \citenamefont {Oohara}, \citenamefont {Yoshizawa}, \citenamefont
  {Nagaosa},\ and\ \citenamefont {Tokura}}]{TOY+01}%
  \BibitemOpen
  \bibfield  {author} {\bibinfo {author} {\bibfnamefont {Y.}~\bibnamefont
  {Taguchi}}, \bibinfo {author} {\bibfnamefont {Y.}~\bibnamefont {Oohara}},
  \bibinfo {author} {\bibfnamefont {H.}~\bibnamefont {Yoshizawa}}, \bibinfo
  {author} {\bibfnamefont {N.}~\bibnamefont {Nagaosa}}, \ and\ \bibinfo
  {author} {\bibfnamefont {Y.}~\bibnamefont {Tokura}},\ }\href {\doibase
  10.1126/science.1058161} {\bibfield  {journal} {\bibinfo  {journal}
  {Science}\ }\textbf {\bibinfo {volume} {291}},\ \bibinfo {pages} {2573}
  (\bibinfo {year} {2001})}\BibitemShut {NoStop}%
\bibitem [{\citenamefont {Tatara}\ and\ \citenamefont {Kawamura}(2002)}]{TK02}%
  \BibitemOpen
  \bibfield  {author} {\bibinfo {author} {\bibfnamefont {G.}~\bibnamefont
  {Tatara}}\ and\ \bibinfo {author} {\bibfnamefont {H.}~\bibnamefont
  {Kawamura}},\ }\href {\doibase 10.1143/JPSJ.71.2613} {\bibfield  {journal}
  {\bibinfo  {journal} {J. Phys. Soc. Japan}\ }\textbf {\bibinfo {volume}
  {71}},\ \bibinfo {pages} {2613} (\bibinfo {year} {2002})}\BibitemShut
  {NoStop}%
\bibitem [{\citenamefont {Onoda}\ and\ \citenamefont {Nagaosa}(2002)}]{ON02}%
  \BibitemOpen
  \bibfield  {author} {\bibinfo {author} {\bibfnamefont {M.}~\bibnamefont
  {Onoda}}\ and\ \bibinfo {author} {\bibfnamefont {N.}~\bibnamefont
  {Nagaosa}},\ }\href {\doibase 10.1143/JPSJ.71.19} {\bibfield  {journal}
  {\bibinfo  {journal} {Journal of the Physical Society of Japan}\ }\textbf
  {\bibinfo {volume} {71}},\ \bibinfo {pages} {19} (\bibinfo {year}
  {2002})}\BibitemShut {NoStop}%
\bibitem [{\citenamefont {Bruno}\ \emph {et~al.}(2004)\citenamefont {Bruno},
  \citenamefont {Dugaev},\ and\ \citenamefont {Taillefumier}}]{BDT04}%
  \BibitemOpen
  \bibfield  {author} {\bibinfo {author} {\bibfnamefont {P.}~\bibnamefont
  {Bruno}}, \bibinfo {author} {\bibfnamefont {V.~K.}\ \bibnamefont {Dugaev}}, \
  and\ \bibinfo {author} {\bibfnamefont {M.}~\bibnamefont {Taillefumier}},\
  }\href {\doibase 10.1103/PhysRevLett.93.096806} {\bibfield  {journal}
  {\bibinfo  {journal} {Phys. Rev. Lett.}\ }\textbf {\bibinfo {volume} {93}},\
  \bibinfo {pages} {096806} (\bibinfo {year} {2004})}\BibitemShut {NoStop}%
\bibitem [{\citenamefont {Tomizawa}\ and\ \citenamefont
  {Kontani}(2010)}]{TK10}%
  \BibitemOpen
  \bibfield  {author} {\bibinfo {author} {\bibfnamefont {T.}~\bibnamefont
  {Tomizawa}}\ and\ \bibinfo {author} {\bibfnamefont {H.}~\bibnamefont
  {Kontani}},\ }\href {\doibase 10.1103/PhysRevB.82.104412} {\bibfield
  {journal} {\bibinfo  {journal} {Phys. Rev. B}\ }\textbf {\bibinfo {volume}
  {82}},\ \bibinfo {pages} {104412} (\bibinfo {year} {2010})}\BibitemShut
  {NoStop}%
\bibitem [{\citenamefont {Schulz}\ \emph {et~al.}(2012)\citenamefont {Schulz},
  \citenamefont {Ritz}, \citenamefont {Bauer}, \citenamefont {Halder},
  \citenamefont {Wagner}, \citenamefont {Franz}, \citenamefont {Pfleiderer},
  \citenamefont {Everschor}, \citenamefont {Garst},\ and\ \citenamefont
  {Rosch}}]{SRB+12}%
  \BibitemOpen
  \bibfield  {author} {\bibinfo {author} {\bibfnamefont {T.}~\bibnamefont
  {Schulz}}, \bibinfo {author} {\bibfnamefont {R.}~\bibnamefont {Ritz}},
  \bibinfo {author} {\bibfnamefont {A.}~\bibnamefont {Bauer}}, \bibinfo
  {author} {\bibfnamefont {M.}~\bibnamefont {Halder}}, \bibinfo {author}
  {\bibfnamefont {M.}~\bibnamefont {Wagner}}, \bibinfo {author} {\bibfnamefont
  {C.}~\bibnamefont {Franz}}, \bibinfo {author} {\bibfnamefont
  {C.}~\bibnamefont {Pfleiderer}}, \bibinfo {author} {\bibfnamefont
  {K.}~\bibnamefont {Everschor}}, \bibinfo {author} {\bibfnamefont
  {M.}~\bibnamefont {Garst}}, \ and\ \bibinfo {author} {\bibfnamefont
  {A.}~\bibnamefont {Rosch}},\ }\href {http://dx.doi.org/10.1038/nphys2231}
  {\bibfield  {journal} {\bibinfo  {journal} {Nature Physics}\ }\textbf
  {\bibinfo {volume} {8}},\ \bibinfo {pages} {301} (\bibinfo {year}
  {2012})}\BibitemShut {NoStop}%
\bibitem [{\citenamefont {Nagaosa}\ and\ \citenamefont {Tokura}(2013)}]{NT13}%
  \BibitemOpen
  \bibfield  {author} {\bibinfo {author} {\bibfnamefont {N.}~\bibnamefont
  {Nagaosa}}\ and\ \bibinfo {author} {\bibfnamefont {Y.}~\bibnamefont
  {Tokura}},\ }\href {http://dx.doi.org/10.1038/nnano.2013.243} {\bibfield
  {journal} {\bibinfo  {journal} {Nature Nanotechnology}\ }\textbf {\bibinfo
  {volume} {8}},\ \bibinfo {pages} {899} (\bibinfo {year} {2013})}\BibitemShut
  {NoStop}%
\bibitem [{\citenamefont {Franz}\ \emph {et~al.}(2014)\citenamefont {Franz},
  \citenamefont {Freimuth}, \citenamefont {Bauer}, \citenamefont {Ritz},
  \citenamefont {Schnarr}, \citenamefont {Duvinage}, \citenamefont {Adams},
  \citenamefont {Bl\"ugel}, \citenamefont {Rosch}, \citenamefont {Mokrousov},\
  and\ \citenamefont {Pfleiderer}}]{FFB+14}%
  \BibitemOpen
  \bibfield  {author} {\bibinfo {author} {\bibfnamefont {C.}~\bibnamefont
  {Franz}}, \bibinfo {author} {\bibfnamefont {F.}~\bibnamefont {Freimuth}},
  \bibinfo {author} {\bibfnamefont {A.}~\bibnamefont {Bauer}}, \bibinfo
  {author} {\bibfnamefont {R.}~\bibnamefont {Ritz}}, \bibinfo {author}
  {\bibfnamefont {C.}~\bibnamefont {Schnarr}}, \bibinfo {author} {\bibfnamefont
  {C.}~\bibnamefont {Duvinage}}, \bibinfo {author} {\bibfnamefont
  {T.}~\bibnamefont {Adams}}, \bibinfo {author} {\bibfnamefont
  {S.}~\bibnamefont {Bl\"ugel}}, \bibinfo {author} {\bibfnamefont
  {A.}~\bibnamefont {Rosch}}, \bibinfo {author} {\bibfnamefont
  {Y.}~\bibnamefont {Mokrousov}}, \ and\ \bibinfo {author} {\bibfnamefont
  {C.}~\bibnamefont {Pfleiderer}},\ }\href {\doibase
  10.1103/PhysRevLett.112.186601} {\bibfield  {journal} {\bibinfo  {journal}
  {Phys. Rev. Lett.}\ }\textbf {\bibinfo {volume} {112}},\ \bibinfo {pages}
  {186601} (\bibinfo {year} {2014})}\BibitemShut {NoStop}%
\bibitem [{\citenamefont {Yin}\ \emph {et~al.}(2015)\citenamefont {Yin},
  \citenamefont {Liu}, \citenamefont {Barlas}, \citenamefont {Zang},\ and\
  \citenamefont {Lake}}]{YLB+15}%
  \BibitemOpen
  \bibfield  {author} {\bibinfo {author} {\bibfnamefont {G.}~\bibnamefont
  {Yin}}, \bibinfo {author} {\bibfnamefont {Y.}~\bibnamefont {Liu}}, \bibinfo
  {author} {\bibfnamefont {Y.}~\bibnamefont {Barlas}}, \bibinfo {author}
  {\bibfnamefont {J.}~\bibnamefont {Zang}}, \ and\ \bibinfo {author}
  {\bibfnamefont {R.~K.}\ \bibnamefont {Lake}},\ }\href {\doibase
  10.1103/PhysRevB.92.024411} {\bibfield  {journal} {\bibinfo  {journal} {Phys.
  Rev. B}\ }\textbf {\bibinfo {volume} {92}},\ \bibinfo {pages} {024411}
  (\bibinfo {year} {2015})}\BibitemShut {NoStop}%
\bibitem [{\citenamefont {Buhl}\ \emph {et~al.}(2017)\citenamefont {Buhl},
  \citenamefont {Freimuth}, \citenamefont {Bl\"ugel},\ and\ \citenamefont
  {Mokrousov}}]{BFBM17}%
  \BibitemOpen
  \bibfield  {author} {\bibinfo {author} {\bibfnamefont {P.~M.}\ \bibnamefont
  {Buhl}}, \bibinfo {author} {\bibfnamefont {F.}~\bibnamefont {Freimuth}},
  \bibinfo {author} {\bibfnamefont {S.}~\bibnamefont {Bl\"ugel}}, \ and\
  \bibinfo {author} {\bibfnamefont {Y.}~\bibnamefont {Mokrousov}},\ }\href
  {\doibase 10.1002/pssr.201700007} {\bibfield  {journal} {\bibinfo  {journal}
  {physica status solidi (RRL) – Rapid Research Letters}\ }\textbf {\bibinfo
  {volume} {11}},\ \bibinfo {pages} {1700007} (\bibinfo {year} {2017})},\
  \bibinfo {note} {1700007}\BibitemShut {NoStop}%
\bibitem [{\citenamefont {Chen}\ \emph {et~al.}(2014)\citenamefont {Chen},
  \citenamefont {Niu},\ and\ \citenamefont {MacDonald}}]{CNM14}%
  \BibitemOpen
  \bibfield  {author} {\bibinfo {author} {\bibfnamefont {H.}~\bibnamefont
  {Chen}}, \bibinfo {author} {\bibfnamefont {Q.}~\bibnamefont {Niu}}, \ and\
  \bibinfo {author} {\bibfnamefont {A.~H.}\ \bibnamefont {MacDonald}},\ }\href
  {\doibase 10.1103/PhysRevLett.112.017205} {\bibfield  {journal} {\bibinfo
  {journal} {Phys. Rev. Lett.}\ }\textbf {\bibinfo {volume} {112}},\ \bibinfo
  {pages} {017205} (\bibinfo {year} {2014})}\BibitemShut {NoStop}%
\bibitem [{\citenamefont {K\"ubler}\ and\ \citenamefont
  {Felser}(2014)}]{KF14a}%
  \BibitemOpen
  \bibfield  {author} {\bibinfo {author} {\bibfnamefont {J.}~\bibnamefont
  {K\"ubler}}\ and\ \bibinfo {author} {\bibfnamefont {C.}~\bibnamefont
  {Felser}},\ }\href {\doibase 10.1209/0295-5075/108/67001} {\bibfield
  {journal} {\bibinfo  {journal} {EPL (Europhysics Letters)}\ }\textbf
  {\bibinfo {volume} {108}},\ \bibinfo {pages} {67001} (\bibinfo {year}
  {2014})}\BibitemShut {NoStop}%
\bibitem [{\citenamefont {Zhang}\ \emph {et~al.}(2017)\citenamefont {Zhang},
  \citenamefont {Sun}, \citenamefont {Yang}, \citenamefont
  {\ifmmode~\check{Z}\else \v{Z}\fi{}elezn\'y}, \citenamefont {Parkin},
  \citenamefont {Felser},\ and\ \citenamefont {Yan}}]{ZSY+17}%
  \BibitemOpen
  \bibfield  {author} {\bibinfo {author} {\bibfnamefont {Y.}~\bibnamefont
  {Zhang}}, \bibinfo {author} {\bibfnamefont {Y.}~\bibnamefont {Sun}}, \bibinfo
  {author} {\bibfnamefont {H.}~\bibnamefont {Yang}}, \bibinfo {author}
  {\bibfnamefont {J.}~\bibnamefont {\ifmmode~\check{Z}\else
  \v{Z}\fi{}elezn\'y}}, \bibinfo {author} {\bibfnamefont {S.~P.~P.}\
  \bibnamefont {Parkin}}, \bibinfo {author} {\bibfnamefont {C.}~\bibnamefont
  {Felser}}, \ and\ \bibinfo {author} {\bibfnamefont {B.}~\bibnamefont {Yan}},\
  }\href {\doibase 10.1103/PhysRevB.95.075128} {\bibfield  {journal} {\bibinfo
  {journal} {Phys. Rev. B}\ }\textbf {\bibinfo {volume} {95}},\ \bibinfo
  {pages} {075128} (\bibinfo {year} {2017})}\BibitemShut {NoStop}%
\bibitem [{\citenamefont {Boldrin}\ and\ \citenamefont {Wills}(2012)}]{BW12}%
  \BibitemOpen
  \bibfield  {author} {\bibinfo {author} {\bibfnamefont {D.}~\bibnamefont
  {Boldrin}}\ and\ \bibinfo {author} {\bibfnamefont {A.~S.}\ \bibnamefont
  {Wills}},\ }\href {\doibase 10.1155/2012/615295} {\bibfield  {journal}
  {\bibinfo  {journal} {Advances in Condensed Matter Physics}\ }\textbf
  {\bibinfo {volume} {2012}},\ \bibinfo {pages} {615295} (\bibinfo {year}
  {2012})}\BibitemShut {NoStop}%
\bibitem [{\citenamefont {S{\"u}rgers}\ \emph {et~al.}(2014)\citenamefont
  {S{\"u}rgers}, \citenamefont {Fischer}, \citenamefont {Winkel},\ and\
  \citenamefont {L{\"o}hneysen}}]{SFWL14}%
  \BibitemOpen
  \bibfield  {author} {\bibinfo {author} {\bibfnamefont {C.}~\bibnamefont
  {S{\"u}rgers}}, \bibinfo {author} {\bibfnamefont {G.}~\bibnamefont
  {Fischer}}, \bibinfo {author} {\bibfnamefont {P.}~\bibnamefont {Winkel}}, \
  and\ \bibinfo {author} {\bibfnamefont {H.~v.}\ \bibnamefont
  {L{\"o}hneysen}},\ }\href {http://dx.doi.org/10.1038/ncomms4400} {\bibfield
  {journal} {\bibinfo  {journal} {Nature Communications}\ }\textbf {\bibinfo
  {volume} {5}},\ \bibinfo {pages} {3400} (\bibinfo {year} {2014})}\BibitemShut
  {NoStop}%
\bibitem [{\citenamefont {Nakatsuji}\ \emph {et~al.}(2015)\citenamefont
  {Nakatsuji}, \citenamefont {Kiyohara},\ and\ \citenamefont {Higo}}]{NKH15}%
  \BibitemOpen
  \bibfield  {author} {\bibinfo {author} {\bibfnamefont {S.}~\bibnamefont
  {Nakatsuji}}, \bibinfo {author} {\bibfnamefont {N.}~\bibnamefont {Kiyohara}},
  \ and\ \bibinfo {author} {\bibfnamefont {T.}~\bibnamefont {Higo}},\ }\href
  {http://dx.doi.org/10.1038/nature15723} {\bibfield  {journal} {\bibinfo
  {journal} {Nature}\ }\textbf {\bibinfo {volume} {527}},\ \bibinfo {pages}
  {212} (\bibinfo {year} {2015})}\BibitemShut {NoStop}%
\bibitem [{\citenamefont {Kiyohara}\ \emph {et~al.}(2016)\citenamefont
  {Kiyohara}, \citenamefont {Tomita},\ and\ \citenamefont {Nakatsuji}}]{KTN16}%
  \BibitemOpen
  \bibfield  {author} {\bibinfo {author} {\bibfnamefont {N.}~\bibnamefont
  {Kiyohara}}, \bibinfo {author} {\bibfnamefont {T.}~\bibnamefont {Tomita}}, \
  and\ \bibinfo {author} {\bibfnamefont {S.}~\bibnamefont {Nakatsuji}},\ }\href
  {\doibase 10.1103/PhysRevApplied.5.064009} {\bibfield  {journal} {\bibinfo
  {journal} {Phys. Rev. Applied}\ }\textbf {\bibinfo {volume} {5}},\ \bibinfo
  {pages} {064009} (\bibinfo {year} {2016})}\BibitemShut {NoStop}%
\bibitem [{\citenamefont {Nayak}\ \emph {et~al.}(2016)\citenamefont {Nayak},
  \citenamefont {Fischer}, \citenamefont {Sun}, \citenamefont {Yan},
  \citenamefont {Karel}, \citenamefont {Komarek}, \citenamefont {Shekhar},
  \citenamefont {Kumar}, \citenamefont {Schnelle}, \citenamefont {K{\"u}bler},
  \citenamefont {Felser},\ and\ \citenamefont {Parkin}}]{NFS+16}%
  \BibitemOpen
  \bibfield  {author} {\bibinfo {author} {\bibfnamefont {A.~K.}\ \bibnamefont
  {Nayak}}, \bibinfo {author} {\bibfnamefont {J.~E.}\ \bibnamefont {Fischer}},
  \bibinfo {author} {\bibfnamefont {Y.}~\bibnamefont {Sun}}, \bibinfo {author}
  {\bibfnamefont {B.}~\bibnamefont {Yan}}, \bibinfo {author} {\bibfnamefont
  {J.}~\bibnamefont {Karel}}, \bibinfo {author} {\bibfnamefont {A.~C.}\
  \bibnamefont {Komarek}}, \bibinfo {author} {\bibfnamefont {C.}~\bibnamefont
  {Shekhar}}, \bibinfo {author} {\bibfnamefont {N.}~\bibnamefont {Kumar}},
  \bibinfo {author} {\bibfnamefont {W.}~\bibnamefont {Schnelle}}, \bibinfo
  {author} {\bibfnamefont {J.}~\bibnamefont {K{\"u}bler}}, \bibinfo {author}
  {\bibfnamefont {C.}~\bibnamefont {Felser}}, \ and\ \bibinfo {author}
  {\bibfnamefont {S.~S.~P.}\ \bibnamefont {Parkin}},\ }\href {\doibase
  10.1126/sciadv.1501870} {\bibfield  {journal} {\bibinfo  {journal} {Sci.
  Adv.}\ }\textbf {\bibinfo {volume} {2}},\ \bibinfo {pages} {e1501870}
  (\bibinfo {year} {2016})}\BibitemShut {NoStop}%
\bibitem [{\citenamefont {S\"{u}rgers}\ \emph {et~al.}(2017)\citenamefont
  {S\"{u}rgers}, \citenamefont {Wolf}, \citenamefont {Adelmann}, \citenamefont
  {Kittler}, \citenamefont {Fischer},\ and\ \citenamefont
  {L\"{o}hneysen}}]{SWA+17}%
  \BibitemOpen
  \bibfield  {author} {\bibinfo {author} {\bibfnamefont {C.}~\bibnamefont
  {S\"{u}rgers}}, \bibinfo {author} {\bibfnamefont {T.}~\bibnamefont {Wolf}},
  \bibinfo {author} {\bibfnamefont {P.}~\bibnamefont {Adelmann}}, \bibinfo
  {author} {\bibfnamefont {W.}~\bibnamefont {Kittler}}, \bibinfo {author}
  {\bibfnamefont {G.}~\bibnamefont {Fischer}}, \ and\ \bibinfo {author}
  {\bibfnamefont {H.~v.}\ \bibnamefont {L\"{o}hneysen}},\ }\href
  {http://dx.doi.org/10.1038/srep42982} {\bibfield  {journal} {\bibinfo
  {journal} {Scientific Reports}\ }\textbf {\bibinfo {volume} {7}},\ \bibinfo
  {pages} {42982} (\bibinfo {year} {2017})}\BibitemShut {NoStop}%
\bibitem [{\citenamefont {Liu}\ \emph {et~al.}(2017)\citenamefont {Liu},
  \citenamefont {Zhang}, \citenamefont {Liu}, \citenamefont {Ding},
  \citenamefont {Liu}, \citenamefont {Jafri}, \citenamefont {Hou},
  \citenamefont {Wang}, \citenamefont {Ma},\ and\ \citenamefont {Wu}}]{LZL+17}%
  \BibitemOpen
  \bibfield  {author} {\bibinfo {author} {\bibfnamefont {Z.~H.}\ \bibnamefont
  {Liu}}, \bibinfo {author} {\bibfnamefont {Y.~J.}\ \bibnamefont {Zhang}},
  \bibinfo {author} {\bibfnamefont {G.~D.}\ \bibnamefont {Liu}}, \bibinfo
  {author} {\bibfnamefont {B.}~\bibnamefont {Ding}}, \bibinfo {author}
  {\bibfnamefont {E.~K.}\ \bibnamefont {Liu}}, \bibinfo {author} {\bibfnamefont
  {H.~M.}\ \bibnamefont {Jafri}}, \bibinfo {author} {\bibfnamefont {Z.~P.}\
  \bibnamefont {Hou}}, \bibinfo {author} {\bibfnamefont {W.~H.}\ \bibnamefont
  {Wang}}, \bibinfo {author} {\bibfnamefont {X.~Q.}\ \bibnamefont {Ma}}, \ and\
  \bibinfo {author} {\bibfnamefont {G.~H.}\ \bibnamefont {Wu}},\ }\href
  {\doibase 10.1038/s41598-017-00621-x} {\bibfield  {journal} {\bibinfo
  {journal} {Scientific Reports}\ }\textbf {\bibinfo {volume} {7}},\ \bibinfo
  {pages} {515} (\bibinfo {year} {2017})}\BibitemShut {NoStop}%
\bibitem [{\citenamefont {Orenstein}\ and\ \citenamefont {Moore}(2013)}]{OM13}%
  \BibitemOpen
  \bibfield  {author} {\bibinfo {author} {\bibfnamefont {J.}~\bibnamefont
  {Orenstein}}\ and\ \bibinfo {author} {\bibfnamefont {J.~E.}\ \bibnamefont
  {Moore}},\ }\href {\doibase 10.1103/PhysRevB.87.165110} {\bibfield  {journal}
  {\bibinfo  {journal} {Phys. Rev. B}\ }\textbf {\bibinfo {volume} {87}},\
  \bibinfo {pages} {165110} (\bibinfo {year} {2013})}\BibitemShut {NoStop}%
\bibitem [{\citenamefont {{Feng}}\ \emph {et~al.}(2015)\citenamefont {{Feng}},
  \citenamefont {{Guo}}, \citenamefont {{Zhou}}, \citenamefont {{Yao}},\ and\
  \citenamefont {{Niu}}}]{FGZ+15}%
  \BibitemOpen
  \bibfield  {author} {\bibinfo {author} {\bibfnamefont {W.}~\bibnamefont
  {{Feng}}}, \bibinfo {author} {\bibfnamefont {G.-Y.}\ \bibnamefont {{Guo}}},
  \bibinfo {author} {\bibfnamefont {J.}~\bibnamefont {{Zhou}}}, \bibinfo
  {author} {\bibfnamefont {Y.}~\bibnamefont {{Yao}}}, \ and\ \bibinfo {author}
  {\bibfnamefont {Q.}~\bibnamefont {{Niu}}},\ }\href
  {http://arxiv.org/abs/1509.02865} {\bibfield  {journal} {\bibinfo  {journal}
  {ArXiv e-prints}\ } (\bibinfo {year} {2015})},\ \bibinfo {note}
  {arXiv:1509.02865 [cond-mat.mtrl-sci]},\ \Eprint
  {http://arxiv.org/abs/1509.02865} {arXiv:1509.02865 [cond-mat.mtrl-sci]}
  \BibitemShut {NoStop}%
\bibitem [{\citenamefont {Wimmer}\ \emph
  {et~al.}(2018{\natexlab{a}})\citenamefont {Wimmer}, \citenamefont {Min\'ar},
  \citenamefont {Mankovsky}, \citenamefont {Yaresko},\ and\ \citenamefont
  {Ebert}}]{WMM+18}%
  \BibitemOpen
  \bibfield  {author} {\bibinfo {author} {\bibfnamefont {S.}~\bibnamefont
  {Wimmer}}, \bibinfo {author} {\bibfnamefont {J.}~\bibnamefont {Min\'ar}},
  \bibinfo {author} {\bibfnamefont {S.}~\bibnamefont {Mankovsky}}, \bibinfo
  {author} {\bibfnamefont {A.~N.}\ \bibnamefont {Yaresko}}, \ and\ \bibinfo
  {author} {\bibfnamefont {H.}~\bibnamefont {Ebert}},\ }\href@noop {} {\enquote
  {\bibinfo {title} {{Magneto-optic and transverse transport properties of
  noncollinear antiferromagnets}},}\ }\bibinfo {howpublished} {unpublished}
  (\bibinfo {year} {2018}{\natexlab{a}})\BibitemShut {NoStop}%
\bibitem [{\citenamefont {Gomonay}(2015)}]{Gom15}%
  \BibitemOpen
  \bibfield  {author} {\bibinfo {author} {\bibfnamefont {O.}~\bibnamefont
  {Gomonay}},\ }\href {\doibase 10.1103/PhysRevB.91.144421} {\bibfield
  {journal} {\bibinfo  {journal} {Phys. Rev. B}\ }\textbf {\bibinfo {volume}
  {91}},\ \bibinfo {pages} {144421} (\bibinfo {year} {2015})}\BibitemShut
  {NoStop}%
\bibitem [{\citenamefont {\ifmmode~\check{Z}\else \v{Z}\fi{}elezn\'y}\ \emph
  {et~al.}(2017)\citenamefont {\ifmmode~\check{Z}\else \v{Z}\fi{}elezn\'y},
  \citenamefont {Zhang}, \citenamefont {Felser},\ and\ \citenamefont
  {Yan}}]{ZZFY17a}%
  \BibitemOpen
  \bibfield  {author} {\bibinfo {author} {\bibfnamefont {J.}~\bibnamefont
  {\ifmmode~\check{Z}\else \v{Z}\fi{}elezn\'y}}, \bibinfo {author}
  {\bibfnamefont {Y.}~\bibnamefont {Zhang}}, \bibinfo {author} {\bibfnamefont
  {C.}~\bibnamefont {Felser}}, \ and\ \bibinfo {author} {\bibfnamefont
  {B.}~\bibnamefont {Yan}},\ }\href {\doibase 10.1103/PhysRevLett.119.187204}
  {\bibfield  {journal} {\bibinfo  {journal} {Phys. Rev. Lett.}\ }\textbf
  {\bibinfo {volume} {119}},\ \bibinfo {pages} {187204} (\bibinfo {year}
  {2017})}\BibitemShut {NoStop}%
\bibitem [{\citenamefont {{Zhang}}\ \emph {et~al.}(2017)\citenamefont
  {{Zhang}}, \citenamefont {{Zelezny}}, \citenamefont {{Sun}}, \citenamefont
  {{van den Brink}},\ and\ \citenamefont {{Yan}}}]{ZZS+17}%
  \BibitemOpen
  \bibfield  {author} {\bibinfo {author} {\bibfnamefont {Y.}~\bibnamefont
  {{Zhang}}}, \bibinfo {author} {\bibfnamefont {J.}~\bibnamefont {{Zelezny}}},
  \bibinfo {author} {\bibfnamefont {Y.}~\bibnamefont {{Sun}}}, \bibinfo
  {author} {\bibfnamefont {J.}~\bibnamefont {{van den Brink}}}, \ and\ \bibinfo
  {author} {\bibfnamefont {B.}~\bibnamefont {{Yan}}},\ }\href@noop {}
  {\bibfield  {journal} {\bibinfo  {journal} {ArXiv e-prints}\ } (\bibinfo
  {year} {2017})},\ \Eprint {http://arxiv.org/abs/1704.03917} {arXiv:1704.03917
  [cond-mat.mtrl-sci]} \BibitemShut {NoStop}%
\bibitem [{\citenamefont {Mendes}\ \emph {et~al.}(2014)\citenamefont {Mendes},
  \citenamefont {Cunha}, \citenamefont {Alves~Santos}, \citenamefont {Ribeiro},
  \citenamefont {Machado}, \citenamefont {Rodr\'{\i}guez-Su\'arez},
  \citenamefont {Azevedo},\ and\ \citenamefont {Rezende}}]{MCA+14}%
  \BibitemOpen
  \bibfield  {author} {\bibinfo {author} {\bibfnamefont {J.~B.~S.}\
  \bibnamefont {Mendes}}, \bibinfo {author} {\bibfnamefont {R.~O.}\
  \bibnamefont {Cunha}}, \bibinfo {author} {\bibfnamefont {O.}~\bibnamefont
  {Alves~Santos}}, \bibinfo {author} {\bibfnamefont {P.~R.~T.}\ \bibnamefont
  {Ribeiro}}, \bibinfo {author} {\bibfnamefont {F.~L.~A.}\ \bibnamefont
  {Machado}}, \bibinfo {author} {\bibfnamefont {R.~L.}\ \bibnamefont
  {Rodr\'{\i}guez-Su\'arez}}, \bibinfo {author} {\bibfnamefont
  {A.}~\bibnamefont {Azevedo}}, \ and\ \bibinfo {author} {\bibfnamefont
  {S.~M.}\ \bibnamefont {Rezende}},\ }\href {\doibase
  10.1103/PhysRevB.89.140406} {\bibfield  {journal} {\bibinfo  {journal} {Phys.
  Rev. B}\ }\textbf {\bibinfo {volume} {89}},\ \bibinfo {pages} {140406}
  (\bibinfo {year} {2014})}\BibitemShut {NoStop}%
\bibitem [{\citenamefont {Zhang}\ \emph {et~al.}(2016)\citenamefont {Zhang},
  \citenamefont {Han}, \citenamefont {Yang}, \citenamefont {Sun}, \citenamefont
  {Zhang}, \citenamefont {Yan},\ and\ \citenamefont {Parkin}}]{ZHY+16a}%
  \BibitemOpen
  \bibfield  {author} {\bibinfo {author} {\bibfnamefont {W.}~\bibnamefont
  {Zhang}}, \bibinfo {author} {\bibfnamefont {W.}~\bibnamefont {Han}}, \bibinfo
  {author} {\bibfnamefont {S.-H.}\ \bibnamefont {Yang}}, \bibinfo {author}
  {\bibfnamefont {Y.}~\bibnamefont {Sun}}, \bibinfo {author} {\bibfnamefont
  {Y.}~\bibnamefont {Zhang}}, \bibinfo {author} {\bibfnamefont
  {B.}~\bibnamefont {Yan}}, \ and\ \bibinfo {author} {\bibfnamefont {S.~S.~P.}\
  \bibnamefont {Parkin}},\ }\href {\doibase 10.1126/sciadv.1600759} {\bibfield
  {journal} {\bibinfo  {journal} {Sci. Adv.}\ }\textbf {\bibinfo {volume}
  {2}},\ \bibinfo {pages} {e1600759} (\bibinfo {year} {2016})}\BibitemShut
  {NoStop}%
\bibitem [{\citenamefont {Guo}\ and\ \citenamefont {Wang}(2017)}]{GW17}%
  \BibitemOpen
  \bibfield  {author} {\bibinfo {author} {\bibfnamefont {G.-Y.}\ \bibnamefont
  {Guo}}\ and\ \bibinfo {author} {\bibfnamefont {T.-C.}\ \bibnamefont {Wang}},\
  }\href {\doibase 10.1103/PhysRevB.96.224415} {\bibfield  {journal} {\bibinfo
  {journal} {Phys. Rev. B}\ }\textbf {\bibinfo {volume} {96}},\ \bibinfo
  {pages} {224415} (\bibinfo {year} {2017})}\BibitemShut {NoStop}%
\bibitem [{\citenamefont {Ikhlas}\ \emph {et~al.}(2017)\citenamefont {Ikhlas},
  \citenamefont {Tomita}, \citenamefont {Koretsune}, \citenamefont {Suzuki},
  \citenamefont {Nishio-Hamane}, \citenamefont {Arita}, \citenamefont {Otani},\
  and\ \citenamefont {Nakatsuji}}]{ITK+17}%
  \BibitemOpen
  \bibfield  {author} {\bibinfo {author} {\bibfnamefont {M.}~\bibnamefont
  {Ikhlas}}, \bibinfo {author} {\bibfnamefont {T.}~\bibnamefont {Tomita}},
  \bibinfo {author} {\bibfnamefont {T.}~\bibnamefont {Koretsune}}, \bibinfo
  {author} {\bibfnamefont {M.-T.}\ \bibnamefont {Suzuki}}, \bibinfo {author}
  {\bibfnamefont {D.}~\bibnamefont {Nishio-Hamane}}, \bibinfo {author}
  {\bibfnamefont {R.}~\bibnamefont {Arita}}, \bibinfo {author} {\bibfnamefont
  {Y.}~\bibnamefont {Otani}}, \ and\ \bibinfo {author} {\bibfnamefont
  {S.}~\bibnamefont {Nakatsuji}},\ }\href {http://dx.doi.org/10.1038/nphys4181}
  {\bibfield  {journal} {\bibinfo  {journal} {Nat. Phys.}\ }\textbf {\bibinfo
  {volume} {13}},\ \bibinfo {pages} {1085} (\bibinfo {year}
  {2017})}\BibitemShut {NoStop}%
\bibitem [{\citenamefont {{\v{S}}mejkal}\ \emph {et~al.}(2018)\citenamefont
  {{\v{S}}mejkal}, \citenamefont {Mokrousov}, \citenamefont {Yan},\ and\
  \citenamefont {MacDonald}}]{SMYM18}%
  \BibitemOpen
  \bibfield  {author} {\bibinfo {author} {\bibfnamefont {L.}~\bibnamefont
  {{\v{S}}mejkal}}, \bibinfo {author} {\bibfnamefont {Y.}~\bibnamefont
  {Mokrousov}}, \bibinfo {author} {\bibfnamefont {B.}~\bibnamefont {Yan}}, \
  and\ \bibinfo {author} {\bibfnamefont {A.~H.}\ \bibnamefont {MacDonald}},\
  }\href {\doibase 10.1038/s41567-018-0064-5} {\bibfield  {journal} {\bibinfo
  {journal} {Nature Physics}\ }\textbf {\bibinfo {volume} {14}},\ \bibinfo
  {pages} {242} (\bibinfo {year} {2018})}\BibitemShut {NoStop}%
\bibitem [{\citenamefont {Kleiner}(1966)}]{Kle66}%
  \BibitemOpen
  \bibfield  {author} {\bibinfo {author} {\bibfnamefont {W.~H.}\ \bibnamefont
  {Kleiner}},\ }\href {\doibase 10.1103/PhysRev.142.318} {\bibfield  {journal}
  {\bibinfo  {journal} {Phys. Rev.}\ }\textbf {\bibinfo {volume} {142}},\
  \bibinfo {pages} {318} (\bibinfo {year} {1966})}\BibitemShut {NoStop}%
\bibitem [{\citenamefont {Seemann}\ \emph {et~al.}(2015)\citenamefont
  {Seemann}, \citenamefont {K\"odderitzsch}, \citenamefont {Wimmer},\ and\
  \citenamefont {Ebert}}]{SKWE15a}%
  \BibitemOpen
  \bibfield  {author} {\bibinfo {author} {\bibfnamefont {M.}~\bibnamefont
  {Seemann}}, \bibinfo {author} {\bibfnamefont {D.}~\bibnamefont
  {K\"odderitzsch}}, \bibinfo {author} {\bibfnamefont {S.}~\bibnamefont
  {Wimmer}}, \ and\ \bibinfo {author} {\bibfnamefont {H.}~\bibnamefont
  {Ebert}},\ }\href {\doibase 10.1103/PhysRevB.92.155138} {\bibfield  {journal}
  {\bibinfo  {journal} {Phys. Rev. B}\ }\textbf {\bibinfo {volume} {92}},\
  \bibinfo {pages} {155138} (\bibinfo {year} {2015})}\BibitemShut {NoStop}%
\bibitem [{\citenamefont {Wimmer}\ \emph {et~al.}(2016)\citenamefont {Wimmer},
  \citenamefont {Chadova}, \citenamefont {Seemann}, \citenamefont
  {K\"odderitzsch},\ and\ \citenamefont {Ebert}}]{WCS+16a}%
  \BibitemOpen
  \bibfield  {author} {\bibinfo {author} {\bibfnamefont {S.}~\bibnamefont
  {Wimmer}}, \bibinfo {author} {\bibfnamefont {K.}~\bibnamefont {Chadova}},
  \bibinfo {author} {\bibfnamefont {M.}~\bibnamefont {Seemann}}, \bibinfo
  {author} {\bibfnamefont {D.}~\bibnamefont {K\"odderitzsch}}, \ and\ \bibinfo
  {author} {\bibfnamefont {H.}~\bibnamefont {Ebert}},\ }\href {\doibase
  10.1103/PhysRevB.94.054415} {\bibfield  {journal} {\bibinfo  {journal} {Phys.
  Rev. B}\ }\textbf {\bibinfo {volume} {94}},\ \bibinfo {pages} {054415}
  (\bibinfo {year} {2016})}\BibitemShut {NoStop}%
\bibitem [{\citenamefont {Wimmer}\ \emph
  {et~al.}(2018{\natexlab{b}})\citenamefont {Wimmer}, \citenamefont {Chadova},\
  and\ \citenamefont {Ebert}}]{WCE18}%
  \BibitemOpen
  \bibfield  {author} {\bibinfo {author} {\bibfnamefont {S.}~\bibnamefont
  {Wimmer}}, \bibinfo {author} {\bibfnamefont {K.}~\bibnamefont {Chadova}}, \
  and\ \bibinfo {author} {\bibfnamefont {H.}~\bibnamefont {Ebert}},\
  }\href@noop {} {\enquote {\bibinfo {title} {{Symmetry and magnitude of the
  direct and inverse Edelstein effect: A KKR-CPA-Kubo approach}},}\ }\bibinfo
  {howpublished} {unpublished} (\bibinfo {year}
  {2018}{\natexlab{b}})\BibitemShut {NoStop}%
\bibitem [{\citenamefont {Stokes}\ \emph {et~al.}()\citenamefont {Stokes},
  \citenamefont {Hatch},\ and\ \citenamefont {Campbell}}]{ISOTROPY}%
  \BibitemOpen
  \bibfield  {author} {\bibinfo {author} {\bibfnamefont {H.~T.}\ \bibnamefont
  {Stokes}}, \bibinfo {author} {\bibfnamefont {D.~M.}\ \bibnamefont {Hatch}}, \
  and\ \bibinfo {author} {\bibfnamefont {B.~J.}\ \bibnamefont {Campbell}},\
  }\href {http://stokes.byu.edu/iso/isotropy.php} {}\bibinfo {howpublished}
  {ISOTROPY Software Suite, \url{iso.byu.edu}.}\BibitemShut {Stop}%
\bibitem [{\citenamefont {Stokes}\ and\ \citenamefont {Hatch}(2005)}]{SH05}%
  \BibitemOpen
  \bibfield  {author} {\bibinfo {author} {\bibfnamefont {H.~T.}\ \bibnamefont
  {Stokes}}\ and\ \bibinfo {author} {\bibfnamefont {D.~M.}\ \bibnamefont
  {Hatch}},\ }\href
  {http://onlinelibrary.wiley.com/doi/10.1111/jcr.2005.38.issue-1/issuetoc}
  {\bibfield  {journal} {\bibinfo  {journal} {J. Appl. Crystallogr,}\ }\textbf
  {\bibinfo {volume} {38}},\ \bibinfo {pages} {237} (\bibinfo {year}
  {2005})}\BibitemShut {NoStop}%
\bibitem [{\citenamefont {Kubo}(1956)}]{Kub56}%
  \BibitemOpen
  \bibfield  {author} {\bibinfo {author} {\bibfnamefont {R.}~\bibnamefont
  {Kubo}},\ }\href {\doibase 10.1139/p56-140} {\bibfield  {journal} {\bibinfo
  {journal} {Can. J. Phys.}\ }\textbf {\bibinfo {volume} {34}},\ \bibinfo
  {pages} {1274} (\bibinfo {year} {1956})}\BibitemShut {NoStop}%
\bibitem [{\citenamefont {Kubo}(1957)}]{Kub57}%
  \BibitemOpen
  \bibfield  {author} {\bibinfo {author} {\bibfnamefont {R.}~\bibnamefont
  {Kubo}},\ }\href {\doibase 10.1143/JPSJ.12.570} {\bibfield  {journal}
  {\bibinfo  {journal} {J. Phys. Soc. Japan}\ }\textbf {\bibinfo {volume}
  {12}},\ \bibinfo {pages} {570} (\bibinfo {year} {1957})}\BibitemShut
  {NoStop}%
\bibitem [{\citenamefont {Butler}(1985)}]{But85}%
  \BibitemOpen
  \bibfield  {author} {\bibinfo {author} {\bibfnamefont {W.~H.}\ \bibnamefont
  {Butler}},\ }\href {\doibase 10.1103/PhysRevB.31.3260} {\bibfield  {journal}
  {\bibinfo  {journal} {Phys. Rev. B}\ }\textbf {\bibinfo {volume} {31}},\
  \bibinfo {pages} {3260} (\bibinfo {year} {1985})}\BibitemShut {NoStop}%
\bibitem [{\citenamefont {Cr\'epieux}\ and\ \citenamefont
  {Bruno}(2001)}]{CB01a}%
  \BibitemOpen
  \bibfield  {author} {\bibinfo {author} {\bibfnamefont {A.}~\bibnamefont
  {Cr\'epieux}}\ and\ \bibinfo {author} {\bibfnamefont {P.}~\bibnamefont
  {Bruno}},\ }\href {\doibase 10.1103/PhysRevB.64.014416} {\bibfield  {journal}
  {\bibinfo  {journal} {Phys. Rev. B}\ }\textbf {\bibinfo {volume} {64}},\
  \bibinfo {pages} {014416} (\bibinfo {year} {2001})}\BibitemShut {NoStop}%
\bibitem [{\citenamefont {K\"{o}dderitzsch}\ \emph {et~al.}(2015)\citenamefont
  {K\"{o}dderitzsch}, \citenamefont {Chadova},\ and\ \citenamefont
  {Ebert}}]{KCE15}%
  \BibitemOpen
  \bibfield  {author} {\bibinfo {author} {\bibfnamefont {D.}~\bibnamefont
  {K\"{o}dderitzsch}}, \bibinfo {author} {\bibfnamefont {K.}~\bibnamefont
  {Chadova}}, \ and\ \bibinfo {author} {\bibfnamefont {H.}~\bibnamefont
  {Ebert}},\ }\href {\doibase 10.1103/PhysRevB.92.184415} {\bibfield  {journal}
  {\bibinfo  {journal} {Phys. Rev. B}\ }\textbf {\bibinfo {volume} {92}},\
  \bibinfo {pages} {184415} (\bibinfo {year} {2015})}\BibitemShut {NoStop}%
\bibitem [{SPR()}]{SPRKKR}%
  \BibitemOpen
  \href {http://olymp.cup.uni-muenchen.de/ak/ebert/SPRKKR} {}\bibinfo
  {howpublished} {{\em The Munich SPR-KKR package},\newline
  \mbox{H.~Ebert~\textit{et~al.}} \newline
  \url{http://olymp.cup.uni-muenchen.de/ak/ebert/SPRKKR}}\BibitemShut {NoStop}%
\bibitem [{\citenamefont {Ebert}(1996)}]{Ebe96}%
  \BibitemOpen
  \bibfield  {author} {\bibinfo {author} {\bibfnamefont {H.}~\bibnamefont
  {Ebert}},\ }\href {\doibase 10.1088/0034-4885/59/12/003} {\bibfield
  {journal} {\bibinfo  {journal} {Rep. Prog. Phys.}\ }\textbf {\bibinfo
  {volume} {59}},\ \bibinfo {pages} {1665} (\bibinfo {year}
  {1996})}\BibitemShut {NoStop}%
\bibitem [{\citenamefont {St\v{r}eda}(1982)}]{Str82}%
  \BibitemOpen
  \bibfield  {author} {\bibinfo {author} {\bibfnamefont {P.}~\bibnamefont
  {St\v{r}eda}},\ }\href {http://stacks.iop.org/0022-3719/15/i=22/a=005}
  {\bibfield  {journal} {\bibinfo  {journal} {J. Phys. C: Solid State Phys.}\
  }\textbf {\bibinfo {volume} {15}},\ \bibinfo {pages} {L717} (\bibinfo {year}
  {1982})}\BibitemShut {NoStop}%
\bibitem [{VES()}]{VESTA}%
  \BibitemOpen
  \href@noop {} {}\bibinfo {howpublished} {This figure has been created using
  the software VESTA\cite{MI11}.}\BibitemShut {Stop}%
\bibitem [{\citenamefont {Zhang}\ \emph {et~al.}(2013)\citenamefont {Zhang},
  \citenamefont {Yan}, \citenamefont {Wu}, \citenamefont {K\"ubler},
  \citenamefont {Kreiner}, \citenamefont {Parkin},\ and\ \citenamefont
  {Felser}}]{ZYW+13}%
  \BibitemOpen
  \bibfield  {author} {\bibinfo {author} {\bibfnamefont {D.}~\bibnamefont
  {Zhang}}, \bibinfo {author} {\bibfnamefont {B.}~\bibnamefont {Yan}}, \bibinfo
  {author} {\bibfnamefont {S.-C.}\ \bibnamefont {Wu}}, \bibinfo {author}
  {\bibfnamefont {J.}~\bibnamefont {K\"ubler}}, \bibinfo {author}
  {\bibfnamefont {G.}~\bibnamefont {Kreiner}}, \bibinfo {author} {\bibfnamefont
  {S.~S.~P.}\ \bibnamefont {Parkin}}, \ and\ \bibinfo {author} {\bibfnamefont
  {C.}~\bibnamefont {Felser}},\ }\href
  {http://stacks.iop.org/0953-8984/25/i=20/a=206006} {\bibfield  {journal}
  {\bibinfo  {journal} {J. Phys.: Cond. Mat.}\ }\textbf {\bibinfo {volume}
  {25}},\ \bibinfo {pages} {206006} (\bibinfo {year} {2013})}\BibitemShut
  {NoStop}%
\bibitem [{ITC(2002)}]{ITCA}%
  \BibitemOpen
  \href@noop {} {\emph {\bibinfo {title} {International Tables for
  Crystallography, Volume A: Space Group Symmetry}}}\ (\bibinfo  {publisher}
  {Springer},\ \bibinfo {year} {2002})\BibitemShut {NoStop}%
\bibitem [{Note1()}]{Note1}%
  \BibitemOpen
  \bibinfo {note} {The sum over all sites is shown, which leaves only the
  z-component non-zero.}\BibitemShut {Stop}%
\bibitem [{\citenamefont {Erskine}\ and\ \citenamefont {Stern}(1975)}]{ES75}%
  \BibitemOpen
  \bibfield  {author} {\bibinfo {author} {\bibfnamefont {J.~L.}\ \bibnamefont
  {Erskine}}\ and\ \bibinfo {author} {\bibfnamefont {E.~A.}\ \bibnamefont
  {Stern}},\ }\href {\doibase 10.1103/PhysRevB.12.5016} {\bibfield  {journal}
  {\bibinfo  {journal} {Phys. Rev. B}\ }\textbf {\bibinfo {volume} {12}},\
  \bibinfo {pages} {5016} (\bibinfo {year} {1975})}\BibitemShut {NoStop}%
\bibitem [{\citenamefont {Thole}\ \emph {et~al.}(1992)\citenamefont {Thole},
  \citenamefont {Carra}, \citenamefont {Sette},\ and\ \citenamefont {van~der
  Laan}}]{TCSL92}%
  \BibitemOpen
  \bibfield  {author} {\bibinfo {author} {\bibfnamefont {B.~T.}\ \bibnamefont
  {Thole}}, \bibinfo {author} {\bibfnamefont {P.}~\bibnamefont {Carra}},
  \bibinfo {author} {\bibfnamefont {F.}~\bibnamefont {Sette}}, \ and\ \bibinfo
  {author} {\bibfnamefont {G.}~\bibnamefont {van~der Laan}},\ }\href {\doibase
  10.1103/PhysRevLett.68.1943} {\bibfield  {journal} {\bibinfo  {journal}
  {Phys. Rev. Lett.}\ }\textbf {\bibinfo {volume} {68}},\ \bibinfo {pages}
  {1943} (\bibinfo {year} {1992})}\BibitemShut {NoStop}%
\bibitem [{\citenamefont {Carra}\ \emph {et~al.}(1993)\citenamefont {Carra},
  \citenamefont {Thole}, \citenamefont {Altarelli},\ and\ \citenamefont
  {Wang}}]{CTAW93}%
  \BibitemOpen
  \bibfield  {author} {\bibinfo {author} {\bibfnamefont {P.}~\bibnamefont
  {Carra}}, \bibinfo {author} {\bibfnamefont {B.~T.}\ \bibnamefont {Thole}},
  \bibinfo {author} {\bibfnamefont {M.}~\bibnamefont {Altarelli}}, \ and\
  \bibinfo {author} {\bibfnamefont {X.}~\bibnamefont {Wang}},\ }\href {\doibase
  10.1103/PhysRevLett.70.694} {\bibfield  {journal} {\bibinfo  {journal} {Phys.
  Rev. Lett.}\ }\textbf {\bibinfo {volume} {70}},\ \bibinfo {pages} {694}
  (\bibinfo {year} {1993})}\BibitemShut {NoStop}%
\bibitem [{\citenamefont {Ebert}\ \emph {et~al.}(2015)\citenamefont {Ebert},
  \citenamefont {Mankovsky}, \citenamefont {Chadova}, \citenamefont {Polesya},
  \citenamefont {Min\'{a}r},\ and\ \citenamefont {K\"odderitzsch}}]{EMC+15}%
  \BibitemOpen
  \bibfield  {author} {\bibinfo {author} {\bibfnamefont {H.}~\bibnamefont
  {Ebert}}, \bibinfo {author} {\bibfnamefont {S.}~\bibnamefont {Mankovsky}},
  \bibinfo {author} {\bibfnamefont {K.}~\bibnamefont {Chadova}}, \bibinfo
  {author} {\bibfnamefont {S.}~\bibnamefont {Polesya}}, \bibinfo {author}
  {\bibfnamefont {J.}~\bibnamefont {Min\'{a}r}}, \ and\ \bibinfo {author}
  {\bibfnamefont {D.}~\bibnamefont {K\"odderitzsch}},\ }\href {\doibase
  http://dx.doi.org/10.1103/PhysRevB.91.165132} {\bibfield  {journal} {\bibinfo
   {journal} {Phys. Rev. B}\ }\textbf {\bibinfo {volume} {91}},\ \bibinfo
  {pages} {165132} (\bibinfo {year} {2015})}\BibitemShut {NoStop}%
\bibitem [{\citenamefont {Berry}(1984)}]{Ber84}%
  \BibitemOpen
  \bibfield  {author} {\bibinfo {author} {\bibfnamefont {M.~V.}\ \bibnamefont
  {Berry}},\ }\href {\doibase 10.1098/rspa.1984.0023} {\bibfield  {journal}
  {\bibinfo  {journal} {Proceedings of the Royal Society of London A:
  Mathematical, Physical and Engineering Sciences}\ }\textbf {\bibinfo {volume}
  {392}},\ \bibinfo {pages} {45} (\bibinfo {year} {1984})}\BibitemShut
  {NoStop}%
\bibitem [{\citenamefont {Xiao}\ \emph {et~al.}(2010)\citenamefont {Xiao},
  \citenamefont {Chang},\ and\ \citenamefont {Niu}}]{XCN10}%
  \BibitemOpen
  \bibfield  {author} {\bibinfo {author} {\bibfnamefont {D.}~\bibnamefont
  {Xiao}}, \bibinfo {author} {\bibfnamefont {M.-C.}\ \bibnamefont {Chang}}, \
  and\ \bibinfo {author} {\bibfnamefont {Q.}~\bibnamefont {Niu}},\ }\href
  {\doibase 10.1103/RevModPhys.82.1959} {\bibfield  {journal} {\bibinfo
  {journal} {Rev. Mod. Phys.}\ }\textbf {\bibinfo {volume} {82}},\ \bibinfo
  {pages} {1959} (\bibinfo {year} {2010})}\BibitemShut {NoStop}%
\bibitem [{\citenamefont {Nakazawa}\ and\ \citenamefont {Kohno}(2014)}]{NK14}%
  \BibitemOpen
  \bibfield  {author} {\bibinfo {author} {\bibfnamefont {K.}~\bibnamefont
  {Nakazawa}}\ and\ \bibinfo {author} {\bibfnamefont {H.}~\bibnamefont
  {Kohno}},\ }\href {\doibase 10.7566/JPSJ.83.073707} {\bibfield  {journal}
  {\bibinfo  {journal} {Journal of the Physical Society of Japan}\ }\textbf
  {\bibinfo {volume} {83}},\ \bibinfo {pages} {073707} (\bibinfo {year}
  {2014})}\BibitemShut {NoStop}%
\bibitem [{\citenamefont {Ishizuka}\ and\ \citenamefont
  {Nagaosa}(2018)}]{IN18}%
  \BibitemOpen
  \bibfield  {author} {\bibinfo {author} {\bibfnamefont {H.}~\bibnamefont
  {Ishizuka}}\ and\ \bibinfo {author} {\bibfnamefont {N.}~\bibnamefont
  {Nagaosa}},\ }\href {\doibase 10.1126/sciadv.aap9962} {\bibfield  {journal}
  {\bibinfo  {journal} {Sci. Adv.}\ }\textbf {\bibinfo {volume} {4}},\ \bibinfo
  {pages} {aap9962} (\bibinfo {year} {2018})}\BibitemShut {NoStop}%
\bibitem [{\citenamefont {Momma}\ and\ \citenamefont {Izumi}(2011)}]{MI11}%
  \BibitemOpen
  \bibfield  {author} {\bibinfo {author} {\bibfnamefont {K.}~\bibnamefont
  {Momma}}\ and\ \bibinfo {author} {\bibfnamefont {F.}~\bibnamefont {Izumi}},\
  }\href {\doibase 10.1107/S0021889811038970} {\bibfield  {journal} {\bibinfo
  {journal} {J. Appl. Crystallogr.}\ }\textbf {\bibinfo {volume} {44}},\
  \bibinfo {pages} {1272} (\bibinfo {year} {2011})}\BibitemShut {NoStop}%
\end{thebibliography}
\end{document}